\documentclass[a4paper,fleqn,usenatbib]{mnras}
\usepackage{aasmacros}
\usepackage{times}
\usepackage{amsmath}
\usepackage{graphicx}
\usepackage{amssymb}
\usepackage[usenames,dvipsnames]{color}
\usepackage{url}
\usepackage{fixltx2e}
\usepackage{mathtools}
\usepackage{amsmath}
\usepackage{float}
\usepackage{cite}
\usepackage{fixltx2e}
\bibpunct{(}{)}{;}{a}{,}{,}
\usepackage{longtable}
\usepackage{hyperref}
\usepackage{morefloats}
\usepackage{gensymb}
\usepackage{enumitem}
\usepackage{tabu}
\usepackage{totpages}
\usepackage{amssymb}
\usepackage{xcolor}
\usepackage[T1]{fontenc}
\usepackage{ae,aecompl}
\usepackage{microtype}
\usepackage{multicol}
\usepackage{placeins}
\usepackage{siunitx}

\let\cite=\citen

\def\apj{ApJ}
\def\aj{AJ}
\def\mnras{MNRAS}
\def\pasa{PASA}

\def\pasp{PASP}

\bibpunct{(}{)}{;}{a}{}{,}

\def\nue{$\nu_{\rm e}$}

\def\lame{$\lambda_{\rm e}$}
\def\lamo{$\lambda_{\rm obs}$}

\newcommand\Tstrut{\rule{0pt}{2.9ex}}         % "top" strut
\newcommand\Bstrut{\rule[-1.2ex]{0pt}{0pt}}   % "bottom" strut
\newcommand\TBstrut{\Tstrut\Bstrut}  

\newcommand{\rowfont}[1]{% Set current row font
   \gdef\rowfonttype{#1}#1%
}

\makeatletter
\newcommand*{\rom}[1]{\expandafter\@slowromancap\romannumeral #1@}
\makeatother

\title[High-$z$ radio polarization]{Radio Polarization Properties of Quasars and Active Galaxies at High Redshifts}

\author[Vernstrom et. al]{T. Vernstrom\thanks{E-mail:vernstrom@dunlap.utoronto.ca}$^1$, B.M. Gaensler$^{1}$, V. Vacca$^2$, J.S. Farnes$^{3}$,\newauthor
M. Haverkorn$^3$, S.P. O'Sullivan$^{4}$ \\
$^1$Dunlap Institute for Astronomy and Astrophysics University of Toronto, Toronto, ON M5S 3H4, Canada\\
$^2$INAF, Osservatorio Astronomico di Cagliari, Via della Scienza 5, 09047 Selargius, Italy\\
$^3$Department of Astrophysics/IMAPP, Radboud University, PO Box 9010, NL-6500 GL Nijmegen, The Netherlands\\
$^4$Hamburger Sternwarte, Universit\"{a}t Hamburg, Gojenbergsweg 112, 21029, Hamburg, Germany\\
 }
\begin{document}

\pagerange{\pageref{firstpage}--\pageref{lastpage}} \pubyear{2016}

\maketitle

\label{firstpage}
\begin{abstract}
We present the largest ever sample of radio polarization properties for $z>4$ sources, with 14 sources having significant polarization detections. Using wideband data from the Karl G. Jansky Very Large Array, we obtained the rest-frame total intensity and polarization properties of 37 radio sources, nine of which have spectroscopic redshifts in the range $1 \le z \le 1.4$, with the other 28 having spectroscopic redshifts in the range $3.5 \le z \le 6.21$. Fits are performed for the Stokes $I$ and fractional polarization spectra, and Faraday rotation measures are derived using Rotation measure synthesis and $QU$ fitting. Using archival data of 476 polarized sources, we compare high redshift ($z>3$) source properties to a $15\,$GHz rest-frame luminosity matched sample of low redshift ($z<3$) sources to investigate if the polarization properties of radio sources at high redshifts are intrinsically different than those at low redshift. We find a mean of the rotation measure absolute values, corrected for Galactic rotation, of $50 \pm 22\,$rad m$^{-2}$ for $z>3$ sources and $57 \pm 4\,$rad m$^{-2}$ for $z<3$. Although there is some indication of lower intrinsic rotation measures at high-$z$ possibly due to higher depolarization from the high density environments, using several statistical tests we detect no significant difference between low and high redshift sources. Larger samples are necessary to determine any true physical difference. 

\end{abstract}


\section{Introduction}
\label{sec:introduction}

Where do magnetic fields come from? What are their strengths in the early Universe? How do they evolve? These are just a few of the unanswered questions regarding cosmic magnetism \citep{Widrow12}. Interstellar and intergalactic magnetic fields at earlier epochs have important implications for the feedback of magnetic energy into the intergalactic medium \citep[IGM,][]{Kronberg01} and for galaxy and large-scale structure evolution \citep{Mestel84,Rees87,Urry95}. At redshifts of $z>2$, typical radio-loud quasars are located in dense environments, where AGN host galaxies are the most massive systems. The study of these systems at early times is necessary for answering open questions on cosmic magnetism and its role in galaxy evolution. 

The Faraday rotation effect is one of the most powerful techniques to detect and probe extragalactic magnetic fields \citep[e.g.][]{Carilli02, Govoni04}. For a source at redshift $z_s$, the rotation measure (RM) is defined as 
\begin{equation}
{\rm RM}(z_s)= 0.81 \, \int_{z_s}^0 \frac{n_{e}(z) B_{||}(z)}{(1+z)^2} \, \frac{dl}{dz} \, dz \,\, {\rm rad} \,\, {\rm m}^{-2},
\label{eq:rm1}
\end{equation}
where $n_e$ is the thermal gas density in cm$^{-3}$ and $B_{||}$ is the magnetic field strength along the line of sight at redshift $z$ in $\mu$Gauss. The RM is a measure of the change in polarization angle $\chi$ with respect to the change in $\lambda^2$ due to a magnetized medium.

We expect high density environments to compress magnetic fields and increase turbulence. If distant quasars reside in such environments, this would lead to higher intrinsic polarized fractions, but also higher depolarization at high $z$. This relationship is complicated due to intervening Faraday screens and could lead to complex spectral energy distributions that require $k$-correction of the polarized fraction \citep{Farnes141}. Recent work has shown that with new broadband radio data and new methods for measuring rotation measures and the Faraday spectrum such as RM synthesis \citep{Brentjens05} and $QU$-fitting \citep[e.g][]{O'Sullivan12,O'Sullivan17,Anderson16} a more detailed and in-depth analysis of polarization properties versus cosmic time is possible. 

It is typically expected that the measured Faraday rotation, or RM, should appear to decrease at early times due to $k$-corrections of the emission at high $z$ (i.e. the $[1+z]^2$ term in the denominator of eq.~\ref{eq:rm1}). To date there is limited information on polarized fraction or Faraday rotation for high-$z$ radio-loud quasars. Current quasar polarization detections have typically only extended out to redshifts of $z\sim 3.5$ \citep{Hammond12}, with only a few having $z>4$ \citep[e.g.][]{O'Sullivan11}. Earlier attempts have been made to detect a $z$-dependence of quasar RMs \citep{Rees72,Kronberg76} using RM data on samples out to $z\simeq1$ and more recently with $z \lessapprox3$ \citep{Kronberg08}, which found evidence for an increase in the observed RM at higher redshifts. However, \citet{Hammond12} and \citet{Bernet12} did not observe this evolution out to redshifts of $z\sim3.5$, and neither was it seen by \citet{Farnes141}, who looked along lines of sight that contain no known intervening objects. \citet{Lamee16} used 222 sources in the redshift range 0 $<$ z $<$ 2.3 and found a weak negative correlation of depolarization with redshift from steep spectrum, depolarized sources.  

These previous studies all only included a handful of sources between $3 \le z \le4$, and one or two at $z>4$ \citep[e.g. ][]{Carilli94,Athreya98,Broderick07a}. The goal of this work is to investigate the rotation measure and polarization fraction properties of a new and larger sample of high-$z$ ($z>3$) sources using new broadband data and  Faraday depth tools. The statistics can then be compared to a low redshift source sample to look for any differences or information about the evolution of cosmic magnetism versus time. In this paper we look at new wideband data from low and high redshift radio sources and analyze the polarization properties. 

In Section~\ref{sec:data} we describe the observation, data reduction, and imaging of 37 new sources, as well as the details on archival data used for a control sample. Section~\ref{sec:fits} details the fitting of the Stokes $I$ and fractional polarization spectra and the rotation measure synthesis and $QU$ fitting. Section~\ref{sec:results} presents the results of the fitting, as well as statistical comparisons of low and high redshift sources. In Section~\ref{sec:discussion} we discuss the results including non-detections and how the results compare to previous findings. Throughout the paper, we assume a concordance cosmology with $H_0=67.3 \,$km s$^{-1}$ Mpc$^{-1}$, $\Omega_m= 0.315$, and $\Omega_{\Lambda}=0.685$ \citep{Planck14a}.

\section{Data}
\label{sec:data}

\subsection{New data}
\label{sec:highzdat}

The high redshift sample of sources was selected from the \citet{Kimball08} catalogue, which cross-matched radio sources from the NVSS \citep{Condon98} and FIRST \citep{Becker95} surveys and with the {\it Sloan Digital Sky Survey} \citep[SDSS,][]{York00} DR6. We selected sources with spectroscopic redshifts $z\ge4$, $1.4\,$GHz flux density $\ge 5\,$mJy, and off the Galactic plane ($|b|>20\degr$). This resulted in a list of 50 high redshift sources. However, after the observations were taken, an updated redshift catalogue was released \citep{Kimball14}, which used updated SDSS data (DR9) for the redshifts. This revealed that 12 of the selected sources had updated redshift values that were less than four ($0.35 \le z \le 3$). Checking with the  Set of Identifications, Measurements and Bibliography for Astronomical Data database\citep[SIMBAD,][]{Wenger00}\footnote{\url{http://simbad.u-strasbg.fr/simbad/}} and the NASA/IPAC Extragalactic Database \citep[NED,][]{Helou91,Helou95}\footnote{\url{http://ned.ipac.caltech.edu/}} showed some at the higher redshift estimates and some at the lower. Since the newer (and lower) estimates came from the newer SDSS release with updated redshift flags, we use the lower values in the following analysis.

We requested observations with the Karl G. Jansky Very Large Array (VLA) for the initial 50 sources and were granted time for 30 of the sources (of which nine had new redshift estimates less than four). We were able to find archival VLA observations for eight additional sources with $z>3$ (with the archive search limited to data since the wide bandwidth upgrade of the VLA), bringing the total to 38 sources (although one of the additional sources is actually two components of the same source). All of the sources are classified as optical quasars according to the Million Optical Radio/X-ray Associations (MORX) catalogue \citep{Flesch16} and The Million Quasars catalog \citep{Flesch15}. The details of the observations and data reduction are discussed below. 

\begin{table*}
\caption{Details of the VLA observations for the 38 observed and processed components of sources. The $1.4\,$GHz flux density values $S_{1.4}$ are from NVSS. The $t_{\rm obs}$ column lists the total on-source time in each observed frequency band. }
\label{tab:vlaobs}
\begin{tabu}{lrrrrccccc}
\hline 
 \multicolumn{1}{c}{Name}  & \multicolumn{1}{c}{RA} & \multicolumn{1}{c}{Dec} & \multicolumn{1}{c}{$z$} & \multicolumn{1}{c}{$S_{1.4}$} & \multicolumn{1}{c}{Date} & \multicolumn{1}{c}{Config} & \multicolumn{1}{c}{Frequency} & \multicolumn{1}{c}{$t_{\rm obs}$}& \multicolumn{1}{c}{Project}  \Tstrut \\
  \rowfont{\tiny}
 & \multicolumn{1}{c}{(J2000)} & \multicolumn{1}{c}{(J2000)} &  & \multicolumn{1}{c}{[mJy]} & MM/YYYY&  & \multicolumn{1}{c}{[GHz]} & \multicolumn{1}{c}{[sec]} & \TBstrut \\
 \hline
J001115+144603 & 00:11:15.34 & +14:46:03.60 &$ 4.97 $&$ 36.0 $& 07/2014 & D & 1-2; 2-4 & 169.6; 1483.9 & 14A-255 \\
J003126+150738 & 00:31:26.79 & +15:07:38.60 &$ 4.29 $&$ 42.0 $& 07/2014 & D & 1-2; 2-4 & 139.6; 1364.4 & 14A-255 \\
J021042$-$001818 & 02:10:43.15 & $-$00:18:18.14 &$ 4.73 $&$ 9.9 $& 07/2014 & D & 1-2; 2-4 & 204.4; 1549.8 & 14A-255 \\
J081333+350812 & 08:13:33.11 & +35:08:12.92 &$ 4.95 $&$ 36.0 $& 07/2014 & D & 2--4 & 1256.4 & 14A-255 \\
J083644+005451 & 08:36:43.90 & +00:54:53.00 &$ 5.77 $&$ 1.1 $& 10/2016 & A & 1--2 & 10858 & 16B-009 \\
J083946+511202 & 08:39:46.20 & +51:12:02.88 &$ 4.40 $&$ 43.0 $& 07/2014 & D & 2--4 & 1256.4 & 14A-255 \\
J085111+142338 & 08:51:11.58 & +14:23:37.86 &$ 4.18 $&$ 12.0 $& 07/2014 & D & 2--4 & 1256.4 & 14A-255 \\
J085853+345826 & 08:58:53.60 & +34:58:26.62 &$ 1.34 $&$ 22.0 $& 07/2014 & D & 2--4 & 1256.4 & 14A-255 \\
J090600+574730 & 09:06:00.06 & +57:47:30.62 &$ 1.34 $&$ 30.0 $& 07/2014 & D & 2--4 & 1256.4 & 14A-255 \\
J091316+591920 & 09:13:16.54 & +59:19:21.61 &$ 5.12 $&$ 18.0 $& 07/2014 & D & 2--4 & 1166.9 & 14A-255 \\
J091824+063653 & 09:18:24.39 & +06:36:53.32 &$ 4.16 $&$ 31.0 $& 07/2014 & D & 2--4 & 1077.1 & 14A-255 \\
J100424+122924 & 10:04:24.87 & +12:29:22.38 &$ 4.52 $&$ 12.0 $& 07/2014 & D & 2--4 & 1256.4 & 14A-255 \\
J100645+462716 & 10:06:45.60 & +46:27:17.42 &$ 4.34 $&$ 6.4 $& 07/2014 & D & 2--4 & 1256.4 & 14A-255 \\
J102551+192314 & 10:25:51.34 & +19:23:13.45 &$ 1.17 $&$ 43.0 $& 07/2014 & D & 2--4 & 1256.4 & 14A-255 \\
J102623+254259 & 10:26:23.62 & +25:42:59.65 &$ 5.27 $&$ 260.0 $& 07/2014 & D & 2--4 & 1525.8 & 14A-255 \\
J103601+500831 & 10:36:01.03 & +50:08:31.78 &$ 4.50 $&$ 11.0 $& 07/2014 & D & 2--4 & 1256.4 & 14A-255 \\
J104624+590524a & 10:46:23.97 & +59:06:06.82 &$ 3.63 $&$ 0.5 $& 03/2012 & C & 2--4 & 36000 & 12A-032 \\
J104624+590524b & 10:46:24.78 & +59:04:45.30 &$ 3.63 $&$ 10.0 $& 03/2012 & C & 2--4 & 36000 & 12A-032 \\
J105320$-$001650 & 10:53:20.43 & $-$00:16:49.58 &$ 4.30 $&$ 9.3 $& 07/2014 & D & 2--4 & 1256.4 & 14A-255 \\
J130738+150752 & 13:07:38.94 & +15:07:58.46 &$ 4.08 $&$ 16.0 $& 07/2014 & D & 2--4 & 1166.5 & 14A-255 \\
J130940+573311 & 13:09:40.70 & +57:33:10.04 &$ 4.28 $&$ 11.0 $& 07/2014 & D & 2--4 & 1256.4 & 14A-255 \\
J132512+112330 & 13:25:12.48 & +11:23:30.01 &$ 4.42 $&$ 81.0 $& 07/2014 & D & 2--4 & 1256.4 & 14A-255 \\
J133342+491625 & 13:33:43.27 & +49:16:23.93 &$ 1.39 $&$ 33.0 $& 07/2014 & D & 2--4 & 1555.8 & 14A-255 \\
J135135+284015 & 13:51:35.69 & +28:40:15.06 &$ 4.73 $&$ 6.1 $& 07/2014 & D & 2--4 & 1256.4 & 14A-255 \\
J142738+331242 & 14:27:38.50 & +33:12:41.00 &$ 6.12 $&$ 1.8 $& 10/2016 & A & 1--2 & 3886 & 16B-009 \\
J142952+544717 & 14:29:52.20 & +54:47:17.99 &$ 6.21 $&$ 3.0 $& 10/2016 & A & 1--2 & 1494 & 16B-009 \\
J151002+570243 & 15:10:02.96 & +57:02:43.62 &$ 4.31 $&$ 200.0 $& 05/2012 & B & 2--4 & 165 & 12A-404 \\
J155633+351757 & 15:56:33.77 & +35:17:57.62 &$ 4.67 $&$ 28.0 $& 01/2013 & D & 1--2; 4.4-6.2 & 269.2; 89.9 & 13A-114 \\
J161105+084437 & 16:11:05.66 & +08:44:35.38 &$ 4.55 $&$ 8.7 $& 10/2012 & A & 1-2; 4.4-6.3 & 93; 209 & 12B-361 \\
J165913+210116 & 16:59:13.24 & +21:01:15.74 &$ 4.89 $&$ 29.0 $& 10/2012 & A & 1-2; 4.4-6.4 & 120; 239 & 12B-361 \\
J221356$-$002457 & 22:13:56.05 & $-$00:24:56.99 &$ 1.06 $&$ 110.0 $& 07/2014 & D & 1--2; 2--2 & 428.2; 2162 & 14A-255 \\
J222032+002535 & 22:20:32.60 & +00:25:35.87 &$ 4.21 $&$ 89.0 $& 07/2014 & D & 1--2; 2--1 & 139.6; 1364.4 & 14A-255 \\
J222235+001536 & 22:22:35.88 & +00:15:36.54 &$ 1.36 $&$ 68.0 $& 07/2014 & D & 1--2; 2--0 & 139.6; 1364.4 & 14A-255 \\
J222843+011032 & 22:28:43.50 & +01:10:32.00 &$ 5.95 $&$ 0.3 $& 10/2016 & A & 1--2 & 8694 & 16B-009 \\
J224924+004750 & 22:49:23.99 & +00:47:52.04 &$ 4.48 $&$ 18.0 $& 07/2014 & D & 1--2; 2--1 & 149.6; 1364.4 & 14A-255 \\
J231443$-$090637 & 23:14:43.21 & $-$09:06:31.54 &$ 1.29 $&$ 5.5 $& 07/2014 & D & 1--2; 2--2 & 159.6; 1364.4 & 14A-255 \\
J232604+001333 & 23:26:04.68 & +00:13:34.39 &$ 1.00 $&$ 9.6 $& 07/2014 & D & 1--2; 2--3 & 169.6; 2321.6 & 14A-255 \\
J235018$-$000658 & 23:50:18.69 & $-$00:06:57.35 &$ 1.36 $&$ 250.0 $& 07/2014 & D & 1--2; 2--4 & 149.6; 1364.4 & 14A-255 \\
\hline
\end{tabu}
\end{table*}

\subsubsection{Observations and calibration}
\label{sec:obs}
Our observations were performed July through September of 2014 in the VLA's D configuration (project code 14A-255). We requested both L (1--$2\,$GHz) and S (2--$4\,$GHz) bands for all of the sources. However, due to limited time on-source only 10 of the 30 sources from the original proposal had any L-band observations taken, and even for those sources only a fraction of the requested L-band time was observed. The details of each observed source are listed in Table~\ref{tab:vlaobs}, including those for the eight sources that were found in archival VLA data (see the project code column), which have different configurations and/or frequency bands.\footnote{Source J104624+590524 is a multi-component source with two lobes. Even though it is only one source its components, labelled ``a" and ``b", are analyzed separately, and thus when referring to the number of sources what is really meant is components.}

All of the data reduction and imaging were done using the Common Astronomy Software Applications (\textsc{CASA}) package.\footnote{\url{http://casa.NRAO.edu/}} The sources 3C286 or 3C138 were observed as primary flux, bandpass, and polarization calibrators, with 3C147 and OQ208 observed as polarization leakage calibrators. The VLA data is separated into 16 sub-bands across each frequency band, with 64 frequency channels per sub-band. Hanning smoothing was performed to suppress Gibbs ringing, and automated radio frequency interference (RFI) detection algorithms were used to flag areas of strong interference. Unfortunately, strong RFI, mainly from satellites, affected several sub-bands requiring the need to flag them entirely, at least in the D-configuration data. These included sub-bands 2, 3, 15, and 16 (mean frequencies 2.15, 2.28, 3.79, and 3.91 GHz, respectively) in the S-band data and 1, 2, 3, 9, and 10 in the L-band data (mean frequencies 1.09, 1.15, 1.22, 1.54 and 1.60 GHz, respectively). After the calibration was applied, additional flagging for each source was performed as needed. Unfortunately one source's data were entirely flagged (source J094224+010858), bringing the number of sources down to 37 (38 components).

\subsubsection{Imaging}
\label{sec:imaging}

For each source, several rounds of phase-only self calibration were performed, cleaning progressively deeper and with shorter calibration intervals with each round. The number of self calibration rounds for each source was determined by examining the image residuals and rms noise after each round. 

In order to investigate the spectral and polarization properties of the sources, image cubes were made for the three Stokes parameters $I$, $Q$, and $U$. The software \textsc{WSCLEAN} \citep[version 2.3,][]{Offringa14} was used for all the imaging and deconvolution. \textsc{WSCLEAN} was used rather than \textsc{CASA} (which has been more traditionally used) for several reasons. First, it has been shown that \textsc{WSCLEAN} performs faster than and may outperform \textsc{CASA} \citep{Offringa17}. Second, and more importantly, \textsc{WSCLEAN} performs polarimetric deconvolution in a more proper way than \textsc{CASA}. The issue of proper treatment for the complex nature of linear polarization deconvolution is discussed in detail by \citet{Pratley16}. Basically, \textsc{CASA} searches for peaks in Stokes $I$ and total polarization $\sqrt{Q^2+U^2+V^2}$ simultaneously, producing individual clean components in each polarization. This approach is designed to constrain peaks so as to select the most highly polarized components associated with a Stokes $I$ peak. \textsc{WSCLEAN}, however, allows for searching of peaks in the sum of squares of $Q^2+U^2$ independent of $I$. Also, \textsc{WSCLEAN} can search for peaks in the sum of squares of the images, rather than the integrated bandwidth, ensuring values with high RM values will not average out.

The image cubes were made by averaging together a set number of spectral channels from the datasets. The number of averaged channels varied  between 10 and 20 channels for S and C-band ($2\,$MHz channels) and 20 to 30 channels in L band ($1\,$MHz channels). The number chosen differed by source and depended on the amount and overall quality of the data and the signal to noise of the source. 

The Stokes $I$ images were deconvolved, or cleaned, separately from the Stokes $Q$ and $U$ images. Peaks were searched for across all spectral images, but then deconvolved separately for each spectral image. The Stokes $Q$ and $U$ data were imaged together with peaks being searched for in the combined polarization $Q^2+U^2$ domain.

\begin{figure}
\includegraphics[scale=.38]{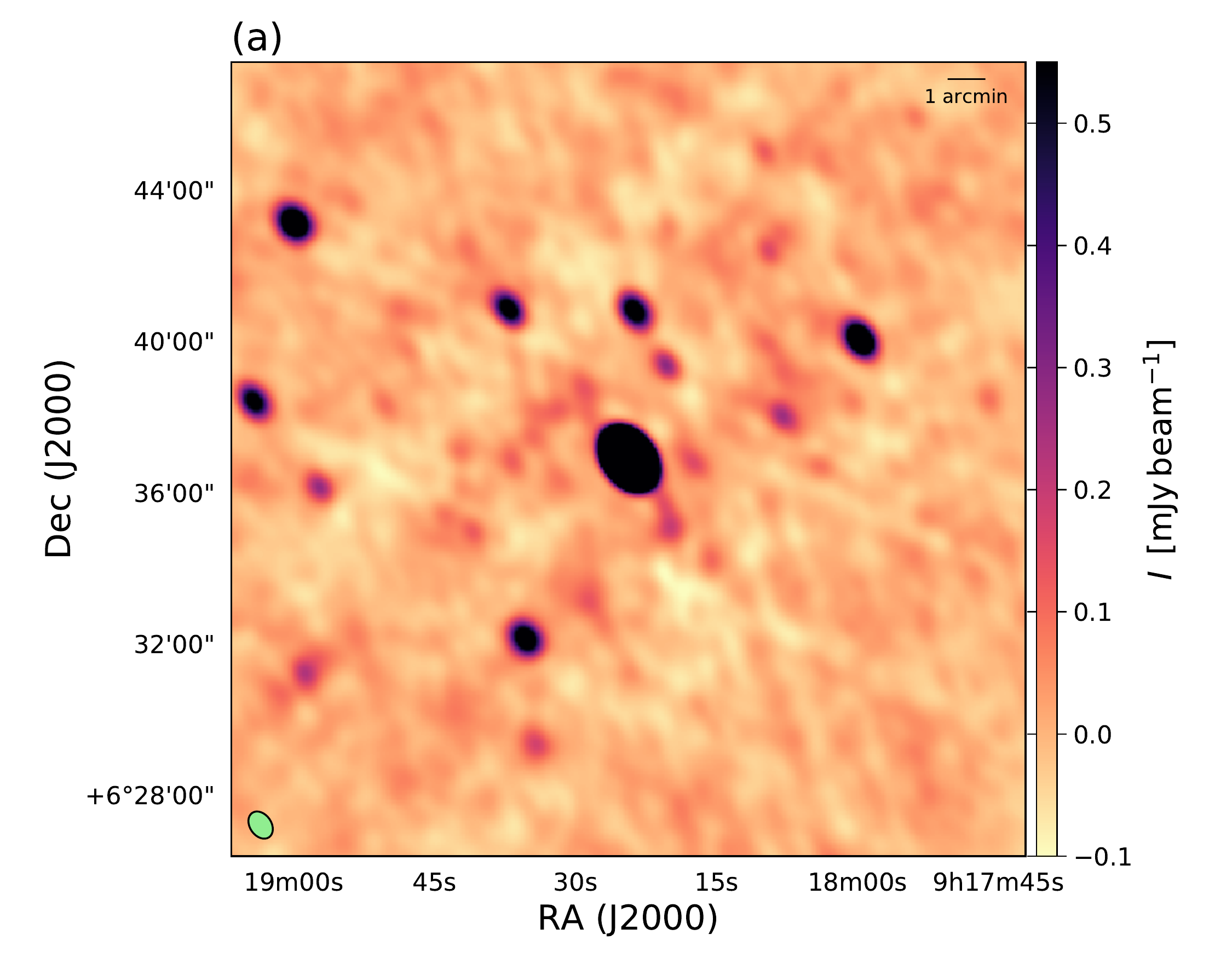}
\includegraphics[scale=.38]{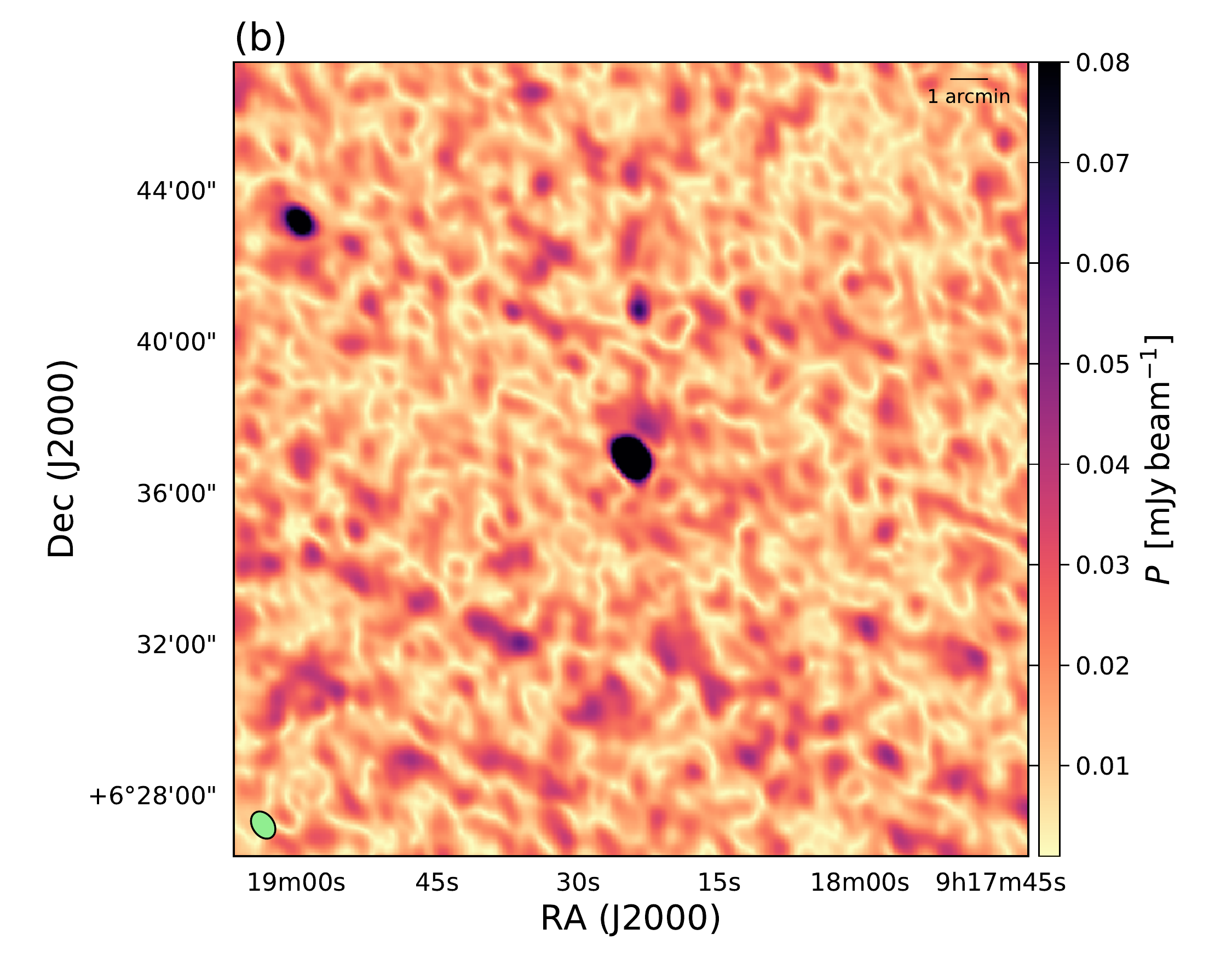}
\caption{Images for source J091824+063653. Panel (a) shows the Stokes $I$ image at the weighted average frequency of $3.09\,$GHz, while panel (b) shows the weighted average polarized intensity, with the average taken over the individual frequency channels. }
\label{fig:implot}
\end{figure}

The resulting image cubes for each source were inspected and any spectral channels which showed large artefacts or a large increase in noise compared to the average were excluded. The number of excluded images varied depending on the source but, was generally only one or two images per source. These excluded channels were usually near the sub-bands that were completely flagged for RFI meaning there may have been some RFI that the automatic flagging routines missed. All of the images in each cube were then convolved to a common resolution, or synthesized beam size; with $B_{\rm maj}$, $B_{\rm min}$, and $B_{\rm PA}$ for the major and minor axis full width half maximum (FWHM) sizes and the position angle, generally matching, or encompassing, that of the lowest frequency image. Table~\ref{tab:vlaims} lists the image details for each source, while Fig.~\ref{fig:implot} shows the Stokes $I$ and polarized intensity weighted average images for one source (J091824+063653). 

The instrumental noise values for the sources were measured in source free regions of the images in the outer parts of the primary beam. The signal to noise ratio in Stokes $I$, S$_I$/N$_I$, for the sources ranges from $\sim5$ to $800$, with a median value of 68. 

All of the sources, with the exception of J104624+590524, were unresolved. The source J104624+590524 is resolved into 3 components, two lobes and a core. The core is too faint for polarization detection but each of the lobes are examined separately, noted with the a and b distinctions, and with the positions set to the locations of the Stokes $I$ peaks for each lobe.

\begin{table*}
 \setlength{\tabcolsep}{4.05pt}
%\scriptsize
\caption{Imaging details for the 38 observed and processed sources. Here $N_{\rm im}$ is the number of spectral images used for each source, and $B_{\rm maj}$, $B_{\rm min}$, and $B_{\rm PA}$ are the common clean synthesized beam major and minor axes FWHM and position angle of the clean synthesized beam. The $I$, $Q$, and $U$ brightnesses and rms values listed are the median values from all the spectral images.}
\label{tab:vlaims}
\begin{tabu}{lcrrrrrrrrr}
\hline \\[-2.25ex]
\multicolumn{1}{c}{Name}  & \multicolumn{1}{c}{$N_{\rm im}$}& \multicolumn{1}{c}{$B_{\rm maj}$}& \multicolumn{1}{c}{$B_{\rm min}$}& \multicolumn{1}{c}{$B_{\rm pa}$}& \multicolumn{1}{c}{$\overline{{I}}$}&\multicolumn{1}{c}{ $\overline{Q}$}& \multicolumn{1}{c}{$\overline{U}$}&\multicolumn{1}{c}{$\overline{\sigma_I}$}&\multicolumn{1}{c}{$\overline{\sigma_Q}$}&\multicolumn{1}{c}{$\overline{\sigma_U}$}\\
 \rowfont{\tiny}
  &  & [arcsec]& [arcsec]& [deg] &[mJy beam$^{-1}$]&[mJy beam$^{-1}$]&[mJy beam$^{-1}$]&[mJy beam$^{-1}$]&[mJy beam$^{-1}$]&[mJy beam$^{-1}$] \TBstrut\\
 \hline
J001115+144603 & 40 &$ 66.0 $&$ 47.0 $&$ 40 $&$ 24.00 $&$ -0.48 $&$ -0.05 $&$ 0.15 $&$ 0.09 $&$ 0.08 $\\
J003126+150738 & 59 &$ 50.0 $&$ 40.0 $&$ 50 $&$ 68.00 $&$ 0.03 $&$ 0.15 $&$ 0.16 $&$ 0.07 $&$ 0.08 $\\
J021042$-$001818 & 64 &$ 65.0 $&$ 45.0 $&$ -9 $&$ 9.80 $&$ -0.32 $&$ 0.07 $&$ 0.23 $&$ 0.08 $&$ 0.09 $\\
J081333+350812 & 56 &$ 48.0 $&$ 28.0 $&$ 84 $&$ 19.00 $&$ -0.60 $&$ 0.95 $&$ 0.28 $&$ 0.13 $&$ 0.12 $\\
J083644+005451 & 80 &$ 2.2 $&$ 1.6 $&$ -1 $&$ 1.20 $&$ -0.03 $&$ 0.01 $&$ 0.11 $&$ 0.04 $&$ 0.04 $\\
J083946+511202 & 55 &$ 61.0 $&$ 47.0 $&$ -4 $&$ 54.00 $&$ 0.69 $&$ -0.36 $&$ 0.34 $&$ 0.10 $&$ 0.09 $\\
J085111+142338 & 39 &$ 45.0 $&$ 30.0 $&$ 44 $&$ 5.10 $&$ -0.04 $&$ -0.03 $&$ 0.21 $&$ 0.10 $&$ 0.11 $\\
J085853+345826 & 35 &$ 36.0 $&$ 26.0 $&$ -70 $&$ 9.40 $&$ -0.01 $&$ -0.01 $&$ 0.12 $&$ 0.06 $&$ 0.06 $\\
J090600+574730 & 31 &$ 55.0 $&$ 40.0 $&$ -74 $&$ 14.00 $&$ -0.53 $&$ -0.61 $&$ 0.17 $&$ 0.05 $&$ 0.05 $\\
J091316+591920 & 45 &$ 54.0 $&$ 39.0 $&$ -79 $&$ 10.00 $&$ 0.02 $&$ -0.01 $&$ 0.14 $&$ 0.06 $&$ 0.06 $\\
J091824+063653 & 53 &$ 47.0 $&$ 34.0 $&$ 34 $&$ 46.00 $&$ -0.43 $&$ -0.20 $&$ 0.12 $&$ 0.07 $&$ 0.08 $\\
J100424+122924 & 26 &$ 56.0 $&$ 30.0 $&$ 42 $&$ 9.70 $&$ 0.01 $&$ 0.03 $&$ 0.14 $&$ 0.06 $&$ 0.06 $\\
J100645+462716 & 29 &$ 45.0 $&$ 36.0 $&$ -63 $&$ 9.50 $&$ 0.04 $&$ 0.01 $&$ 0.15 $&$ 0.06 $&$ 0.06 $\\
J102551+192314 & 44 &$ 45.0 $&$ 39.0 $&$ 43 $&$ 25.00 $&$ -0.29 $&$ -0.82 $&$ 0.22 $&$ 0.06 $&$ 0.07 $\\
J102623+254259 & 62 &$ 40.0 $&$ 37.0 $&$ -10 $&$ 150.00 $&$ 11.00 $&$ 6.50 $&$ 0.18 $&$ 0.07 $&$ 0.07 $\\
J103601+500831 & 24 &$ 65.0 $&$ 50.0 $&$ -63 $&$ 6.10 $&$ 0.01 $&$ -0.01 $&$ 0.44 $&$ 0.13 $&$ 0.13 $\\
J104624+590524a & 31 &$ 16.0 $&$ 14.0 $&$ 69 $&$ 0.55 $&$ -0.04 $&$ 0.01 $&$ 0.04 $&$ 0.02 $&$ 0.02 $\\
J104624+590524b & 31 &$ 16.0 $&$ 14.0 $&$ 69 $&$ 7.10 $&$ -0.42 $&$ -0.25 $&$ 0.04 $&$ 0.02 $&$ 0.03 $\\
J105320$-$001650 & 29 &$ 55.0 $&$ 36.0 $&$ 18 $&$ 8.50 $&$ 0.01 $&$ 0.01 $&$ 0.22 $&$ 0.05 $&$ 0.05 $\\
J130738+150752 & 39 &$ 35.0 $&$ 33.0 $&$ -20 $&$ 6.50 $&$ 0.05 $&$ -0.25 $&$ 0.10 $&$ 0.05 $&$ 0.06 $\\
J130940+573311 & 26 &$ 44.0 $&$ 34.0 $&$ -26 $&$ 11.00 $&$ 0.01 $&$ -0.04 $&$ 0.13 $&$ 0.06 $&$ 0.05 $\\
J132512+112330 & 38 &$ 44.0 $&$ 34.0 $&$ -20 $&$ 56.00 $&$ 0.06 $&$ 0.36 $&$ 0.16 $&$ 0.08 $&$ 0.09 $\\
J133342+491625 & 29 &$ 50.0 $&$ 45.0 $&$ -39 $&$ 19.00 $&$ 0.77 $&$ 0.77 $&$ 0.23 $&$ 0.08 $&$ 0.09 $\\
J135135+284015 & 27 &$ 55.0 $&$ 36.0 $&$ -49 $&$ 1.60 $&$ 0.01 $&$ 0.01 $&$ 0.24 $&$ 0.13 $&$ 0.12 $\\
J142738+331242 & 48 &$ 3.0 $&$ 2.5 $&$ 60 $&$ 1.50 $&$ -0.01 $&$ -0.01 $&$ 0.37 $&$ 0.06 $&$ 0.06 $\\
J142952+544717 & 39 &$ 2.7 $&$ 2.5 $&$ 80 $&$ 3.20 $&$ -0.01 $&$ -0.02 $&$ 0.73 $&$ 0.11 $&$ 0.11 $\\
J151002+570243 & 26 &$ 14.0 $&$ 7.0 $&$ -79 $&$ 250.00 $&$ -7.10 $&$ -2.70 $&$ 0.61 $&$ 0.16 $&$ 0.17 $\\
J155633+351757 & 27 &$ 25.0 $&$ 21.0 $&$ 37 $&$ 22.00 $&$ -0.08 $&$ 2.20 $&$ 0.15 $&$ 0.12 $&$ 0.12 $\\
J161105+084437 & 65 &$ 2.0 $&$ 1.5 $&$ 29 $&$ 13.00 $&$ -0.07 $&$ 0.14 $&$ 0.60 $&$ 0.30 $&$ 0.27 $\\
J165913+210116 & 45 &$ 1.7 $&$ 1.4 $&$ 29 $&$ 13.00 $&$ -0.17 $&$ 0.08 $&$ 0.66 $&$ 0.25 $&$ 0.25 $\\
J221356$-$002457 & 42 &$ 52.0 $&$ 44.0 $&$ 79 $&$ 55.00 $&$ 0.29 $&$ -0.09 $&$ 0.18 $&$ 0.09 $&$ 0.09 $\\
J222032+002535 & 100 &$ 143.0 $&$ 65.0 $&$ 49 $&$ 54.00 $&$ -1.90 $&$ 3.00 $&$ 0.64 $&$ 0.16 $&$ 0.21 $\\
J222235+001536 & 52 &$ 150.0 $&$ 75.0 $&$ 49 $&$ 51.00 $&$ -1.60 $&$ -2.20 $&$ 0.63 $&$ 0.20 $&$ 0.19 $\\
J222843+011032 & 16 &$ 4.2 $&$ 1.9 $&$ 45 $&$ 0.28 $&$ 0.02 $&$ -0.01 $&$ 0.05 $&$ 0.04 $&$ 0.04 $\\
J224924+004750 & 48 &$ 93.8 $&$ 32.7 $&$ 49 $&$ 9.30 $&$ -0.70 $&$ 0.50 $&$ 0.26 $&$ 0.10 $&$ 0.10 $\\
J231443$-$090637 & 43 &$ 105.0 $&$ 50.0 $&$ 34 $&$ 3.60 $&$ -0.03 $&$ 0.03 $&$ 0.27 $&$ 0.11 $&$ 0.11 $\\
J232604+001333 & 68 &$ 79.0 $&$ 50.0 $&$ 39 $&$ 5.80 $&$ 0.01 $&$ -0.02 $&$ 0.21 $&$ 0.11 $&$ 0.10 $\\
J235018$-$000658 & 81 &$ 90.0 $&$ 63.0 $&$ 24 $&$ 130.00 $&$ 1.50 $&$ 3.40 $&$ 0.37 $&$ 0.16 $&$ 0.30 $\\
\hline
\end{tabu}
\normalsize
\end{table*}

\subsection{Archival data}
\label{sec:control}
We need a sample of sources at lower redshifts to compare to our high redshift sample. We decided to create the control sample based on sources with similar rest frame luminosities. To create the control sample, we need sources with redshifts and Stokes $I$, as well as polarization data available at multiple wavelengths, in order to fit for or find the rest-frame luminosities as well as the polarization properties (fractional polarization and rotation measure). For this we used the catalogues of \citet{Klein03} and \citet{Farnes14}. \citet{Klein03} provided Stokes $I$ and polarization data for sources from the B3-VLA sample at $1.4$, $2.7$, $4.8$, and $10.5\,$GHz, as well as redshift and rotation measure information. \citet{Farnes14} provided data at multiple frequencies for sources from a variety of surveys such as AT20G \citep{Murphy10}, GB6 \citep{Gregory96}, NORTH6CM \citep{Becker91}, Texas \citep{Douglas96}, and WENSS \citep{Rengelink97}. 

For both catalogues we initially looked for additional redshifts for any sources where a redshift was not provided using the Simbad and NED databases. We also cut out any sources which only had data at two or less frequencies or had a $|$Galactic latitude$|< 20\degr$, as well as eliminating any duplicates. If a duplicate for the new sample was found the archival data for that source was not included, and if there was a duplicate between the Klein and Farnes catalogs the one providing more frequency points per source was kept. This left us with $502$ sources from \citet{Farnes14} and $33$ sources from \citet{Klein03} for a total of 535 sources.

A list of all the archival sources used is given in Appendix~\ref{sec:appendA}. There are a total of four sources from these two catalogues which have $z>3$, which are thus included in the high redshift sample for analysis. The sample of low $z$ ($z<3$) archival sources is used as a control sample by defining by a rest-frame luminosity range, which is described in more detail in Section~\ref{sec:stksi}.

\section{Source Properties}
\label{sec:fits}

\subsection{Stokes $I$ spectrum}
\label{sec:stksi}

For the majority of sources a simple power-law model was used of the form
\begin{equation}
I(\nu) \, = \,k \left ( \frac{\nu}{1 \, {\rm GHz}} \right )^{\alpha_1},
\label{eq:stksi1}
\end{equation}
where $k$ is a constant with units of mJy, $\alpha_1$ is the spectral index, and $\nu$ is in GHz, with $k$ and $\alpha_1$ being solved for in the fitting process. However, for sources where there was a clear turnover in the spectrum a broken power-law model was used,
\begin{equation}
\begin{split}
I(\nu<\nu_{\rm peak})&  =  k \left ( \frac{\nu}{ 1 \, {\rm GHz}} \right )^{\alpha_2} \\
I(\nu>\nu_{\rm peak}) & =  k \, \nu_{\rm peak}^{ -\alpha_1+\alpha_2}\, \left ( \frac{\nu}{1 \, {\rm GHz}} \right )^{\alpha_1},
\end{split}
\label{eq:stksi2}
\end{equation}
where $\nu_{\rm peak}$ is the peak of the spectrum in GHz and $k$, $\alpha_1$, $\alpha_2$, and $\nu_{\rm peak}$ are all solved for in the fitting. The fitting was performed for all of our observed sources as well as all of the sources from the control sample, except for control sources from \citet{Farnes14} as the fitted parameters for those sources were provided. The fitting results for our sample are given in Table~\ref{tab:stksif}. 

The resulting fitted models are used to find the Stokes I flux density at the chosen rest-frame frequency \nue$=15\,$GHz, or rather the observed frequency that translates to the rest-frame frequency such that $\nu_{\rm obs} (1+z) = \nu_{\rm e}$. The value of $15\,$GHz was chosen as it lies within (or just outside of) the rest-frame frequencies for all of our new high-$z$ sources, and the corresponding $\nu_{\rm obs}$ for the archival sources is generally within or near the observed range of frequencies of each source; only 190 of the 502 sources needed extrapolation to $\nu_{\rm obs}$, the rest had at least one frequency point on either side of the $15\,$GHz rest-frame frequency. 

\begin{figure}
\includegraphics[scale=0.365]{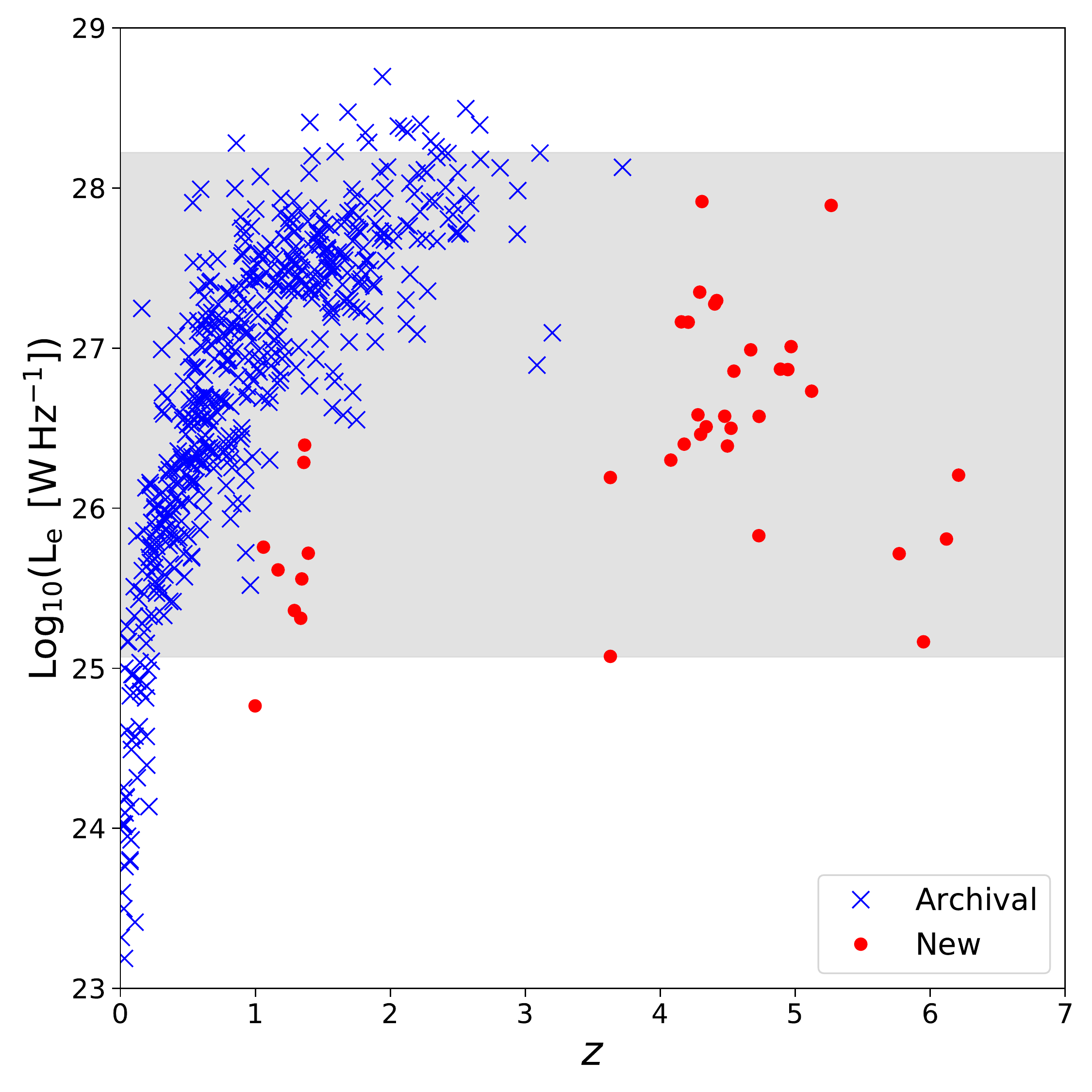}
\caption{Rest-frame $15\,$GHz luminosity vs redshift for the control and new data samples. The grey region shows the range of luminosities considered for the analysis. The blue crosses are the archival data (with those inside the grey region defined as the control sample) and the red circles are the new data presented in this paper. }
\label{fig:lumz}
\end{figure}

We used the rest frame flux density from the Stokes $I$ fitting to compute the rest-frame luminosities such that
\begin{equation}
L_{{\rm e}}=\frac{I(\nu_{\rm e}) \, 4 \, \pi D_{L}^2}{(1+z)},
\label{eq:lum}
\end{equation}
where $I(\nu_{\rm e})$ is the flux density in W m$^{-2}$ Hz$^{-1}$ and $D_{L}$ is the luminosity distance. It has been shown that the polarization properties of low-luminosity sources appear to be different than for high-luminosity sources \citep[e.g.][]{Pshirkov15}. Therefore, we used the minimum and maximum values of the luminosity for the high-$z$ sample to define the luminosity limits of the control sample. The luminosity limits are $25.1 \le \log_{10}[L_{\rm e}] \le 28.3$. This cut the archival sample size from 535 sources to 476.  Figure~\ref{fig:lumz} shows the luminosities vs. redshift for both samples. 

\subsection{Fractional polarization spectrum}
\label{sec:stksp}

The fractional polarization is defined as $\Pi=P/I$, where $I$ are the Stokes $I$ fitted, or model, vaules. In order to fit the fractional polarization spectral energy distribution (SED) and obtain values for the fractional polarization at the rest-frame frequency, $\Pi_{\rm e}$, we follow the method laid out by \citet{Farnes14}. \citet{Farnes14} opted to use non-physical models that allow for smoothly interpolating fits that are functionally similar to physical models. Three models were fit to the data as a function of $\lambda$: a power-law model, a Gaussian model, and an offset Gaussian model. These models are given by
\begin{equation}
\Pi(\lambda) = c_1 \left ( \frac{\lambda}{1 \, {\rm cm}} \right )^{\beta},
\label{eq:pfrac1}
\end{equation} 
\begin{equation}
\Pi(\lambda)  = c_2 e^{-(c_3-\lambda)^2/(2 c_4^2)},
\label{eq:pfrac2}
\end{equation}
and
\begin{equation}
\Pi(\lambda)  = c_2 e^{-(c_3-\lambda)^2/(2 c_4^2)}+c_5.
\label{eq:pfrac3}
\end{equation}
Here $c_1$, $c_2$, and $c_5$ are dimensionless constants, $\beta$ is a polarization spectral index, $c_3$ is the peak wavelength in cm and $c_4$ is the Gaussian width, also in cm. In this case $c_1$, $c_2$, $c_3$, $c_4$, $c_5$, and $\beta$ are all solved in the fitting. 

All three models are fit for each source, with the model having the lowest $\chi^2$ per degrees of freedom being chosen as the best fitting model. Again, the fitting was performed for all (new and archival) sources, with the \citet{Farnes14} sources having the fitted parameters provided. The fitting was performed on the debiased polarization values (for details on the debiasing procedure used see Appendix~\ref{sec:debias}). 

We chose to adopt these non-physical models for several reasons. First, given that the majority of the archival sources are from \citet{Farnes14}, using the same models means the fit parameters for these sources are already provided. Also, as stated by \citet{Farnes14}, the physical models all assume an optically-thin emitting region, which may not be the case. Additionally, all of the physical models make either the critical assumption that the polarized fraction is a meaningful quantity with the measured peak in polarized intensity on the sky coming from the same region as the total intensity peak, or that we detect the same emitting region at each frequency. By choosing non-physical models that have similarities to the physical models, we avoid such assumptions. 

With the right constants, the Gaussian model of eq.~(\ref{eq:pfrac2}) has a very similar wavelength-dependence of fractional polarization to a `Burn' law \citep{Burn66}, or to a `Spectral Depolarizer' \citep{Conway74}. The power-law model is akin to a `Tribble' law \citep{Tribble91}, and is able to fit a `repolarizer' \citep{Homan02, Mantovani09, Hovatta12}. The offset Gaussian of eq.~(\ref{eq:pfrac3}) is similar to the `Rossetti--Mantovani' law \citep{Rossetti08, Mantovani09}. For more description and discussion on the physical models and laws mentioned here see Appendix~\ref{sec:append1}.

The best-fitting models were used to find the polarization fraction at the chosen rest-frame wavelength \lame$=2\,$cm (\nue$=15\,$GHz), i.e. at the observed wavelength that translates to the rest-frame wavelength such that \lamo$/(1+z) =\lambda_{\rm e}$. The fitting results for our sample are given in Table~\ref{tab:stkspf}. 

\subsection{Rotation measures}
\label{sec:rms}
Faraday rotation causes a change to the intrinsic polarization angle $\chi_0$ by an amount that depends on the wavelength of the radiation such that after Faraday rotation
\begin{equation}
\chi\,=\,\chi_0+\phi\,\lambda^2,
\label{eq:RM0}
\end{equation}
where $\lambda$ is the wavelength, $\chi$ is the observed polarization angle, and $\phi$ is known as the Faraday depth.
The values of $\chi$ can be found from
\begin{equation}
\chi=\frac{1}{2}\tan^{-1}\left ( \frac{U}{Q} \right).
\label{eq:chii}
\end{equation}

The value of $\phi$, is related to the properties of the Faraday rotating plasma (at $z=0$) by the equation
\begin{equation}
\phi (L) \, = \, 0.81 \int_{\mathrm{L}}^{\mathrm{telescope}}n_e\,B_{\parallel}\,dl \,\,\, {\rm rad} \,\,\, {\rm m}^{-2},
\label{eq:RM1}
\end{equation}
where $n_e$ (cm$^{-3}$) is the thermal electron density, $B_{\parallel}$ ($\mu$G) is the magnetic field, and $l$ (pc) is the distance along the line of sight (LOS). It is only the magnetic field component parallel to the LOS ($B_{||}$) that contributes. Equation~\ref{eq:RM1} for the Faraday depth is similar to eq.~\ref{eq:rm1} for the rotation measure RM, but the Faraday depth at which all polarized emission is produced is equal to the RM if there is only one emitting source along the line of sight, which has no internal Faraday rotation, and is not affected by beam depolarization \citep[for further detail, see][]{Brentjens05}.

\subsubsection{RM synthesis}
\label{sec:rmsynth}

If the polarization vector is expressed as an exponential ($P=p\,e^{2i\chi}$), and eq.~(\ref{eq:RM0}) is used for $\chi$, when integrating over all possible Faraday depths, \citet{Burn66} showed that
\begin{equation}
P(\lambda^2)=\int_{-\infty}^{+\infty}F(\phi)\,e^{2i\phi\lambda^2}\,d\phi,
\label{eq:phi}
\end{equation}
where $P(\lambda^2)$ is the (complex) observed polarization vector and $F(\phi)$ is the Faraday dispersion function (FDF), which describes the {\em intrinsic} polarization vector at each Faraday depth.

RM synthesis \citep{Brentjens05} is a technique for calculating $F(\phi)$ directly from observations of $P(\lambda^2)$ using a Fourier transform-like equation. The rotation measure response function (RMSF), similar to the synthesized image beam, is determined by the total bandwidth, or $\Delta\lambda^2$, with the full width at half maximum (FWHM) of the RMSF ,$\Phi$, given by
 \begin{equation}
 \Phi\simeq \frac{2 \sqrt{3}}{\Delta \lambda^2}. 
 \label{eq:rmsf}
 \end{equation}
 The RMSF is used in a procedure called \textsc{RMCLEAN} \citep{Heald09}. \textsc{RMCLEAN} is similar to interferometric imaging cleaning and deconvolution, where a peak is found in the dirty FDF and a percentage of the peak multiplied by the dirty RMSF is iteratively subtracted with the final clean components convolved with the clean RMSF and added back to the residual FDF. \textsc{RMCLEAN} is applied to $F(\phi)$ to remove artefacts caused by the $\lambda^2$ sampling. 

For the RM synthesis and cleaning we used a pipeline (Purcell et al., in prep) being developed for use with the future Australian SKA Pathfinder (ASKAP) polarization Sky Survey of the Universe's Magnetism \citep[POSSUM;][]{Gaensler10} survey.\footnote{\url{https://github.com/crpurcell/RM-tools}} The $I$, $Q$, and $U$ images are read in for each source. The pipeline measures the noise in each spectral and polarization image as well as the mean flux density in a small region around the source position. The polarization vector is created from measuring the $Q$ and $U$ intensities of the source, which is then transformed to $\phi$ space. 

Either uniform or variance based weighting can be applied in the transformation. The polarization vector noise $\sigma_{QU}$ is taken as  $(\sigma_Q+\sigma_U)/2$ for each frequency. Since it is unknown exactly the effect of the different weighting schemes we performed RM synthesis and cleaning on each source using both types of weighting; with the uniform weighted results hereafter referred to as ``no-wt" and variance weighted referred to as ``sd-wt". The type of weighting does not seem to have a large impact on the results, with the largest effect being seen for those sources with more than one frequency band and the two bands have largely different noise measurements (e.g. L-band and S-band data where the L-band data consists of significantly less time).

RM cleaning is then performed down to a $5\sigma_{\rm FDF}$ level, where
\begin{equation}
\sigma_{\rm FDF}=\left ( {\frac{1}{\sum{\sigma_{QU}^{-2}}}} \right )^{1/2}.
\label{eq:sigfdf}
\end{equation}
Once a peak in the FDF is detected, the region around the peak is oversampled and the position of the peak (the RM) and the peak value ($A$) are fit for, with these values being reported in Table~\ref{tab:rmsynf}. The uncertainties in the fit parameters are given by
\begin{equation}
\Delta A \propto \frac{\sqrt{\delta \phi} \, {\sigma_{FDF}}}{\sqrt{\Phi}},
\label{eq:daa}
\end{equation}
and
\begin{equation}
\Delta {\rm RM}\propto  \frac{\sqrt{\Phi \, \delta \phi} \, {\sigma_{FDF}} }{A}, 
\label{eq:drm}
\end{equation}
where $\delta \phi$ is the $\phi$ spacing. The expressions for the uncertainties come from the derived Gaussian fitting parameter uncertainties \citep[e.g.][]{Landman82}. The signal-to-noise ratio (S/N) is then defined as $A/{\sigma_{FDF}}$.

\begin{figure}
\includegraphics[scale=0.365]{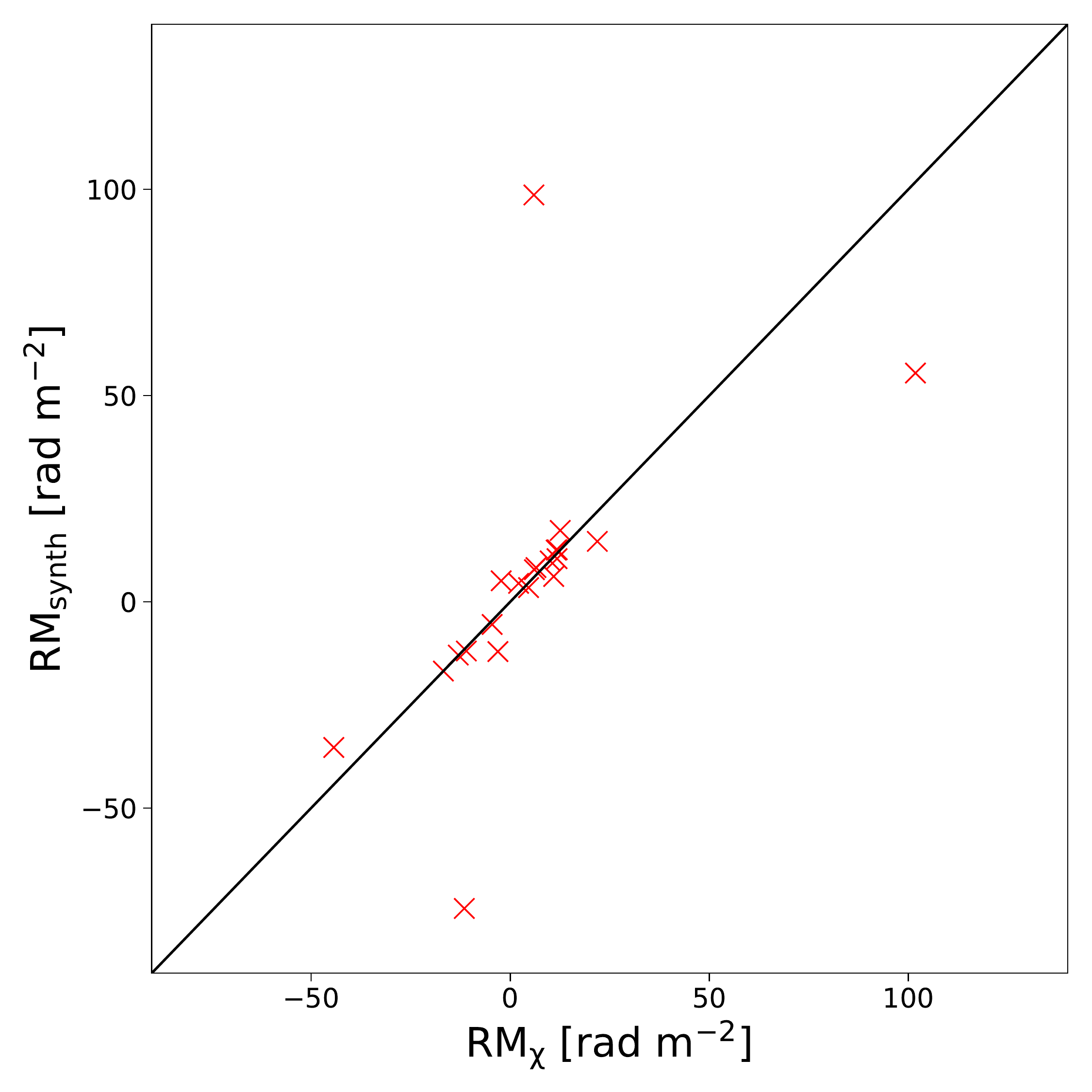}
\caption{Rotation measures measured from RM synthesis compared to those from fitting the slope of $\chi$ vs $\lambda^2$ for those sources with a RM synthesis detection. The black line shows a one-to-one correspondence. }
\label{fig:rmvsrm}
\end{figure}

The RM synthesis was only performed on the new sample of sources presented here, not for any of the archival sample sources (for the archival source RM values see Appendix~\ref{sec:appendA}). Since the archival source RMs were obtained by fitting the $\chi$ vs $\lambda^2$ slope, rather than RM synthesis, we also fit for the $\chi$ slope of each of the new sources. The RM obtained via this method is designated RM$_{\chi}$ and is also reported in Table~\ref{tab:rmsynf}. The median ratio of RM$_{\chi}$ to RM$_{\rm synth}$ is 1.08 for those sources with a detection from the RM synthesis, with the two plotted against each other in Fig.~\ref{fig:rmvsrm}.

\subsubsection{$QU$ fitting}
\label{sec:qufit}
Another way to fit the Faraday dispersion function is $QU$ fitting. In this approach the FDF is forward modelled by assuming a $F(\phi)$ model, transforming it to the complex vector $\vec{P}(\lambda^2)$ and fitting the Stokes $Q$ (real part) and $U$ data (imaginary part) directly; as opposed to RM synthesis which transforms $\vec{P}(\lambda^2)$ directly to $F(\phi)$. 

The advantages to this approach are that one is able to fit more complex or specific models to the FDF. RM synthesis will often fail to find the underlying Faraday structure, even in the simple case of two components \citep[e.g.][]{Farnsworth11, O'Sullivan12}. $QU$ fitting allows for multiple RMs or a range of RMs to be found, rather than simply finding the RM as the peak $\phi$, as in the RM synthesis case. 

The disadvantages of $QU$ fitting compared to RM synthesis are that the results are model dependent and that there may be degeneracies between different models. Also, the case of multiple or complex components can lead to more than one RM value, which can make things more difficult to interpret when looking at group statistics, for example \citep[see e.g.][ for more discussion on the differences of QU fitting and RM synthesis]{Farnsworth11,Sun15}. 

\begin{figure}
\includegraphics[width=3.285in,height=3.285in]{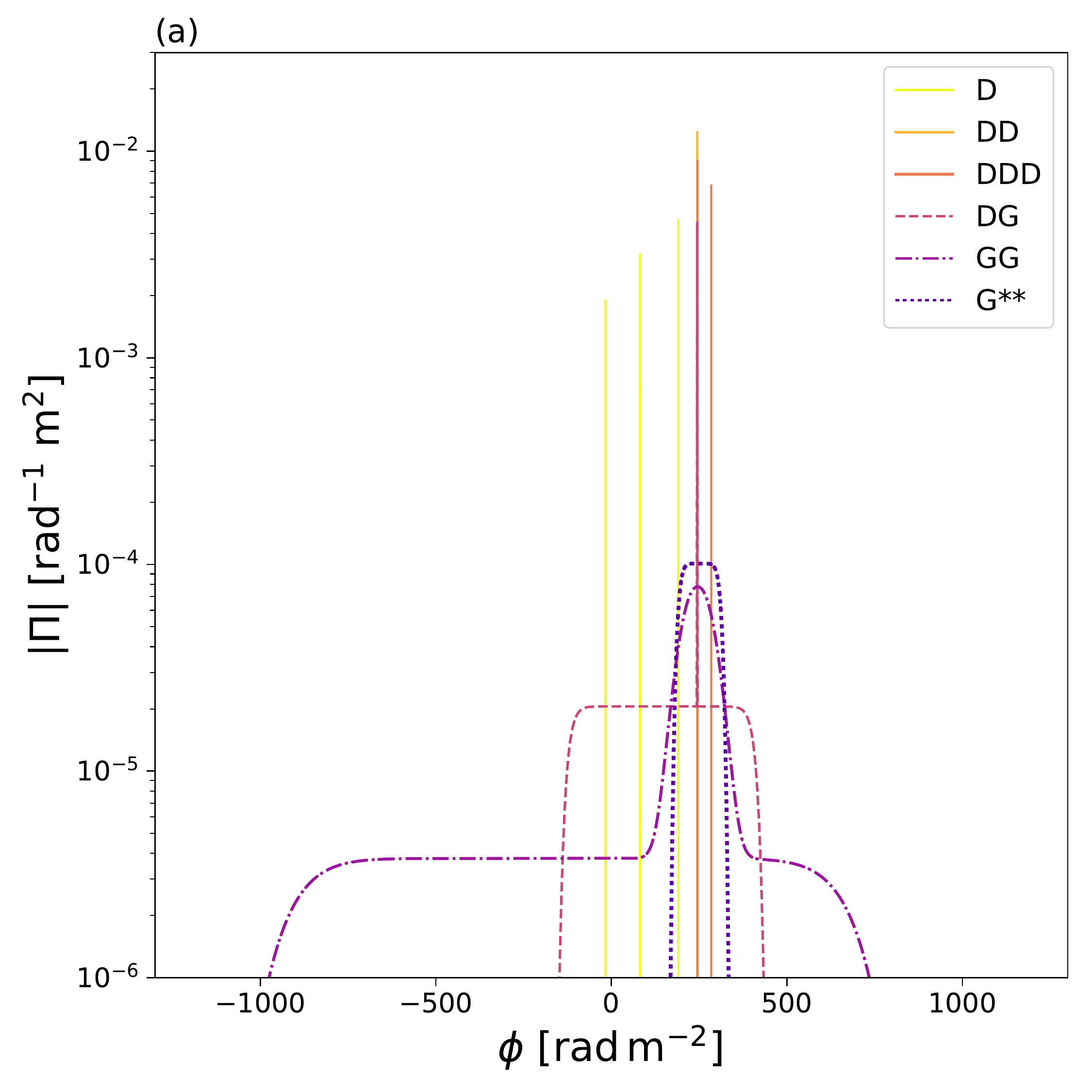}
\includegraphics[width=3.285in,height=3.285in]{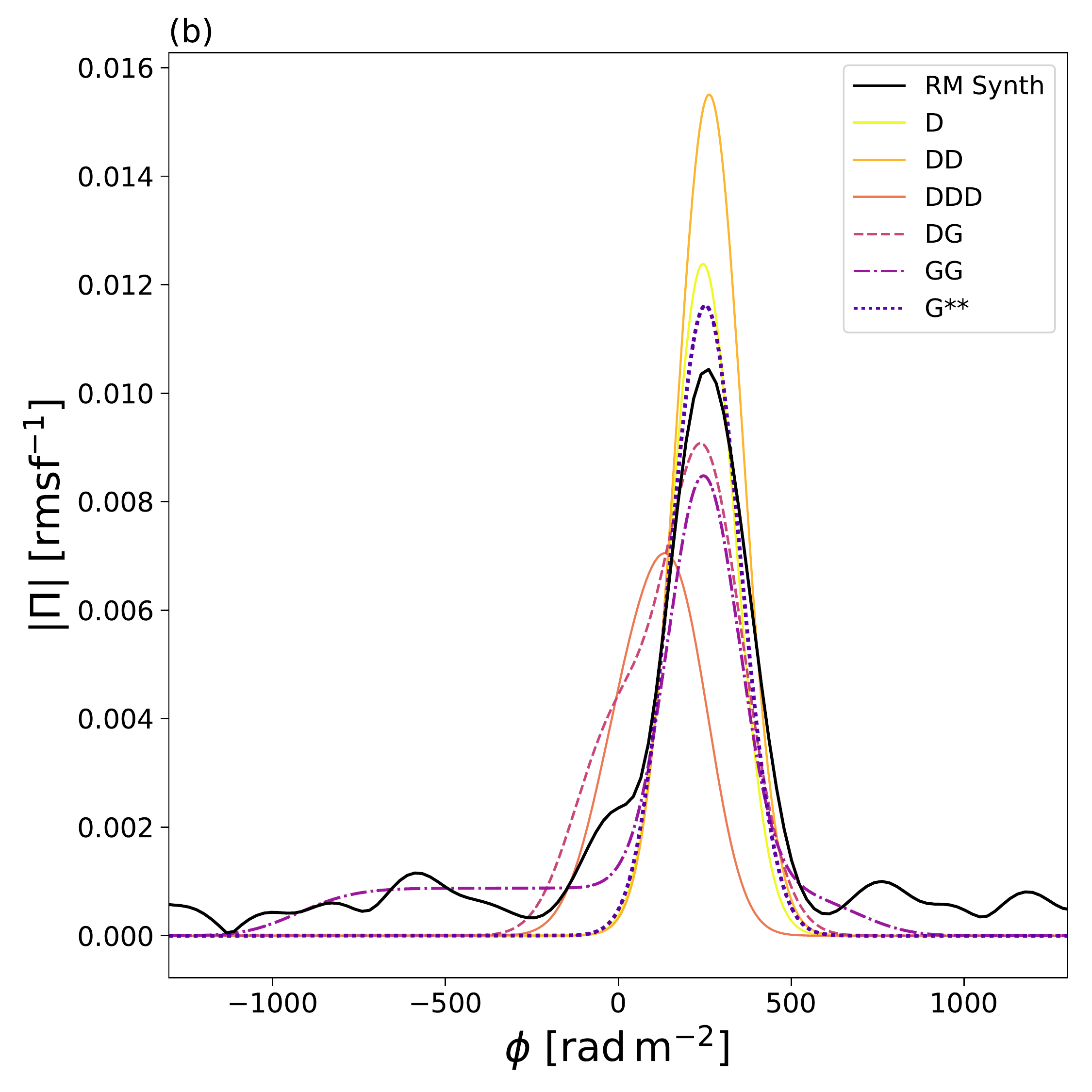}
\caption{$QU$-fitting absolute value Faraday dispersion functions for source J091824+063653. Both panels show the lowest $\chi^2$ result for each of the six models, where ``D" represents a delta function, or thin component, and ``G" is a modified Gaussian, or thick component. The number of letters in each model is the number of thin and thick components. The model marked $**$ (in this case G) is the best-fit of the six, i.e. the chosen model with the lowest $\chi^2$ per degrees of freedom. Panel (b) is the same as panel (a) except convolved with the RMSF from the RM synthesis and plotted against the RM synthesis output (solid black line).}
\label{fig:qufdf}
\end{figure}

We decided to do $QU$ fitting in addition to RM synthesis in order to compare the results from the two methods and look for more complex Faraday structure.\footnote{The $QU$ fitting was only performed on the new sample of sources presented here, not for any of the archival sample sources. } For models we followed the example of \citet{Anderson16}, where a $\delta$ function in $\phi$ space is used to represent a Faraday ``thin" component with polarization angle $\psi_0$ and modulus $p$ at a Faraday depth of $\phi_0$, with these components denoted as ``D" for delta function. A modified, or super, Gaussian is used for a Faraday ``thick" component with the form
\begin{equation}
{\bf p}(\phi)=-\frac{p}{\sqrt{2 \pi} \sigma_{\phi}} {\rm exp} \left ( 2i\psi_0 + \frac{1}{2} \left [ \frac{- | \phi - \phi_{0}|}{ \sigma_{\phi}} \right ]^N \right ),
\label{eq:qugaus}
\end{equation}
where $p$ is the peak, $\phi_0$ is the Faraday depth, $\sigma_{\phi}$ is the width, $\psi_0$ is the polarization angle, and $N$ is the shape parameter which controls the deviation from a standard Gaussian function. The thick components are denoted as ``G" for Gaussian.

We fit models consisting of one, two, and three thin components (``D", ``DD", and ``DDD"), a thin and thick component (``DG"), two thick components (``GG"), and one thick component (``G"). All six models were fit to our source sample using Monte Carlo Markov Chains (MCMC) to find the  parameters and uncertainties for each model and source (10,000 steps in a chain per model per source). For each source the parameters that yielded the lowest $\chi^2$ for each model were used to compute the best-fitting values for each model. Then those six models were compared and the one with the lowest $\chi^2$ per degrees of freedom was chosen as the reported best-fit model and parameters. Figure~\ref{fig:qufdf} compares all six best-fit models for one source.

For the sources with RM S/N$<8$ from the RM synthesis, the S/N was too low for accurate QU fitting as well. The more complex models were not converging, and all of these sources have a single thin component, or delta function, listed as the chosen model. 

\section{Results}
\label{sec:results}

\subsection{Fitting results}
\label{sec:result}

\begin{figure*}
\includegraphics[scale=0.42]{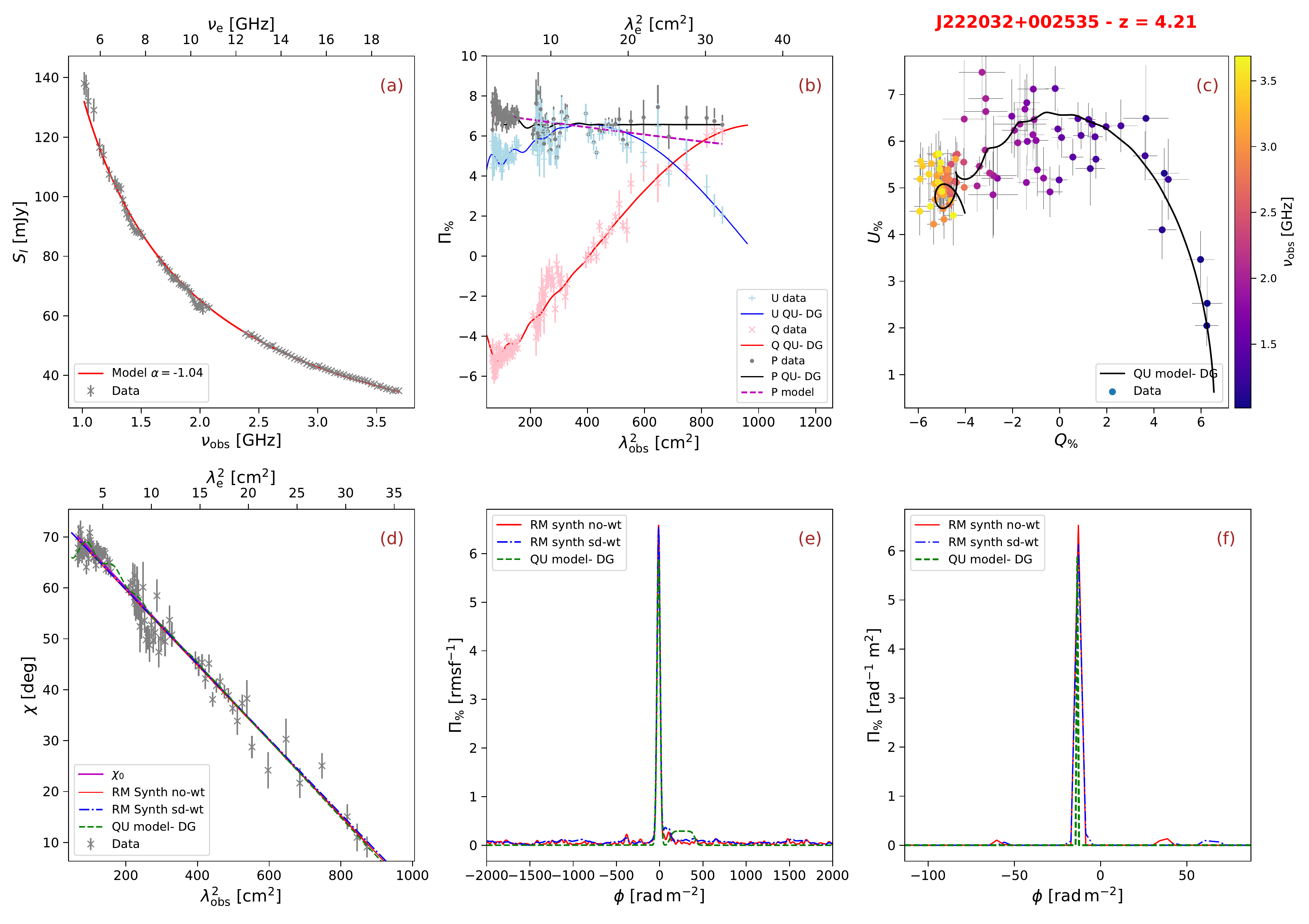}
\caption{Spectra and Faraday dispersion functions for source J222032+002535. Panel (a) is the Stokes $I$ spectrum vs frequency, with grey points showing the data and the red solid line showing the fitted model from eq.(\ref{eq:stksi1}) or (\ref{eq:stksi2}). Panel (b) shows the fractional Stokes $Q$ (red) and $U$ (blue) and $P$ (gray) spectra vs $\lambda^2$, with the points showing the image data and the lines showing the best-fit models from the $QU$ fitting, the dashed purple line shows the best fit $P$ model from eq.(\ref{eq:pfrac1}), (\ref{eq:pfrac2}), or (\ref{eq:pfrac3}). Panel (c) shows the fractional Stokes $Q$ vs $U$ with the solid line being the best-fit $QU$-fitting model and the colouring of the points showing the frequency. Panel (d) shows the polarization angle vs $\lambda^2$, with the red solid and blue dot-dashed lines from the RM synthesis (eq.~\ref{eq:RM0} and \ref{eq:chii}), the green dashed line from the QU fitting, and the purple solid line from the slope fitting. Panel (e)  shows the fractional (cleaned) absolute value Faraday dispersion functions for ``no-wt" (red solid) and ``sd-wt" (blue dot-dashed) weighting and the $QU$-fitting (green dashed line) convolved with a Gaussian with width equal to the mean of the two RMSFs. Panel (f) shows the un-convolved (fractional) absolute value Faraday dispersion functions (or \textsc{RMCLEAN} clean components) from the ``no-wt" RM synthesis (red solid), ``sd-wt" (blue dot-dashed), and $QU$ fitting (green dashed). }
\label{fig:explot1}
\end{figure*}

\begin{table*}
\caption{Fitted and derived parameters for the new sources. Here $I_{\rm e}$ is the rest-frame stokes $I$ flux density, $\alpha_1$ is the spectra index, $\Pi_{\rm e}$ is the rest-frame polarization fraction, $L_{\rm e}$ is the rest-frame luminosity in units of W m$^{-2}$ Hz$^{-1}$, RM$_{\rm no-wt}$ and RM$_{\rm sd-wt}$ are the RM synthesis rotation measures from uniform and variance weighting, respectively. The flag column indicates a S/N>8 detection in the RM synthesis (1=detection, 0=no detection). The RM$_{\rm <QU>}$ values for sources with multiple components are the mean RMs of those components, while RM$_{\rm QU^*} $ are the max, or peak location, RMs. }
\label{tab:fvals}
%\begin{tabu}{lrrrrrrrrrr}
\begin{tabular}{lScSSSSSSSS}
\hline
\multicolumn{1}{c}{Name}  & \multicolumn{1}{c}{$z$}& \multicolumn{1}{c}{flag}&\multicolumn{1}{c}{$I_{\rm e}$}&\multicolumn{1}{c}{$\alpha_1$}& \multicolumn{1}{c}{$\log_{10}[L_{\rm e}] $}& \multicolumn{1}{c}{$\Pi_{\rm e} $}& \multicolumn{1}{c}{RM$_{\rm no-wt}$}& \multicolumn{1}{c}{RM$_{\rm sd-wt} $}& \multicolumn{1}{c}{RM$_{\rm <QU>} $}& \multicolumn{1}{c}{RM$_{\rm QU^*} $}\\
 & & & \multicolumn{1}{c}{[mJy]} & &  & \multicolumn{1}{c}{[$\%$]} & \multicolumn{1}{c}{[rad m$^{-2}$]}& \multicolumn{1}{c}{[rad m$^{-2}$]}& \multicolumn{1}{c}{[rad m$^{-2}$]} & \multicolumn{1}{c}{[rad m$^{-2}$]}\\
\hline
\scriptsize J001115+144603 & 4.97 & 1 & 24.3 & -0.42 & 27.0 & 2.63 & -11.9 & -6.5 & 6.2 & -6.8  \\
\scriptsize J003126+150738 & 4.29 & 1 & 66.6 & 0.4 & 27.3 & 0.31 & 15 & 10 & -61 & 7  \\
\scriptsize J021042$-$001818 & 4.73 & 1 & 9.6 & -0.37 & 26.6 & 3.41 & 5.1 & -4.5 & 15.0 & -8.7  \\
\scriptsize J081333+350812 & 4.95 & 1 & 23.7 & -0.93 & 27.0 & 4.73 & 12.6 & 11.9 & 4.0 & 6.5  \\
\scriptsize J083644+005451 & 5.77 & 0 & 0.99 & -0.5 & 25.7 & 4.79 & -8682 & -8688 & 12 & 12  \\
\scriptsize J083946+511202 & 4.40 & 1 & 54.0 & -0.19 & 27.3 & 1.56 & 10.5 & 13.5 & 8.8 & 2.1  \\
\scriptsize J085111+142338 & 4.18 & 0 & 7.8 & -0.50 & 26.4 & 2.29 & -60 & -3 & 30 & 30  \\
\scriptsize J085853+345826 & 1.34 & 0 & 4.56 & -0.94 & 25.3 & 0.73 & 8480 & 8720 & -903 & -903  \\
\scriptsize J090600+574730 & 1.34 & 1 & 7.94 & -0.70 & 25.6 & 3.47 & -5.5 & -5.2 & -35.0 & 1.7  \\
\scriptsize J091316+591920 & 5.12 & 0 & 12.24 & -0.99 & 26.7 & 0.76 & -7810 & -7810 & -1010 & -1010  \\
\scriptsize J091824+063653 & 4.16 & 1 & 45.7 & -0.14 & 27.2 & 1.31 & 256.2 & 259.0 & 251.4 & 251.4  \\
\scriptsize J100424+122924 & 4.52 & 0 & 8.67 & 0.54 & 26.5 & 0.73 & 150 & 150 & 120 & 120  \\
\scriptsize J100645+462716 & 4.34 & 0 & 9.4 & -0.40 & 26.5 & 0.66 & -2860 & -2860 & 69 & 69  \\
\scriptsize J102551+192314 & 1.17 & 1 & 11.82 & -0.8 & 25.6 & 4.86 & 17.3 & 16.0 & -2.5 & 14.5  \\
\scriptsize J102623+254259 & 5.27 & 1 & 169.8 & -0.60 & 27.9 & 8.12 & 9.92 & 9.90 & 10.10 & -8.31  \\
\scriptsize J103601+500831 & 4.50 & 0 & 6.8 & -0.83 & 26.4 & 1.99 & 2310 & 2330 & -1100 & -1100  \\
\scriptsize J104624+590524a & 3.63 & 1 & 0.460 & -1.52 & 25.1 & 9.93 & -12 & -7 & -7 & -7  \\
\scriptsize J104624+590524b & 3.63 & 1 & 6.03 & -1.44 & 26.2 & 7.02 & 8.3 & 8.3 & -55.0 & 6.5  \\
\scriptsize J105320$-$001650 & 4.30 & 0 & 8.6 & -0.62 & 26.5 & 0.79 & -1160 & -6690 & 275 & 275  \\
\scriptsize J130738+150752 & 4.08 & 1 & 6.44 & -0.61 & 26.3 & 4.13 & 6 & 11 & 10 & -50  \\
\scriptsize J130940+573311 & 4.28 & 0 & 11.5 & -0.54 & 26.6 & 0.50 & 9370 & 30 & 35 & 35  \\
\scriptsize J132512+112330 & 4.42 & 1 & 56.5 & -0.34 & 27.3 & 0.76 & 55 & 78 & 170 & 65  \\
\scriptsize J133342+491625 & 1.39 & 1 & 10.75 & -0.80 & 25.7 & 6.04 & 12.6 & 12.7 & 21.9 & 15.2  \\
\scriptsize J135135+284015 & 4.73 & 0 & 1.73 & -1.23 & 25.8 & 8.05 & -1980 & 2690 & -1530 & -1530  \\
\scriptsize J142738+331242 & 6.12 & 0 & 1.12 & -0.9 & 25.8 & 7.87 & 75 & 4802 & -560 & -560  \\
\scriptsize J142952+544717 & 6.21 & 0 & 2.7 & -0.6 & 26.2 & 4.10 & -6891 & -6880 & -1013 & -1013  \\
\scriptsize J151002+570243 & 4.31 & 1 & 243.3 & -0.39 & 27.9 & 3.79 & -74.3 & -82.9 & -36.2 & -127.0  \\
\scriptsize J155633+351757 & 4.67 & 1 & 25.5 & -0.20 & 27.0 & 9.12 & 7.8 & 7.4 & 59.0 & 7.0  \\
\scriptsize J161105+084437 & 4.55 & 0 & 19.6 & -0.4 & 26.9 & 3.56 & 4601 & 4547 & 1061 & 1061  \\
\scriptsize J165913+210116 & 4.89 & 1 & 18.02 & -0.69 & 26.9 & 2.04 & 99 & 96 & 190 & 92  \\
\scriptsize J221356$-$002457 & 1.06 & 1 & 19.8 & -1.12 & 25.8 & 0.40 & -35 & -34 & -30 & -31  \\
\scriptsize J222032+002535 & 4.21 & 1 & 44.61 & -1.04 & 27.2 & 7.04 & -12.90 & -12.83 & 62.34 & -13.24  \\
\scriptsize J222235+001536 & 1.36 & 1 & 41.5 & -0.26 & 26.3 & 4.30 & -16.78 & -16.92 & -15.94 & -23.16  \\
\scriptsize J222843+011032 & 5.95 & 0 & 0.27 & -0.1 & 25.2 & 17.90 & -7123 & -7123 & -9 & -9  \\
\scriptsize J224924+004750 & 4.48 & 1 & 9.31 & -0.68 & 26.5 & 9.51 & 4.5 & 4.1 & 12.0 & 5.3  \\
\scriptsize J231443$-$090637 & 1.29 & 0 & 5.5 & 0.5 & 25.4 & 2.17 & 7 & 8910 & 1000 & 1000  \\
\scriptsize J232604+001333 & 1.00 & 0 & 2.27 & -0.90 & 24.8 & 1.46 & -8174 & 2280 & 398 & 398  \\
\scriptsize J235018$-$000658 & 1.36 & 1 & 52.83 & -1.00 & 26.4 & 3.94 & 3.45 & 5.93 & 13.50 & 17.20  \\
%\scriptsize J091824+063653 &$ 4.16 $& 1 &$ 45.69 $&$ -0.14 $&$ 27.2 $&$ 1.22 $&$ 256.0 $&$ 259.0 $&$ 250.0 $&$ 250.0 $ \\

\hline
\end{tabular}
\normalsize
\end{table*}

The full results for all the fitting and RM synthesis are given in Appendix~\ref{sec:aptables} with the Stokes $I$ fitting given in Table~\ref{tab:stksif}, the polarized fraction in Table~\ref{tab:stkspf}, the RM synthesis in Table~\ref{tab:rmsynf}, and the $QU$ fitting in Table~\ref{tab:quff}. Plots of the spectra and FDF, with fitting results, for one example source (J222032+002535) are shown in Fig.~\ref{fig:explot1}, with similar plots for all of the new sources imaged in this work given in Appendix~\ref{sec:appendB}. 

Only one source, J161105+084437, showed a turnover in the Stokes $I$ spectrum, requiring the use of eq.(\ref{eq:stksi2}). The mean spectral index from the sample is $\langle \alpha_1 \rangle = -0.59 \pm 0.03$. The majority of sources are steep spectrum, with 30 of the 38 sources having $\alpha<-0.3$, three sources have $\alpha_1>0$, or increasing intensity with frequency, and six sources show ultra steep spectra with $\alpha_1<-1.0$ (two of those being the AGN lobe components of J104624+590524). When considering all of the new and control sources, the mean spectral index is $\langle \alpha \rangle =-0.47 \pm 0.02$, with $60\,$per cent having $\alpha \le -0.3$, $15\,$per cent with $\alpha\le -1.0$, and $21\,$per cent having $\alpha \ge 0$. 

For the polarized fraction, there are 19 sources fit with the power-law model of eq.(\ref{eq:pfrac1}), 13 sources fit with the Gaussian model of eq.(\ref{eq:pfrac2}), and 6 sources fit with the offset Gaussian of eq.(\ref{eq:pfrac3}). The mean rest-frame polarization fraction $\langle \Pi_{\lambda_{\rm e}} \rangle = 4.2 \pm 0.6 \,$ per cent for all of the new sources, and $4.5 \pm 0.18\,$per cent for all sources including the control sample. 

From the RM synthesis we set a S/N cutoff of 8, which leaves us with 22 sources with a RM detection (those with a flag value of ``1" in Table~\ref{tab:fvals}), with 16 of those sources having $z\ge3$ (Those with non-detections are discussed further in Section~\ref{sec:nondet}). For those with S/N<8, an RM value (and peak) is still fit and reported in Table~\ref{tab:fvals} and Table~\ref{tab:rmsynf}, which is the Faraday depth at the peak amplitude of the FDF; it is just not a significant peak and is therefore a non-detection and the values are not used in further analysis. With the additional four sources from the control sample with RM measurements and $z\ge 3$ we have a total of 20 high-$z$ sources with RM values and 478 sources with RM values and $z<3$. 

\begin{figure}
\includegraphics[scale=0.35]{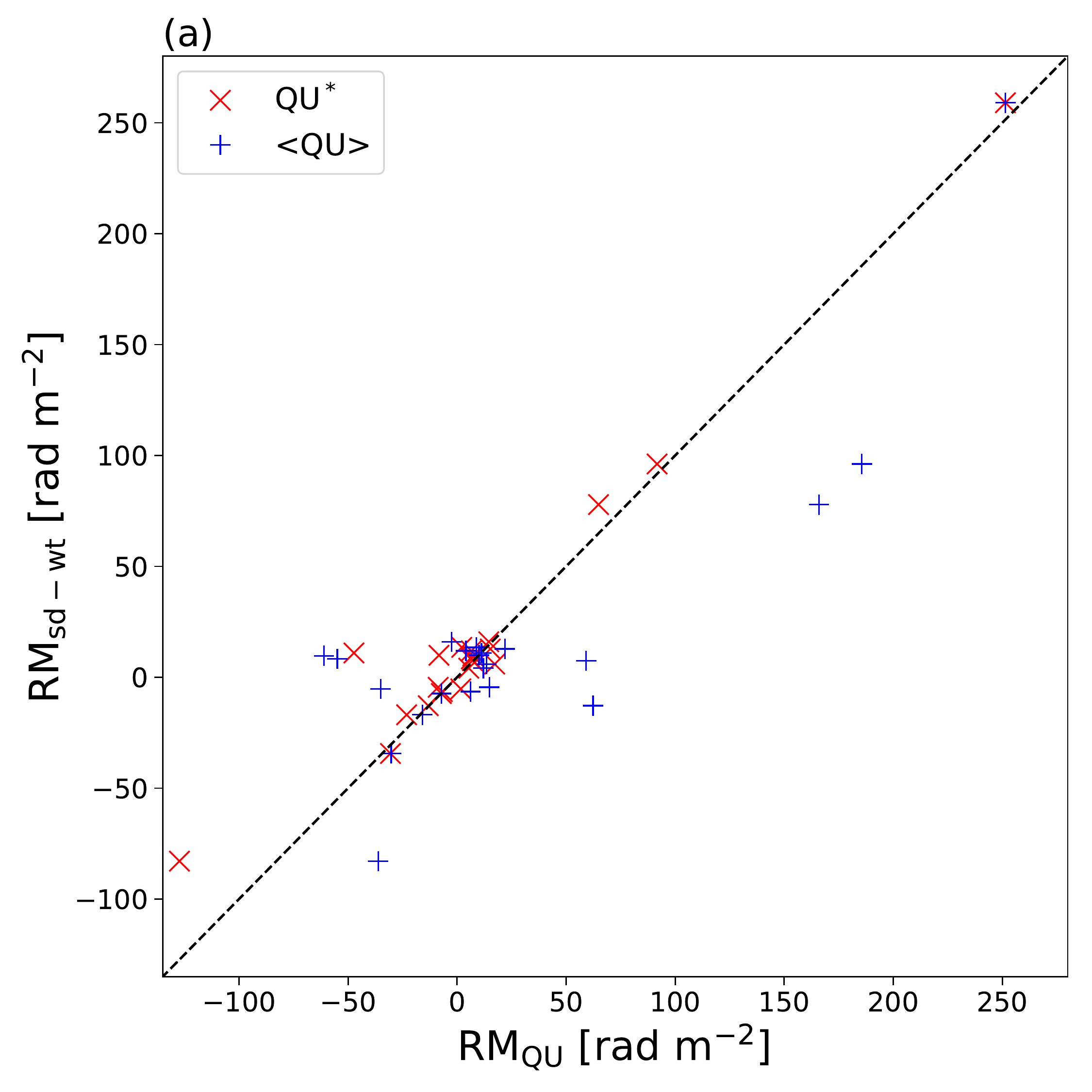}
\includegraphics[scale=0.35]{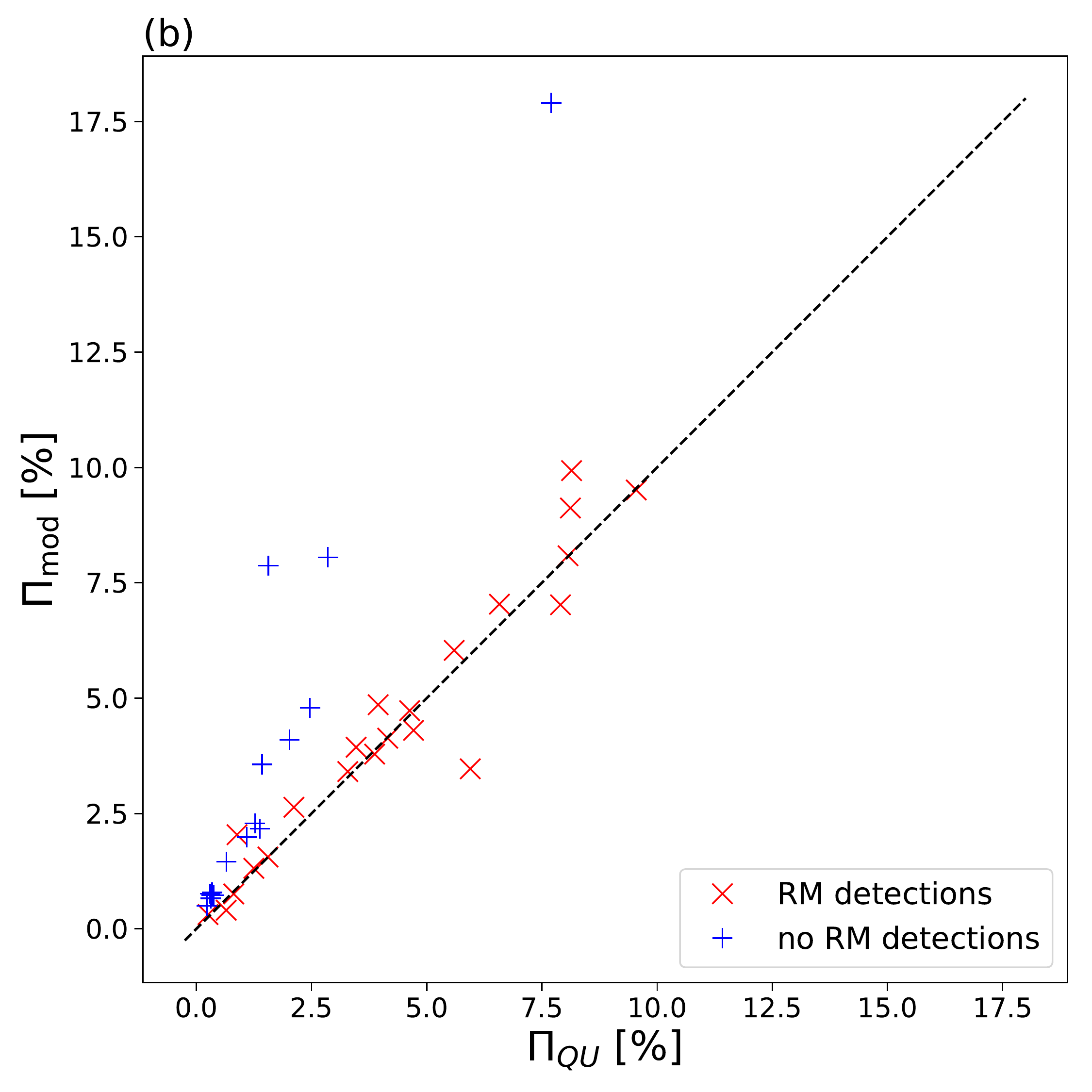}
\caption{Comparison of RM and $\Pi_{\rm e}$ values from QU fitting with those from RM synthesis and $\Pi_{\rm e}$ model fitting. Panel (a) shows the QU fitting RMs vs. the sd-wt RMs from RM synthesis. The red crosses are the RM values for the peak amplitudes for those sources with multiple components fit in the QU fitting, whereas the blue pluses are the mean of the multiple component RMs. Panel (b) shows the $\Pi_{\rm e}$ value from the QU fitting model at the rest-frame wavelength vs. the values obtained from fitting models in Section~\ref{sec:stksp}. The red crosses are the sources with RM detections from the RM synthesis (S/N$\ge8$),while the blue pluses are the sources with RM synthesis S/N<8. The black dashed lines show a one-to-one relation.}
\label{fig:quvsrm}
\end{figure}

From the $QU$ fitting there are 16,  9,  5,  4,  2,  and 2 sources for which the best-fit models were respectively ``D", ``DD", ``DDD", ``DG", ``GG", and ``G". For those 21 sources with detected peaks from the RM synthesis the results are 0, 9,  5,  4,  2,  and 2 sources. When considering all of the sources, a single delta function, or ``D", model is the most common, however, when looking at the sources with RM synthesis detections, two thin components is the most common, with three thin components being nearly as common. As previously mentioned, for the low S/N sources the more complex models were not converging, and all of these sources have a single thin component, or delta function, listed as the chosen model. We report the fit parameters for these sources, but still consider them a non-detection, even in the QU fitting. 

Figure~\ref{fig:quvsrm} compares the values from QU fitting to those from RM synthesis (panel a) and the $\Pi_{\rm e}$ fitting (panel b). From this we can see that for sources with a detection (red crosses in panel b) the $\Pi_{\rm e}$ from the QU model is quite close to the values obtained from the $\Pi_{\rm e}$ model fitting, with a median ratio of 1.008. However, for the no-detection sources the median ratio changes to 0.45, with the model fit values being higher than from QU fitting. 

When comparing the RMs we can see that the max RM (the RM associated with the maximum amplitude component) from the QU fitting, RM$_{QU^*}$, is a closer match to the RM synthesis value than the mean RM from the QU fitting, RM$_{<QU>}$ (the mean of the RMs from the multiple components). The mean absolute difference, $|({\rm RM}_{QU}-{\rm RM}_{sd-wt})|$, for RM$_{QU^*}$ is $20\,$rad m$^{-2}$ and for RM$_{<QU>}$ is $60\,$rad m$^{-2}$. This makes sense as the RM synthesis only reports the value of the peak (even if more clean components are found during the RM cleaning). The two sources with the largest max QU differences, J132512+112330 and J221356-002457, look to have more complicated Faraday spectra that may be better fit with more complex QU-fitting models (see Figs.~\ref{fig:spec22} and \ref{fig:spec31} for plots of these two sources). 

\begin{figure}
\includegraphics[scale=0.67]{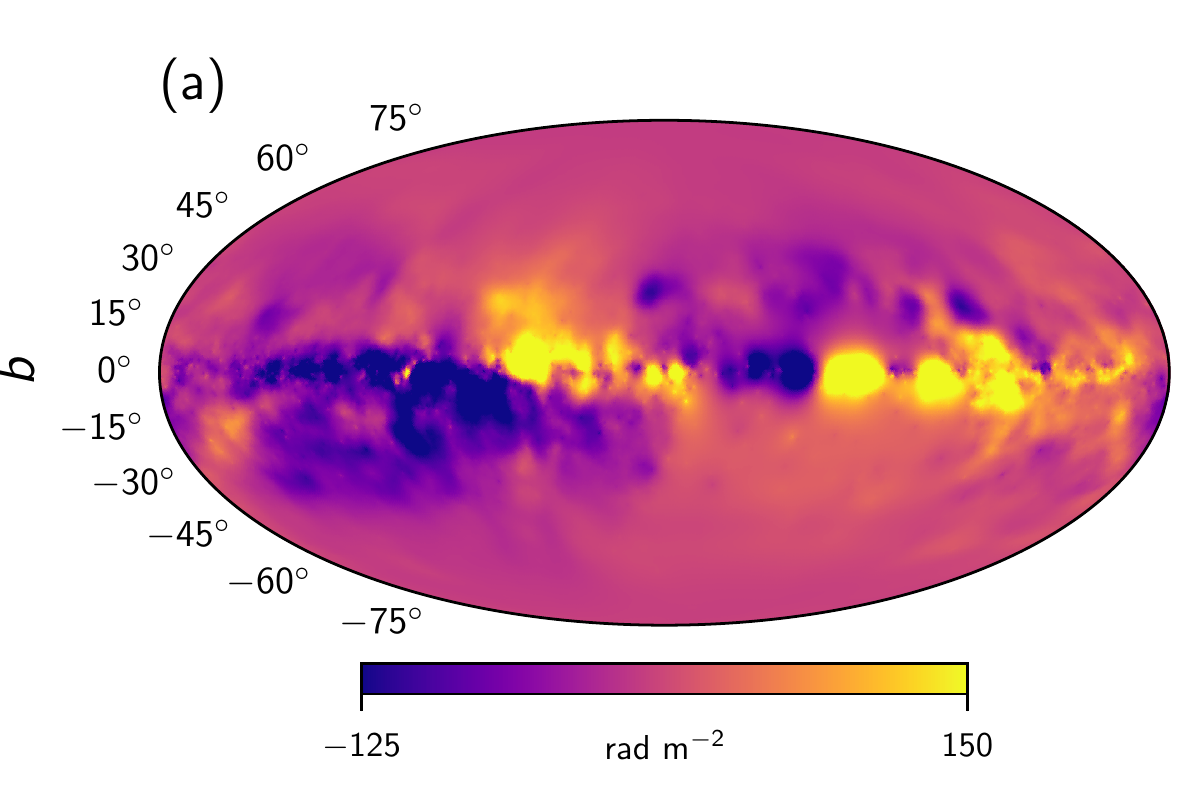}
\includegraphics[scale=0.67]{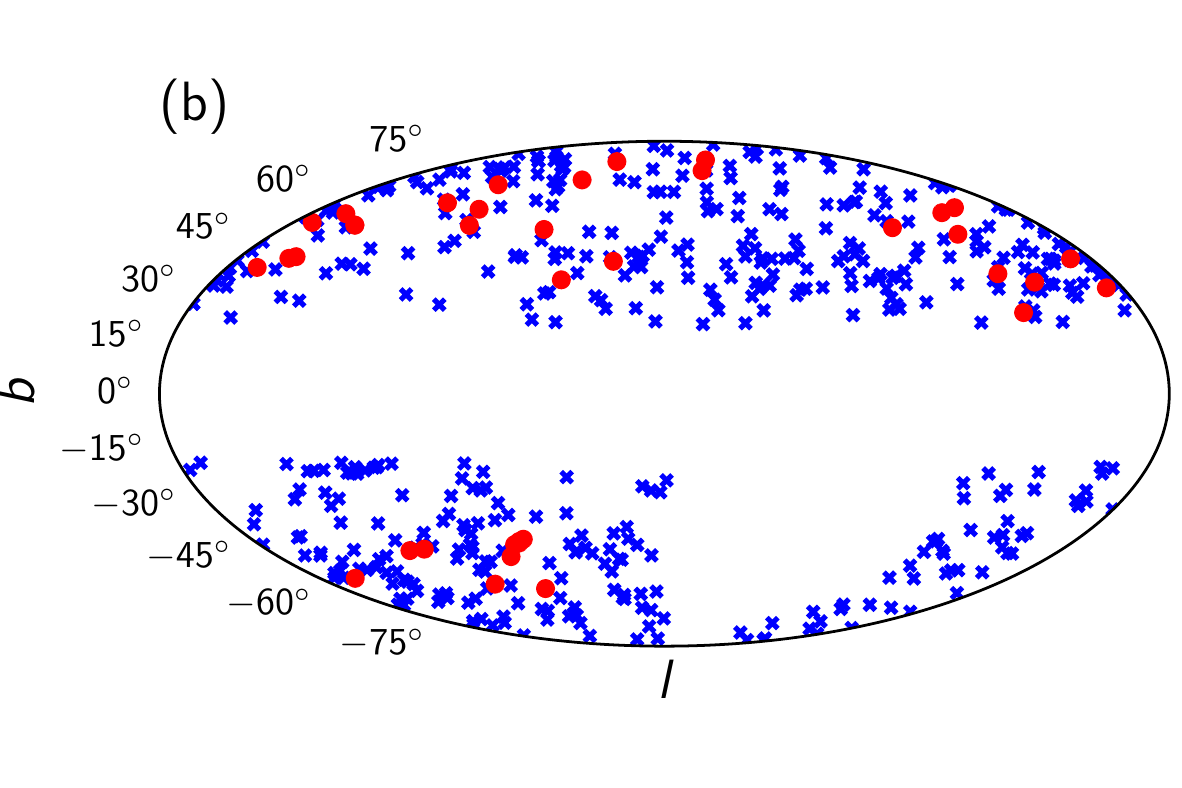}
\caption{Map of Galactic rotation measures and source positions. The top panel shows the Galactic rotation measures. The bottom panel shows the positions in Galactic coordinates of the new sample of sources in this work (red circles) and the archival sample of sources (blue crosses).}
\label{fig:galmap}
\end{figure}

The observed RM is a combination of the Galactic contribution (GRM), the extragalactic residual RM (RRM), and the measurement uncertainty $N$ such that
\begin{equation}
{\rm RM}={\rm RRM} + {\rm GRM} + N.
\label{eq:rrm} 
\end{equation} 
In order to examine just the extragalactic RM we applied the algorithm developed by \citet{Oppermann15} \citep[see also][]{Oppermann12}. This entailed using the same code and same sample of RM sources as Oppermann (41632 RMs) and including the 22 new RMs measured by our RM synthesis, and then iteratively solving for the Galactic RM. For the full details about the procedure please refer to \citet{Oppermann12,Oppermann15}. The effect of adding the new sources is small, particularly as they reside off the Galactic plane in regions where the Galactic RM contribution is more comparable. However, the 22 new RM sources have smaller uncertainties than the majority of the original Oppermann sources, which come largely from \citet{Taylor09}, and therefore increase the overall accuracy of the estimation. The Galactic RM output from this procedure is shown in Fig.~\ref{fig:galmap}. 

The extragalactic residual can be obtained by subtracting the Galactic RM from the observed RM, RM-GRM. It should be noted that simply taking the difference of observed and Galactic RMs yields an extragalactic contribution combined with the uncertainty (RRM+$N$). In \citet{Oppermann15} a method is shown for separating the uncertainty and extragalactic estimates. However, this technique is only valid under certain assumptions for the posterior distribution, which may not hold for sources over a range of redshifts and is therefore not optimal to apply when studying cosmic RM evolution \citep[for more discussion on this see appendix C of ][]{Oppermann15}. 

\begin{figure*}
\includegraphics[scale=0.26]{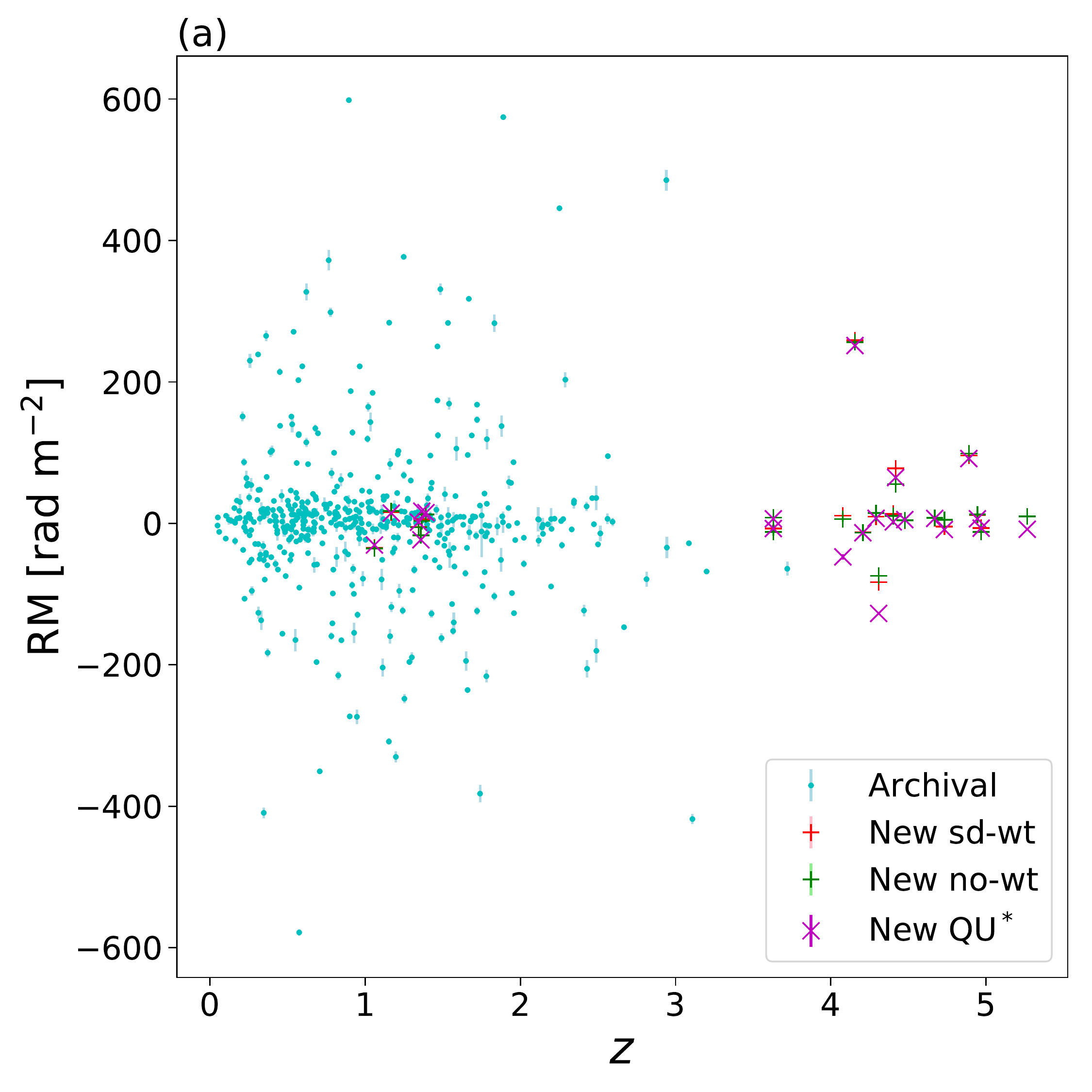}\includegraphics[scale=0.26]{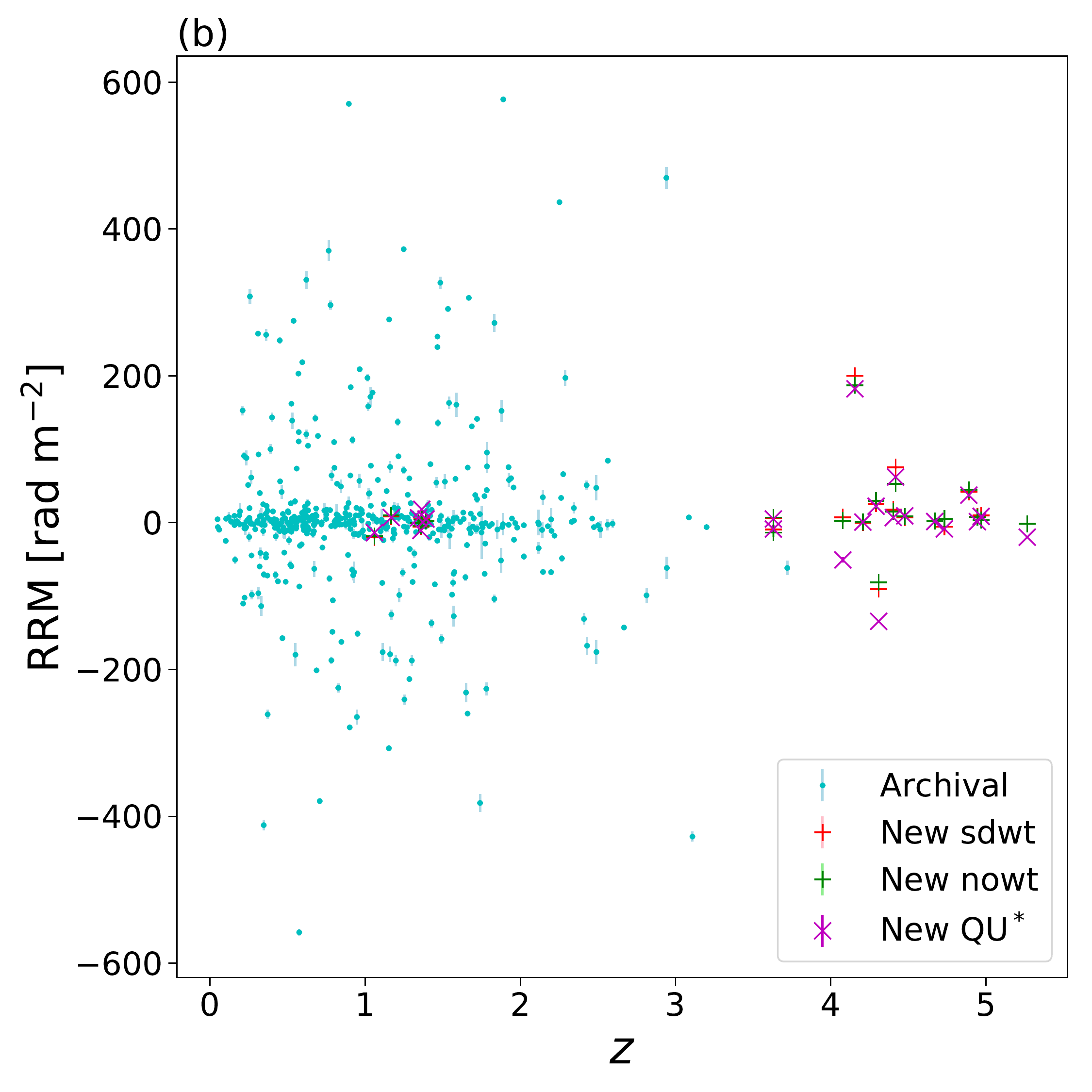}\includegraphics[scale=0.26]{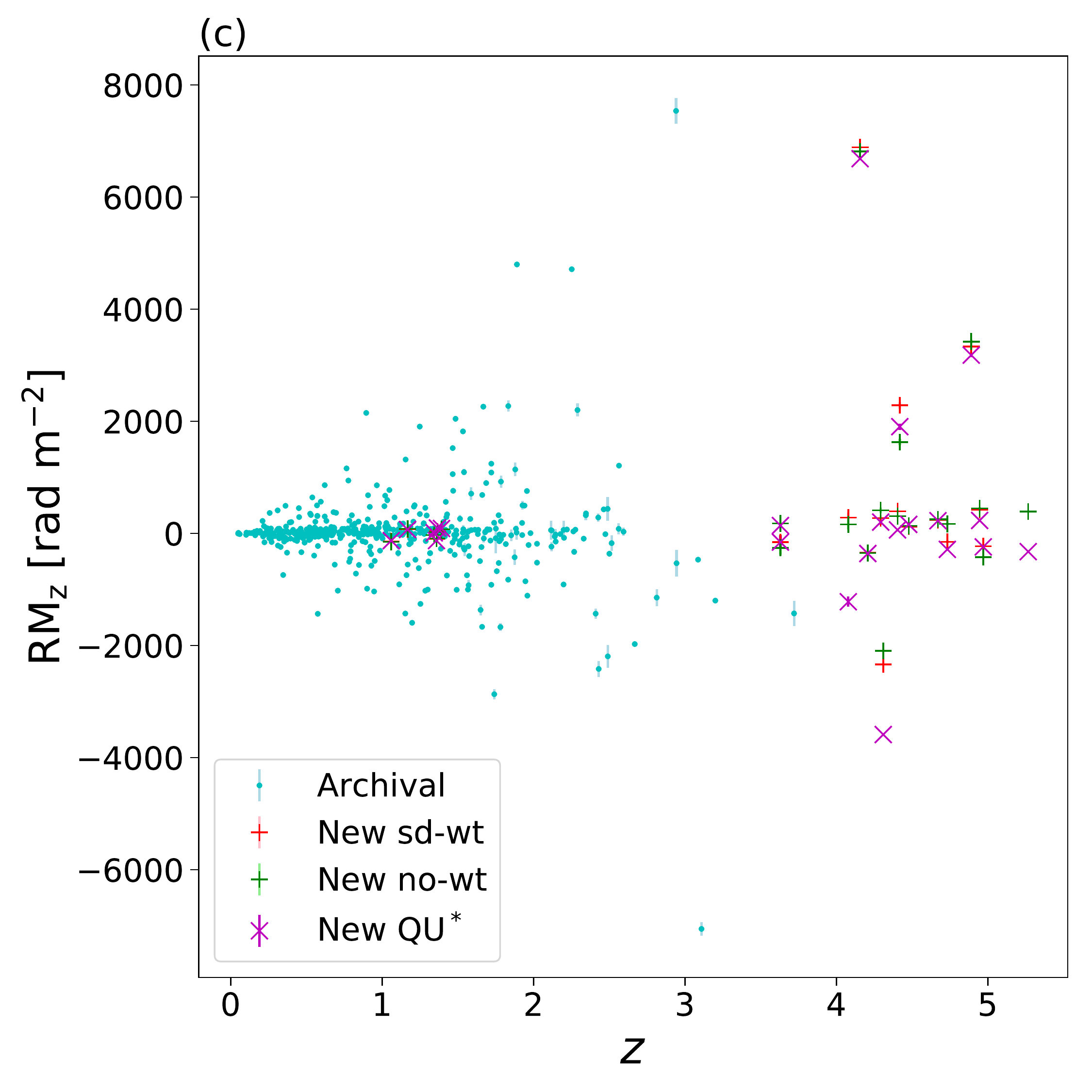}
\includegraphics[scale=0.26]{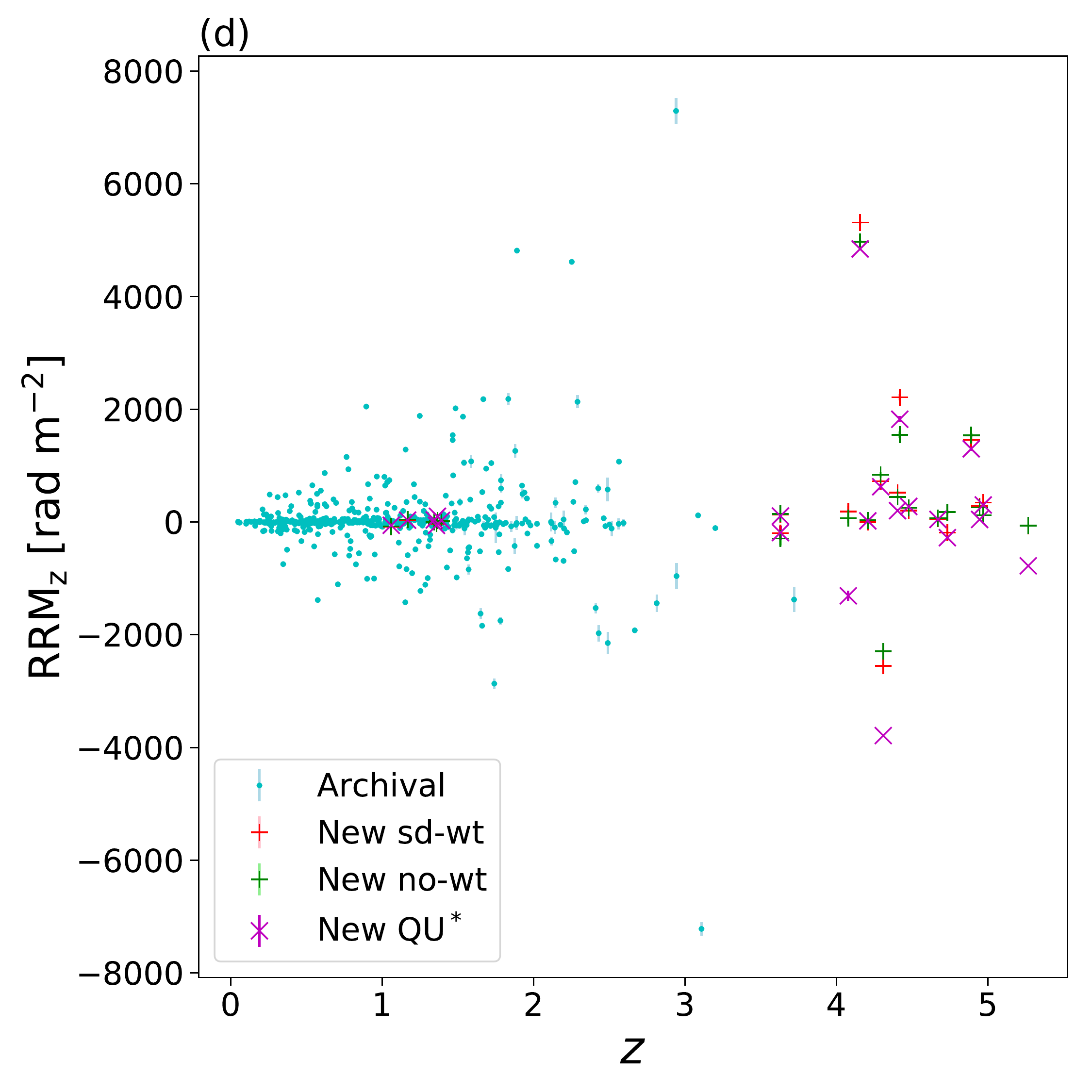}\includegraphics[scale=0.26]{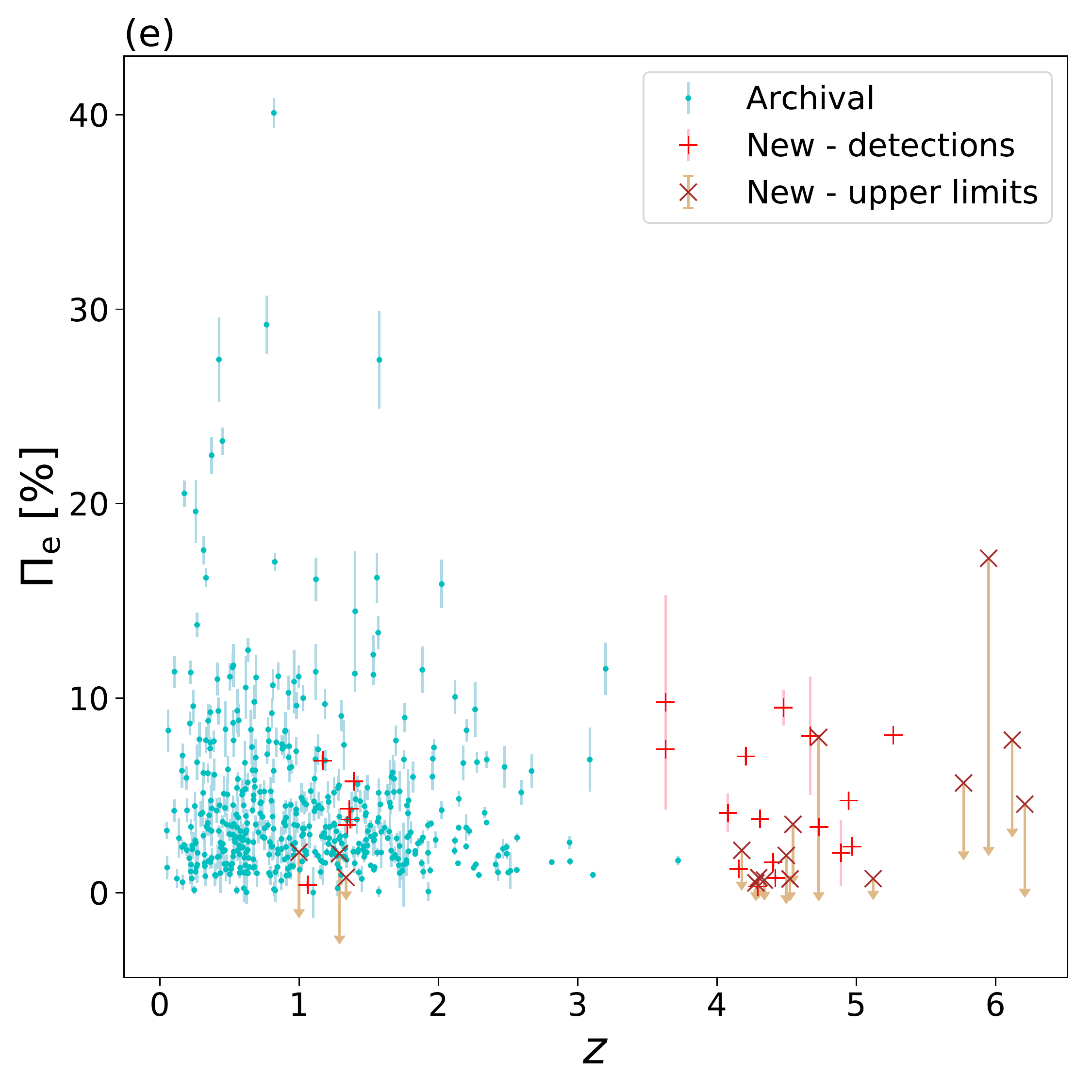}\includegraphics[scale=0.26]{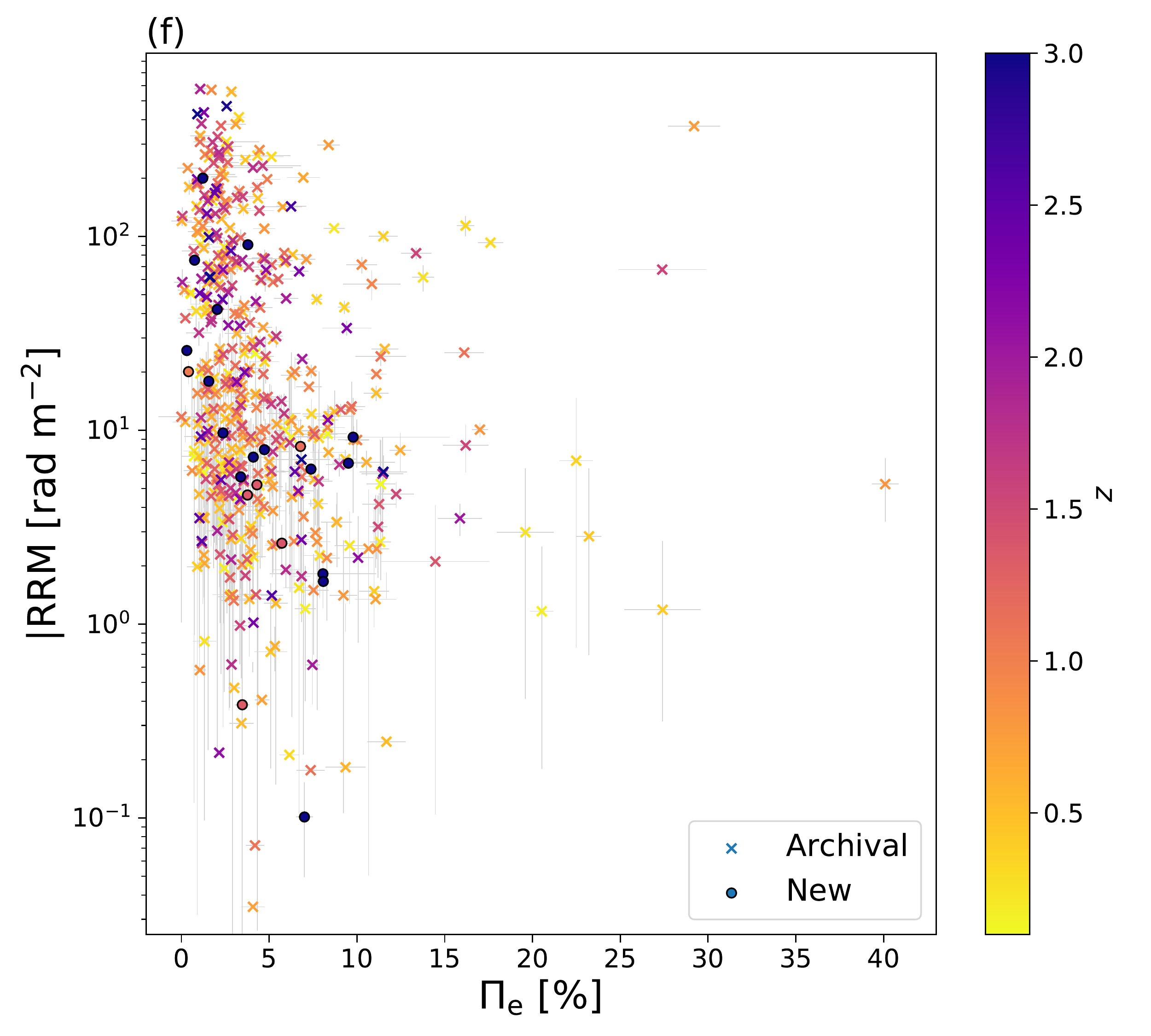}
\caption{Rotation measures, polarization fractions, and redshifts compared against each other. From left to right top to bottom the panels are: (a) RM, (b) RRM, (c) RM$_z$, (d) RRM$_z$, (e) $\Pi_{\rm e}$ vs. $z$, and (f) $|$RRM$|$ vs. $\Pi_{\rm e}$ colour coded by redshift. The blue circles are from the archival sample (with luminosity limits imposed), and for panels (a)-(d),  the red pluses are the new data using the ``sd-wt" weighting in the RM synthesis, the green pluses are the new data using the ``no-wt" RM synthesis weighting, and the magenta crosses are the peak RMs from the QU fitting. In panel (e), the red pluses are the polarization fractions from those sources with RM detections, whereas the brown pluses are sources denoted as upper limits meaning there was no significant RM detection (those with flag=0 in Table~\ref{tab:fvals}). In panel (f) the crosses are the new sample with the RRM from the sd-wt RM synthesis values and the circles are the archival sources. }
\label{fig:rmvsz}
\end{figure*}

While it is possible for the RRM of a source to be due to intervening magnetic fields, we can also look at the rest-frame RRM, or RRM$_z$. Equation~(\ref{eq:RM1}) can be rewritten as eq.(~\ref{eq:rm1}). From that we can see that if we assume that all the RRM comes from magnetized plasma in the vicinity of the source then 
\begin{equation}
{\rm RRM}_z={\rm RRM}(1+z)^2,
\label{eq:rrmz}
\end{equation}
and similarly ${\rm RM}_z={\rm RM} (1+z)^2$. The values of GRM, RRM, and RRM$_{z}$ are given in Table~\ref{tab:resrms}, with the values of GRM obtained from the source locations on the Galactic RM map. Figure~\ref{fig:rmvsz} shows the RMs and $\Pi_{\rm e}$ against redshift, and the RRMs against $\Pi_{\rm e}$. Table~\ref{tab:rmstats} provides the mean and uncertainties values for the different RMs and $\Pi_{\rm e}$, divided into high and low redshift.

\begin{table*}
 \setlength{\tabcolsep}{4.pt}
%\scriptsize
\caption{Galactic, residual, and residual rest-frame rotation measures. The Galactic rotation measures, or GRM, are computed as described in Section~\ref{sec:results}, using the algorithm of \citet{Oppermann12,Oppermann15} using both of the RMs from the RM synthesis and the mean RM from the QU fitting, as described in Section~\ref{sec:rmsynth} and Section~\ref{sec:qufit}. The intrinsic, or rest-frame, RMs, RRM$_{z}$ are computed from eq.~(\ref{eq:rrmz}). The flag column is the same as that from Table~\ref{tab:fvals}, where 1 indicates an RM detection and a 0 indicates no detection, only values for sources with a detection are reported.}
\label{tab:resrms}
%\begin{tabu}{lrrrrrrrrrr}
\begin{tabular}{lcSSSSSSSSS}
\hline %\\[-2.25ex]
 %\rowfont{\scriptsize}
  & & \multicolumn{3}{c}{no-wt} & \multicolumn{3}{c}{sd-wt} &\multicolumn{3}{c}{QU$^*$}\\
\multicolumn{1}{c}{Name} & \multicolumn{1}{c}{flag}& \multicolumn{1}{c}{GRM}&\multicolumn{1}{c}{RRM}&\multicolumn{1}{c}{RRM$_{z}$} &\multicolumn{1}{c}{GRM}&\multicolumn{1}{c}{RRM}&\multicolumn{1}{c}{RRM$_{z}$} &\multicolumn{1}{c}{GRM}&\multicolumn{1}{c}{RRM}&\multicolumn{1}{c}{RRM$_{z}$}\\
 \rowfont{\tiny}
  &  & \multicolumn{1}{c}{\scriptsize [rad m$^{-2}$]}& \multicolumn{1}{c}{\scriptsize [rad m$^{-2}$]}& \multicolumn{1}{c}{\scriptsize [rad m$^{-2}$]} &\multicolumn{1}{c}{\scriptsize [rad m$^{-2}$]}&\multicolumn{1}{c}{\scriptsize [rad m$^{-2}$]}&\multicolumn{1}{c}{\scriptsize [rad m$^{-2}$]}&\multicolumn{1}{c}{\scriptsize [rad m$^{-2}$]}&\multicolumn{1}{c}{\scriptsize [rad m$^{-2}$]}&\multicolumn{1}{c}{\scriptsize [rad m$^{-2}$]} \\%\TBstrut\\
\hline
\scriptsize J001115+144603 & 1 & -15.3 & 3.4 & 122.0 & -16.2 & 9.8 & 350.0 & -15.8 & 8.6 & 310.0  \\
\scriptsize J003126+150738 & 1 & -15.1 & 30 & 830 & -16.1 & 30 & 700 & -15.6 & 20 & 600  \\
\scriptsize J021042$-$001818 & 1 & -0.3 & 5.4 & 180.0 & 1.2 & -5.7 & -190.0 & 0.4 & -8.4 & -280.0  \\
\scriptsize J081333+350812 & 1 & 5.1 & 7.5 & 266.0 & 3.9 & 8.0 & 283.0 & 4.5 & 1.4 & 50.0  \\
\scriptsize J083644+005451 & 0 & 6.4 & {--} & {--} & 6.3 & {--} & {--} & 6.3 & {--} & {--}  \\
\scriptsize J083946+511202 & 1 & -4.7 & 15.2 & 443.0 & -4.4 & 17.8 & 521.0 & -4.5 & 6.8 & 200.0  \\
\scriptsize J085111+142338 & 0 & 24.2 & {--} & {--} & 24.0 & {--} & {--} & 24.1 & {--} & {--}  \\
\scriptsize J085853+345826 & 0 & 15.0 & {--} & {--} & 15.0 & {--} & {--} & 15.0 & {--} & {--}  \\
\scriptsize J090600+574730 & 1 & -4.9 & -0.6 & -3.3 & -4.9 & -0.4 & -2.0 & -4.9 & 6.6 & 36.0  \\
\scriptsize J091316+591920 & 0 & -5.2 & {--} & {--} & -5.2 & {--} & {--} & -5.2 & {--} & {--}  \\
\scriptsize J091824+063653 & 1 & 69.5 & 186.7 & 4964.0 & 59.4 & 199.6 & 5307.0 & 64.4 & 181.9 & 4837.0  \\
\scriptsize J100424+122924 & 0 & -2.1 & {--} & {--} & -2.2 & {--} & {--} & -2.2 & {--} & {--}  \\
\scriptsize J100645+462716 & 0 & 8.7 & {--} & {--} & 8.6 & {--} & {--} & 8.7 & {--} & {--}  \\
\scriptsize J102551+192314 & 1 & 7.6 & 9.7 & 45.5 & 7.9 & 8.1 & 38.1 & 7.7 & 6.9 & 32.4  \\
\scriptsize J102623+254259 & 1 & 11.5 & -1.57 & -61.60 & 11.6 & -1.66 & -65.20 & 11.5 & -19.80 & -777.00  \\
\scriptsize J103601+500831 & 0 & 10.7 & {--} & {--} & 10.7 & {--} & {--} & 10.7 & {--} & {--}  \\
\scriptsize J104624+590524a & 1 & 1.5 & -14 & -290 & 1.8 & -9 & -200 & 1.7 & -9 & -200  \\
\scriptsize J104624+590524b & 1 & 1.6 & 6.7 & 140.0 & 2.0 & 6.2 & 130.0 & 1.8 & 4.9 & 100.0  \\
\scriptsize J105320$-$001650 & 0 & 5.4 & {--} & {--} & 5.4 & {--} & {--} & 5.4 & {--} & {--}  \\
\scriptsize J130738+150752 & 1 & 3.5 & 3 & 70 & 3.7 & 7 & 190 & 3.6 & -51 & -1300  \\
\scriptsize J130940+573311 & 0 & 7.0 & {--} & {--} & 7.0 & {--} & {--} & 7.0 & {--} & {--}  \\
\scriptsize J132512+112330 & 1 & 2.7 & 53 & 1500 & 2.5 & 75 & 2200 & 2.6 & 62 & 1800  \\
\scriptsize J133342+491625 & 1 & 10.3 & 2.3 & 12.9 & 10.1 & 2.6 & 14.9 & 10.2 & 4.9 & 28.2  \\
\scriptsize J135135+284015 & 0 & 3.4 & {--} & {--} & 3.5 & {--} & {--} & 3.4 & {--} & {--}  \\
\scriptsize J142738+331242 & 0 & 2.3 & {--} & {--} & 2.2 & {--} & {--} & 2.2 & {--} & {--}  \\
\scriptsize J142952+544717 & 0 & 12.8 & {--} & {--} & 12.8 & {--} & {--} & 12.8 & {--} & {--}  \\
\scriptsize J151002+570243 & 1 & 7.1 & -81.4 & -2290.0 & 7.6 & -90.6 & -2550.0 & 7.4 & -134.0 & -3790.0  \\
\scriptsize J155633+351757 & 1 & 5.6 & 2.2 & 71.0 & 5.6 & 1.8 & 59.0 & 5.6 & 1.4 & 46.0  \\
\scriptsize J161105+084437 & 0 & 6.0 & {--} & {--} & 5.7 & {--} & {--} & 5.8 & {--} & {--}  \\
\scriptsize J165913+210116 & 1 & 54.2 & 44 & 1500 & 54.1 & 42 & 1500 & 54.2 & 37 & 1300  \\
\scriptsize J221356$-$002457 & 1 & -17.0 & -18 & -77 & -14.4 & -20 & -85 & -15.7 & -14 & -57  \\
\scriptsize J222032+002535 & 1 & -14.0 & 1.14 & 30.90 & -12.8 & -0.08 & -2.14 & -13.4 & 0.80 & 21.77  \\
\scriptsize J222235+001536 & 1 & -12.6 & -4.14 & -23.06 & -11.8 & -5.16 & -28.72 & -12.2 & -10.53 & -58.57  \\
\scriptsize J222843+011032 & 0 & -5.7 & {--} & {--} & -5.5 & {--} & {--} & -5.6 & {--} & {--}  \\
\scriptsize J224924+004750 & 1 & -3.9 & 8.3 & 250.0 & -2.6 & 6.7 & 200.0 & -3.3 & 9.2 & 280.0  \\
\scriptsize J231443$-$090637 & 0 & -6.8 & {--} & {--} & -6.9 & {--} & {--} & -6.9 & {--} & {--}  \\
\scriptsize J232604+001333 & 0 & -15.8 & {--} & {--} & -15.9 & {--} & {--} & -15.8 & {--} & {--}  \\
\scriptsize J235018$-$000658 & 1 & -0.9 & 4.39 & 24.50 & 1.2 & 4.69 & 26.20 & 0.2 & 18.20 & 102.05  \\
\hline
\end{tabular}
\normalsize
\end{table*}

\begin{table*}
 \setlength{\tabcolsep}{4.pt}
\caption{RM statistics for sources for high and low redshift sources. All values were calculated using the absolute values of the RMs and only using values with an RM S/N$>8$. Here $\sigma_{\rm SE}$ is the standard error on the mean. The mean$^*$ and $\sigma_{\rm{SE}}^*$ are the weighted mean and weighted variance.}
\label{tab:rmstats}
\begin{tabu}{lrrrrrrrrrrrrr}
\hline 
 & & \multicolumn{4}{c}{no-wt}&\multicolumn{4}{c}{ sd-wt}&\multicolumn{4}{c}{QU$^*$}  \\ 
 
 & $\Pi_{\rm e}$ & $|\rm{RM}|$ & $|\rm{RRM}|$ & $|\rm{RM}_z|$ & $|\rm{RRM}_z|$ &$|\rm{RM}|$ & $|\rm{RRM}|$ & $|\rm{RM}_z|$ & $|\rm{RRM}_z|$ &$|\rm{RM}|$ & $|\rm{RRM}|$ & $|\rm{RM}_z|$ & $|\rm{RRM}_z|$ \\
 \rowfont{\tiny}
  & [$\%$]& [rad m$^{-2}$] & [rad m$^{-2}$] & [rad m$^{-2}$] & [rad m$^{-2}$] & [rad m$^{-2}$] & [rad m$^{-2}$] & [rad m$^{-2}$] & [rad m$^{-2}$] & [rad m$^{-2}$] & [rad m$^{-2}$] & [rad m$^{-2}$] & [rad m$^{-2}$] \\
 \hline
 \multicolumn{14}{c}{$z>3$}\\
 \hline 
mean &$ 4.5 $&$ 60 $&$ 50 $&$ 1400 $&$ 1100 $&$ 60 $&$ 50 $&$ 1400 $&$ 1200 $&$ 60 $&$ 50 $&$ 1500 $&$ 1200 $\\
median &$ 3.5 $&$ 10 $&$ 8 $&$ 420 $&$ 260 $&$ 10 $&$ 9 $&$ 390 $&$ 240 $&$ 10 $&$ 9 $&$ 340 $&$ 290 $\\
mean$^*$ &$ 1.5 $&$ 10 $&$ 2 $&$ 380 $&$ 74 $&$ 10 $&$ 3 $&$ 400 $&$ 90 $&$ 10 $&$ 2 $&$ 370 $&$ 40 $\\
$\sigma_{\rm{SE}}$ &$ 0.7 $&$ 20 $&$ 20 $&$ 450 $&$ 410 $&$ 20 $&$ 20 $&$ 460 $&$ 420 $&$ 20 $&$ 20 $&$ 460 $&$ 420 $\\
$\sigma_{\rm{SE}}^*$ &$ 0.4 $&$ 2 $&$ 2 $&$ 70 $&$ 70 $&$ 3 $&$ 3 $&$ 90 $&$ 90 $&$ 2 $&$ 2 $&$ 70 $&$ 50 $\Bstrut\\

\hline
\multicolumn{14}{c}{$z<3$}\\
\hline
mean &$ 4.5 $&$ 60 $&$ 60 $&$ 300 $&$ 290 $&$ 60 $&$ 60 $&$ 300 $&$ 290 $&$ 60 $&$ 60 $&$ 300 $&$ 290 $\\
median &$ 3.2 $&$ 24 $&$ 15 $&$ 80 $&$ 60 $&$ 20 $&$ 15 $&$ 80 $&$ 60 $&$ 20 $&$ 15 $&$ 84 $&$ 60 $\\
mean$^*$ &$ 5.8 $&$ 20 $&$ 12 $&$ 96 $&$ 50 $&$ 20 $&$ 15 $&$ 110 $&$ 70 $&$ 30 $&$ 20 $&$ 140 $&$ 90 $\\
$\sigma_{\rm{SE}}$ &$ 0.2 $&$ 4 $&$ 4 $&$ 30 $&$ 30 $&$ 4 $&$ 4 $&$ 30 $&$ 30 $&$ 4 $&$ 4 $&$ 30 $&$ 30 $\\
$\sigma_{\rm{SE}}^*$ &$ 0.3 $&$ 2 $&$ 2 $&$ 10 $&$ 9 $&$ 2 $&$ 2 $&$ 11 $&$ 11 $&$ 2 $&$ 2 $&$ 10 $&$ 10 $\Bstrut\\
\hline
\end{tabu}
\normalsize
\end{table*}

\subsection{Effect of redshift}
\label{sec:zdis}
We want to know if there is a difference between the polarization and Faraday rotation of high and low redshift sources. There are several ways we can attempt to answer this question using our current data. 

\subsubsection{RM vs. $z$}
\label{sec:zvsrm}
\begin{figure}
\includegraphics[scale=0.35]{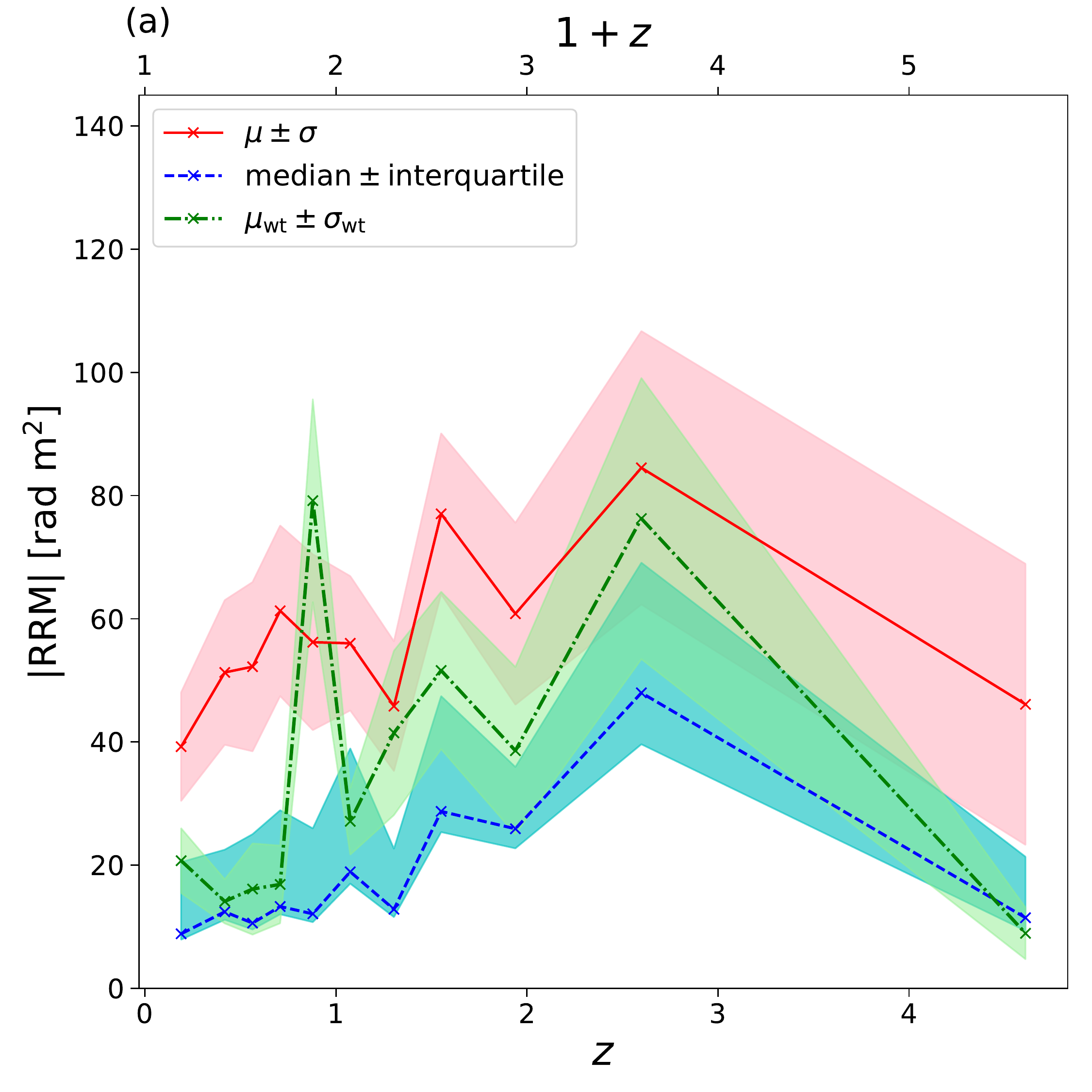}
\includegraphics[scale=0.35]{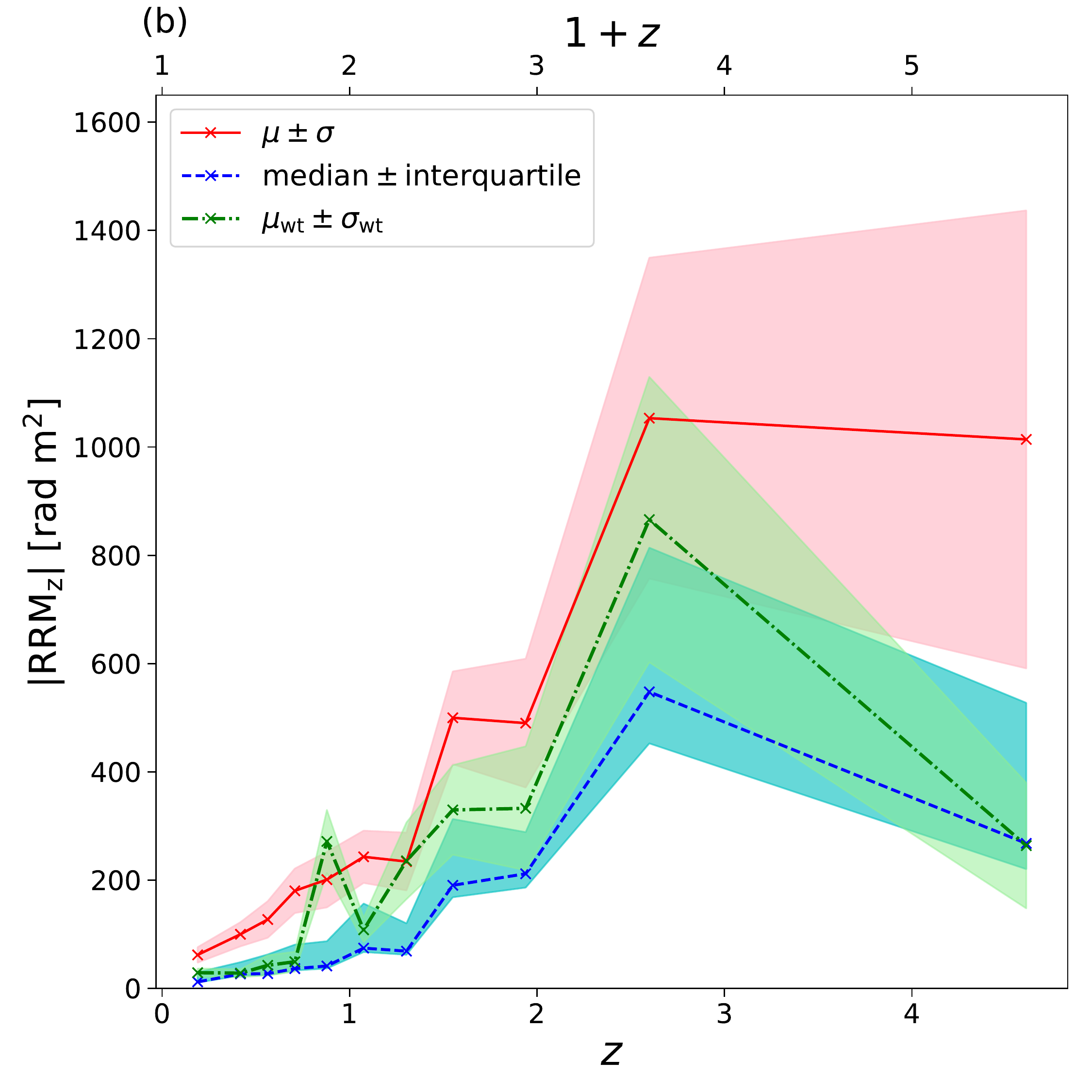}
\caption{Rotation measures in bins of redshift. The top panel shows $|{\rm RRM}|$, while the bottom panel shows $|{\rm RRM}|_z$, both using the sd-wt RM synthesis values for the new sources. The mean per $z$ bin is shown by the red solid lines and the $1\sigma$ uncertainties the red shaded regions. The blue dashed lines show the medians, with the blue shaded regions being the interquartile ranges. The green dot-dashed lines are the weighted means with the green regions showing the weighted $1\sigma$ uncertainties. }
\label{fig:rmvszbins}
\end{figure}

The first thing we can do is look at RM as a function of $z$. We computed the mean, median, and weighted mean (along with the standard deviation, interquartile range, and weighted standard deviation) for $|{\rm RM}|$, $|{\rm RRM}|$, $|{\rm RM}_z|$ and $|{\rm RRM}_z|$ in bins of $z$. The $z$ bins were chosen to give the same number of sources in each bin to within a factor of two. The results for $|{\rm RRM}|$ and $|{\rm RRM}_z|$ are shown in Fig.~\ref{fig:rmvszbins}.  

If there is no dependence on redshift, we would expect $|{\rm RRM}|$ to remain flat as a function of z, and $|{\rm RRM}_z|$ to change as $(1+z)^2$. From Fig.~\ref{fig:rmvszbins} it appears that for both RRM and RRM$_z$ there is a decrease in the mean (and medians and weighted averages) RRM for the sources at $z>3$. Fitting a function of the form $(1+z)^{\kappa}$ shows that regardless of which RM is used (no-wt,sd-wt, QU) or which statistic (mean, median, weighted mean) for $|{\rm RRM}_z|$ or $|{\rm RM}_z|$, $\kappa > 2$ if the $z>3$ sources are not included ($2.4 \le \kappa \le 3.4$), but this drops when including the highest redshift bin ($1.1\le \kappa \le 2.1$).

This seems to indicate that the higher redshift sources might have intrinsically lower RMs. However it is unclear from simply examining these plots if this difference is significant (or how significant). Further tests are required to quantitatively determine the significance of any differences, which are discussed below in Sections~\ref{sec:boots} and \ref{sec:kss}.

\subsubsection{Bootstrap tests}
\label{sec:boots}
A bootstrap test can be used to test the hypothesis that two samples are from the same population \citep{Efron79}. In general there are two samples $X_1$ and $X_2$ of size $n_1$ and $n_2$. The test statistic is computed, generally the difference in the means of the two samples, $\mu^*=\mu_1^*-\mu_2^*$. The bootstrap procedure is as follows:
\begin{enumerate}[label={\arabic{enumi}.},leftmargin=*]
\item The two samples are combined into one sample $X$ of size $n=n_1+n_2$. 
\item Two new samples are drawn randomly with replacement from $X$ of size $n_1$ and $n_2$
\item Recompute the test statistic  $\mu=\mu_1-\mu_2$ 
\item Repeat steps 2 and 3 $N_B$ times (500 to several thousand) to obtain $N_B$ values of the test statistic
\item The p-value, p$^*$ is then calculated from the distribution of $\mu$'s, for a two-tailed probability
\begin{equation}
{\rm p}^* = 2 \, \times \, {\rm min} \left [ \frac{N_{\mu^* > \mu}}{N_B} \, ,\,\frac{N_{\mu^* < \mu}}{N_B} \right ],  
\label{eq:boot}
\end{equation}
where $N_{\mu^* > \mu}$ is the number is trials where $\mu^*$ is greater than $\mu$ and $N_{\mu^* < \mu}$ is the number of trials where $\mu^*$ is less than $\mu$.

\end{enumerate}
The hypothesis that the samples are from the same population can be rejected with $\alpha$ significance (usually $\alpha =0.05$ ) if ${\rm p}^*<\alpha$. The idea is that if the two samples are from the same population, $\mu^*$ should be fairly common, whereas if the two samples are genuinely different, then a value of $\mu^*$ should not happen frequently with the resampling. 

In our case, the test statistic is the difference in the means of the absolute value of the RMs from each sample, $\mu^*=\langle |{\rm RM}_1| \rangle^* -\langle |{\rm RM}_2| \rangle ^*$ (or RRM, RM$_z$, RRM$_z$), with $n_1=478$ ($z<3$) and $n_2=20$ ($z\ge 3$). 

We performed 10,000 bootstrap trials for $|{\rm RM}|$, $|{\rm RRM}|$, $|{\rm RM}_z|$  and $|{\rm RRM}_z|$ with the no-wt and sd-wt RM synthesis values, the peak RM QU fitting values, and the model fit $\Pi_{\rm e}$ values using the means, weighted means, and medians for calculating the test statistics. For the $|{\rm RM}_z|$ and $|{\rm RRM}_z|$ cases, rather than drawing randomly with replacement from sources already corrected for the redshift, the RM and RRM values were randomized with respect to their redshift values before selection such that RM$_z$ and RRM$_z$ were recomputed for the new samples of sizes $n_1$ and $n_2$. This is to ensure any difference that might be detected is not simply an effect of the true $n_2$ sample being multiplied by higher values. 

\begin{figure*}
\includegraphics[scale=0.26]{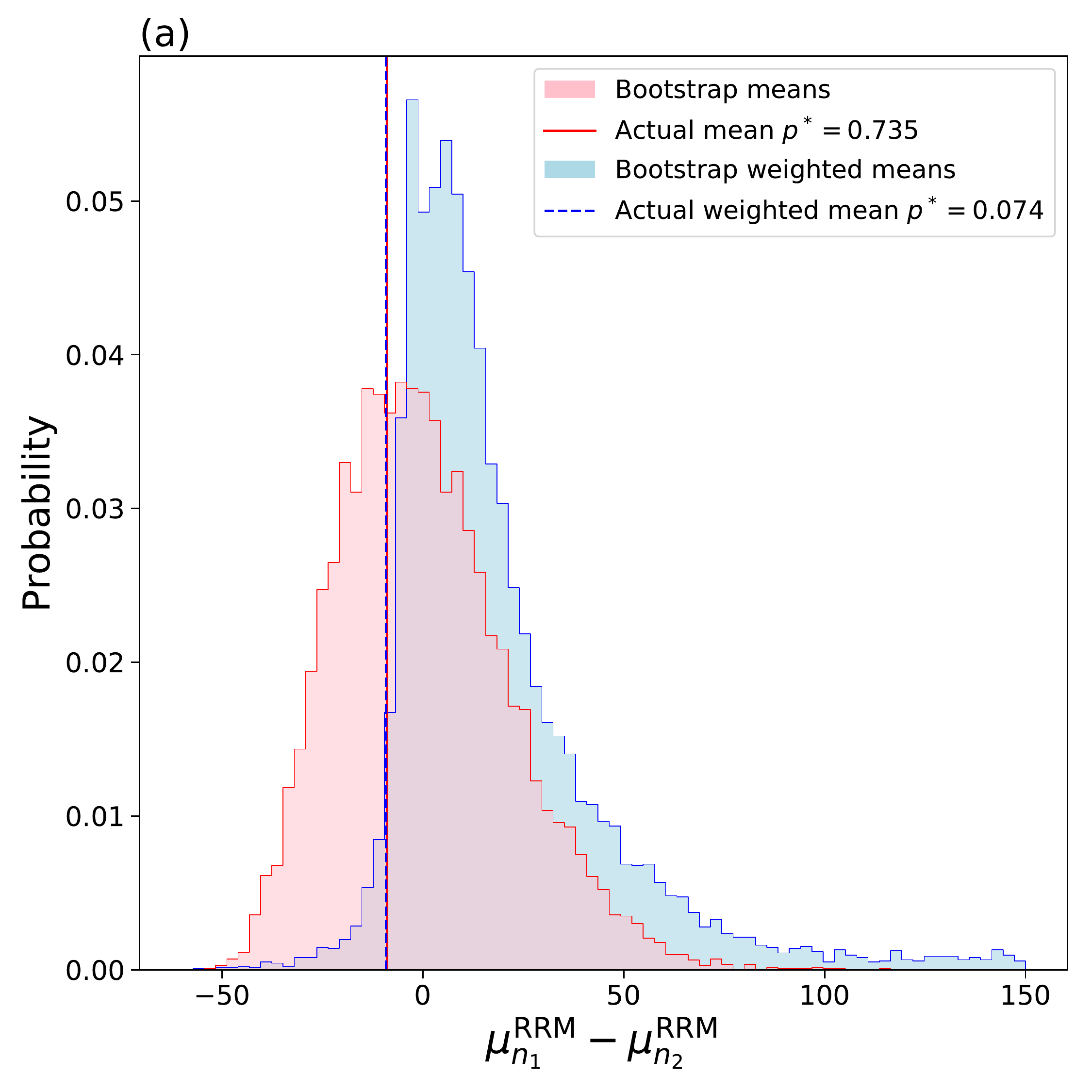}\includegraphics[scale=0.26]{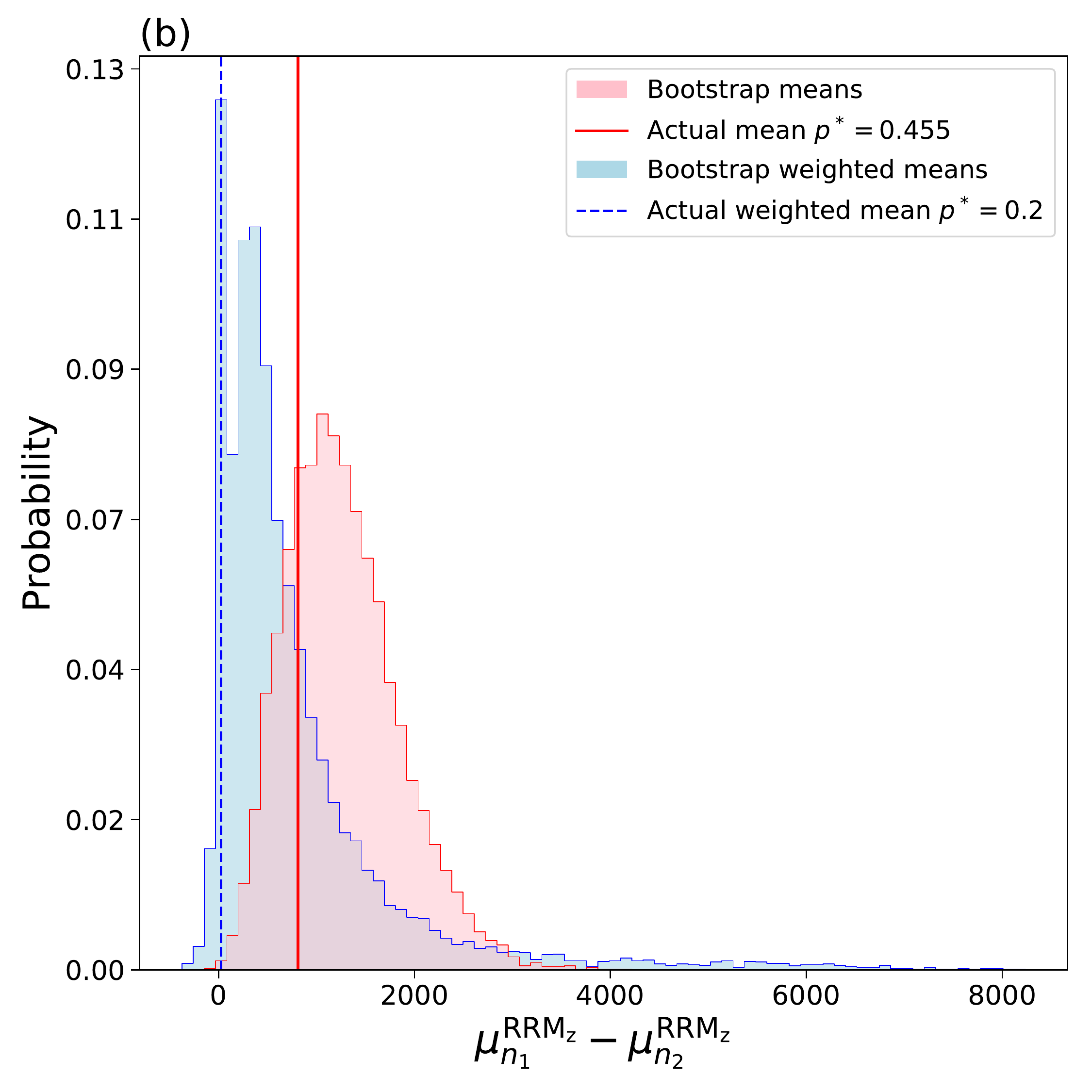}\includegraphics[scale=0.26]{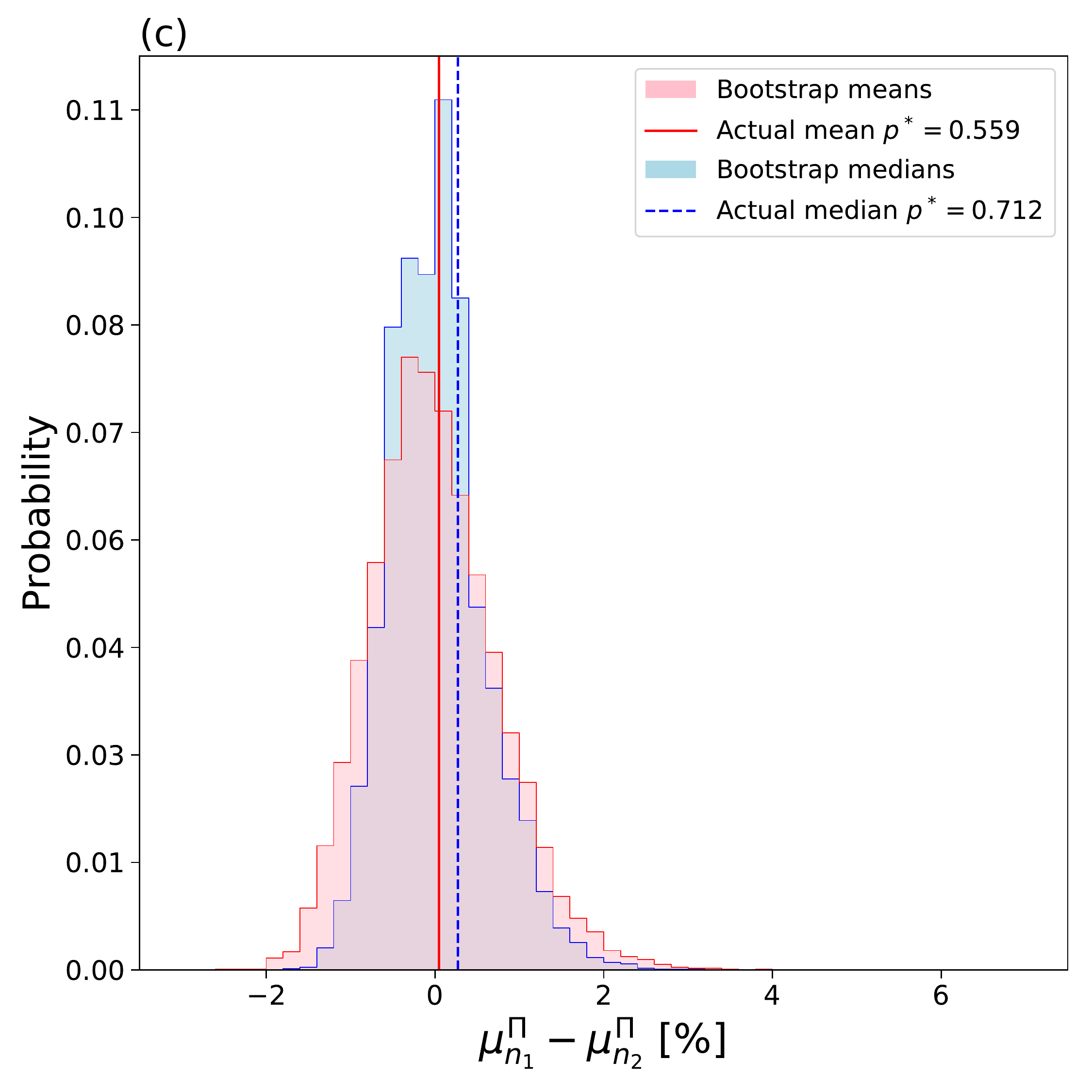}
\caption{Bootstrap test results from resampling the Galactic corrected RRM, the Galactic and redshift corrected RRM$_z$, and rest-frame polarization fraction. Panels (a) and (b) show results using the ``no-wt" RM values. Panel (a) shows the values and distributions using the RRM values, while panel (b) uses RRM$_z$ values. Panel (c) shows the values and distributions for the polarization fraction. The probability distributions are the results from 10,000 trials of random resampling. The solid and dashed vertical lines show the locations of the actual values using the low redshift ($n_1=478$) and high redshift ($n_2=20$) samples. The red distributions and solid lines show the results when the test statistic is calculated using the mean of the RRM absolute values (or mean of the polarization fraction), whereas the blue distributions and dashed lines show the results when the test statistic is calculated using the weighted means. See Table~\ref{tab:rmstats} for the mean and median RMs and polarization fractions.}
\label{fig:boots}
\end{figure*}

Some of the results from the bootstrap tests are shown in Fig.~\ref{fig:boots} for the rotation measures and the polarization fraction. From all of the RM cases the minimum p$^*$ values were 0.1, 0.07, and 0.06 for the means, weighted means, and medians, respectively. For the $\Pi_{\rm e}$ trials the p$^*$ values were 0.9, 0.7, and 0.6 for the means, weighted means, and medians. None of the bootstrap tests result in a statistically significant difference (p$^*<0.05$) between the two samples. 

It is possible that the choice of $z=3$ as a cutoff may affect the result. We did rerun the bootstrap tests, redefining the high-$z$ cutoff as $z=$1.5, 2., and 2.5. In all cases no significant difference was found between the high-$z$ and low-$z$ sources. %The original sample of high-$z$ sources were chosen to be predominantly at $z\ge4$, thus the redshift distribution for sources at intermediate redshifts may not be optimized for a lower redshift cutoff. 

\subsubsection{KS $\&$ AD tests}
\label{sec:kss}
The Kolmogorov--Smirnov test (or KS test) is a nonparametric test of equality used to compare a sample with a reference probability distribution or, as in our case, to compare two different samples and test the hypothesis that they are from the same parent population. The KS test quantifies a distance between the cumulative distribution functions (CDF) of the two samples. The main advantage of the KS test is its sensitivity to the shape of a distribution because it can detect differences everywhere along the scale. For formulae and details on the KS test see Appendix~\ref{sec:ksad}. 

We performed a KS test on the $|{\rm RM}|$ and  $|{\rm RRM}|$ distributions, with the CDFs for the sd-wt $|{\rm RRM}|$. The KS$_{n_1,n_2}$ values translate into p$^*$ values of 0.2, 0.1, and 0.1 for $|{\rm RM}|$, no-wt, sd-wt, and peak QU respectively, and 0.4, 0.4, and 0.5 for $|{\rm RRM}|$. 

We also performed a KS test on the rest-frame polarization fractions. The test was performed once using all available fitted polarized fractions ($n_1=480$ and $n_2=33$) and once excluding those that did not have a detection in the RM synthesis ($n_1=478$ and $n_2=20$). When considering all the fitted sources the p$^*=0.19$, but when only considering those with RM detections it increases to p$^*=0.27$. Using this test no significance difference is found between the high and low redshift sources. 

An alternative to the KS test is the two-sample Anderson Darling (AD) test. Both the KS and the AD test are based on the cumulative probability distribution of data. They are both based on calculating the distance between distributions at each unit of the scale. The AD test has the same advantages as mentioned for the KS test, with the additional advantages that it is more sensitive towards differences at the tails of the distributions and the AD test is better at detecting very small differences. For formulae and details on the AD test see Appendix~\ref{sec:ksad}.

The p$^*$ values for $|{\rm RM}|$ are 0.4, 0.3, and 0.3 and for $|{\rm RRM}|$  they are 0.6, 0.6, 0.8 for the no-wt , sd-wt, and peak QU RMs, respectively. The AD p$^*$ values for the polarization fraction are 0.06 and 0.3 for all sources ($n_1=480$ and $n_2=33$) and only those with RM detections ($n_1=478$ and $n_2=20$), respectively. 

The KS and AD test p-values can be difficult to compute and or unreliable with small sample sizes. The p$^*$ values presented above were determined from the distributions of the 10,000 bootstrap resampling iterations. The statistics are all largely dominated by relatively low sample sizes, $n\sim20$. Some work has been done on Bayesian statistical tests for such small samples of Faraday rotation data \citep[e.g.][]{Farnes17a} and hierarchical Bayesian methods have also been used to look for the magnetized large-scale structure \citep{Vacca16}, in order to enable reliable statistical frameworks for large surveys. Bayesian non-parametric two-sample tests may be more powerful than standard frequentist tests like the KS test \citep[e.g.][]{Labadi14} when it comes to small sample sizes. However, they require assumptions for the prior distributions, which can affect the conclusions if the assumed models are incorrect. 

\section{Discussion}
\label{sec:discussion}

\subsection{Depolarization and non-detections}
\label{sec:nondet}

The binned RMs as a function of redshift seem to imply possible lower intrinsic RM values at $z>3$. Given that the average RRM$_z$s at $z<3$ seem to increase as $(1+z)^2$ and a change is only seen after $z=3$, there is a possible indication of a change in the sources or environment at high redshifts. However, none of the statistical tests indicate a statistically significant difference between sources with $z>3$ and those with $z<3$. 

We can likely rule out any issues from the Galactic foreground correction, as the same results are seen for the RMs with and without the subtraction. It is possible that the results are due to an inadvertent selection effect, that we unintentionally selected high-$z$ sources with lower RMs. This is discussed further in the following subsections (Sections \ref{sec:subsamp} and \ref{sec:compare}), but to answer with higher certainty would be a larger sample of both high and low $z$ sources selected with the same criterion and observed and processed in the same way. 

The fact that the polarization fraction shows no difference between high and low redshift could indicate no intrinsic difference in the sources and or their environments. However, the apparent lack of difference in the high redshift source's polarizations fractions could also be explained if the high-$z$ sources are in high density environments that lead to larger polarized fractions but more depolarization. To estimate the amount of depolarization one can look at the ratio of $\Pi_{\rm e}$ at different frequencies \citep[as was done in e.g. ][]{Lamee16}. This is more complicated for our sample as the archival sources all have different frequency coverage. However, using the $\Pi_{\rm e}$ models we can compute the ratio using frequencies on either side of our chosen rest frequency of $15\,$GHz, such that $D_{\rm e}=\Pi_{\rm e}(20 \, {\rm GHz})/\Pi_{\rm e}(10 \, {\rm GHz})$. The median $D_{\rm e}$ for low-$z$ sources of 0.99, whereas for high-$z$ sources it is 1.45. This suggests that high-$z$ may be more depolarized. 

%The depolarization can change the depth into which one can see into a source. If at low redshifts we can see all the way through the source this results in a higher RM. However, if there is higher depolarization at high $z$ we are not seeing all the way into it (it is akin to the source becoming optically thick), resulting in a lower rotation measure. Thus if the depolarization for the high-$z$ sources is higher it could make sense for the RM to be smaller. 
Rotation measures depend on the strength of the magnetic field, the density of the environment, and the degree of order in the magnetic field in the local Faraday screen. Which effect dominates cannot be decided from RM measurements alone. \citet{Goodlet05} found no correlation of RM with redshift, but found a strong correlation of the RM dispersion across a source, or a source's multiple components, with redshift. This result is indicative of more chaotic field structures at higher redshifts, however, it requires higher resolution to resolve the sources (higher than our current data). Thus while it is possible for the higher density and or stronger magnetic field to lead to higher polarization and depolarization, the ordering of the magnetic field, or amount of turbulence, could explain the lack of higher intrinsic RMs seen at the higher redshifts. 

%However, the higher density leading to higher polarization assumes coherent magnetic fields. If there are disordered, or turbulent fields, then the polarization should be lower. Turbulent incoherent fields would also lead to lower RMs. 

Of the 29 sources we imaged and analyzed with $z>3$, there are 13 which had no detectable peaks in their Faraday dispersion functions (those with flag=0 in Table~\ref{tab:fvals}), and three low redshift sources that had no detections. This is because the signal-to-noise of the data was too low to get a proper detection and they have been depolarized (either physical depolarization and or beam depolarization). 

Given the low angular resolution of the majority of our observations, beam depolarization may be affecting the RM measurements. There are several sources with multiple FIRST sources, or multiple components, within the VLA D-configuration beam. Complex source structure is likely causing significant wavelength-independent depolarization, and or multiple emission components with different amounts of Faraday rotation are causing wavelength-dependent depolarization.

All of the sources with no RM detection have maximum Stokes $I$ brightnesses less than $20\,$mJy and a median S$_I$/N$_I$ of 20, whereas those with detections have a median S$_I$/N$_I$ of 150. Nearly $45\,$per cent of the $z>3$ sources did not have a detection, whereas only $33\,$per cent of the $z<3$ sources had no detection (although the $z<3$ group is a smaller sample). The S/N from RM synthesis for sources with non-detections has a mean of 5.2 for sources with $z<3$ and a mean of 4.1 for sources with $z>3$. The mean upper limit on the peak polarization fraction from RM synthesis ($8\sigma_{FDF}$) for non-detections with $z>3$ is $2.1\,$per cent and $4.7\,$per cent for $z>3$. However, for sources with detections the mean upper limit polarization fraction for $z<3$ is $0.3\,$per cent and $0.9\,$ per cent for $z>3$. Of the high-$z$ sources, the mean redshift of those with no detections is slightly higher than the mean redshift for those with detections ($\langle z \rangle=4.9$ compared to $\langle z \rangle =4.4$). 

If the high-$z$ sources are more depolarized, or less intrinsically polarized, then it is not unexpected that there would be more non-detections for those sources. However, fainter sources are more sensitive to calibration errors or artefacts (such as low-level ripples) that can interfere with a polarization. It is unclear at this time if the non-detections are more due to the source(s) just being to faint overall to be detected with our observations, effected by errors or artefacts, or if something physical in or around the source or intervening medium has depolarized it below the detection limit. Obviously a larger sample, from either targeted observations or surveys, of both high and low redshift sources would help to distinguish whether high-$z$ sources are more depolarized, and or if it is an observational or physical effect.

\subsection{Subsamples}
\label{sec:subsamp}

Thus far in our analysis, we have compared the RMs of high-$z$ to low-$z$ sources with the samples defined only by luminosity and Galactic latitude. For a proper analysis, the samples should be further subdivided by other characteristics such as galaxy type (e.g. AGN, quasar, X-ray loud, etc.), multicomponent sources, spectral index, those that are absorbers and those that aren't, and or those that are behind or near clusters and those that are not.  However, given that our high-$z$ sample size is already small (20 sources), further division into smaller groups would result in the loss of meaningful statistical power.

For example, we did cross match both the new and archival sources with the NED databases looking for those with clusters or groups within $2\,$Mpc, where the cluster redshift is less than the source redshift. This returned 83 sources, 15 of which are from the new sample of sources. Only eight of those 15 have RMs with a high enough significance from the RM synthesis and only four of those eight have $z>3$.\footnote{The four new high-z sources with RM detections matched to clusters are J021042-001818, J081333+350812, J165913+210116, and J222032+002535.} Four high-$z$ sources is not a large enough sample size to discern if they are statistically different from the low redshift sources. From this we can see that much larger samples are needed.

\subsection{Previous results}
\label{sec:compare}
This work has presented the largest number of $z>4$ RMs and polarization fractions yet determined. However, there have been previous studies that looked at the effect of redshift on polarization properties. 

Polarization data for greater than 40 high redshift ($z>2$) radio galaxies \citep{Carilli94,Carilli97,Athreya98,Pentericci00,Broderick07a,O'Sullivan11,Liu17} showed that several sources have rest-frame RMs values greater than $1000\,$rad m$^{-2}$. High resolution (mas) imaging of these sources showed large variation of the RM across the source and or multiple components. One interpretation of the finding of these high RMs is that these sources are located in cluster environments at high redshift \citep{Miley08}. \citet{Athreya98} pointed out that cluster cooling flows are unlikely to have a large role in forming deep Faraday screens at $z>2$, and suggest that for these sources with high RMs the Faraday screens are other collapsed galactic or sub-galactic sized objects in the environment of the sources. 

\citet{Kronberg08} looked at a sample of 268 sources out to $z\sim 3.7$ (with only two sources at $z\ge3$). They found that beyond $z\sim 2$ progressively fewer sources are found with a ``small" RM in the observer's frame, or rather RMs increase with increasing redshift, which would indicate significantly magnetized environments at high redshifts. This result was found by others as well, albeit with the use of smaller data sets \citep[][]{Welter84,You03}.

\citet{Hammond12} used 4003 sources with RMs from NVSS that they matched with spectroscopic redshifts with $z\le 5.3$, which resulted in 19 sources with $z\ge3$, but only two with $z\ge4$. They found no significant evolution of RMs with redshift, but found an anti-correlation of the extragalactic rotation measure with the fractional polarization of the source. They argue their findings require a population of magnetized intervening objects that lie outside our Galaxy in the foreground to the emitting sources and result from beam depolarization from small-scale fluctuations in the foreground magnetic fields or electron densities. \citet{Bernet12}, using a smaller sample of NVSS sources, was also unable to reproduce the evolution found by earlier works. They explain the discrepancy between their work and previous studies as due to severe depolarization induced by inhomogeneous Faraday screens on high wavelength radiation.

Four of our high redshift sources, and two of the archival sources with $z>3$, show $|{\rm RRM}_z|$ values greater than $1000\,$rad m$^{-2}$. However, it is difficult to directly compare our sample of high-$z$ sources with those of these previous works as the sample selections and RM computation methods differ. The majority of previously published high-$z$ RMs come from fitting the $\lambda^2$ slope with narrowband data, rather than RM synthesis (or QU fitting) with wideband data. Additionally, several of the previously published sources come from much brighter samples \citep[e.g.][which had sources brighter than $1\,$Jy]{Athreya98}. 

Many of these previous high-$z$ large RMs come from better resolution and or higher frequency data \citep[for example][which presented 10 high-$z$ sources at milliarcsec resolution at 5 and $8\,$GHz]{O'Sullivan11}. At higher frequencies, emission from AGN flat-spectrum cores tends to dominate over the steeper spectrum jets or lobes at lower frequencies. Similarly with higher resolution data, the core tends to be targeted. whereas with our lower resolution data, more diffuse emission is blended with the compact core emission, which can result in more depolarization. As discussed in Section~\ref{sec:subsamp}, it is necessary to compare matched samples in order to draw valid conclusions. Ideally the high-$z$ sources observed at higher resolutions and frequencies with narrowband data would be re-observed with matching observational setup and RM synthesis and or QU fitting done for a proper comparison.  

This is a good demonstration of why a new broadband large survey (deeper and higher resolution than NVSS) is needed. Surveys such as the new VLA Sky Survey \citep[VLASS,][]{Mao14} or ASKAP's POSSUM will produce millions of new rotation measures from which well-matched high and low redshift samples (and subsamples) can be analyzed.

\section{Conclusions}
\label{sec:conclusions}
We have presented the Stokes $I$ and linear polarization properties of a sample of 37 radio sources (38 source components), 29 of which have $z>3$ (27 with $z>4$). We performed fitting of the Stokes $I$ and polarization fraction spectra, which we used to obtain the $15\,$GHz rest-frame luminosities and polarization fractions. RM synthesis and $QU$ fitting were also performed to obtain rotation measures. This is the largest sample of RMs from $z>3.5$ sources. Using a map of Galactic rotation measures, we found the residual (or extragalactic) rotation measures, RRM, and the intrinsic, or redshift corrected RM$_z$ and RRM$_z$ values.

Using RM synthesis, we obtained significant RM detections for 16 of the 29 high-$z$ sources and six of the nine low-$z$ sources. $QU$ fitting was also performed on all sources using models with varying number of thin (delta functions) and thick (Gaussian) components. We found that for the sources with an RM detection from RM synthesis, the best-fitting $QU$ model was more complex than a single component, with the most common being a combination of two thin components. 

Using archival data, we created a luminosity matched control sample of 472 sources with $z<3$, also adding an additional four archival sources with $z>3$. We also fit for their rest-frame luminosity and polarization fractions. This allowed for a comparison of low versus high redshift polarization properties. We found a mean |RRM|$=55 \pm 23\,$rad m$^{-2}$ (depending on the type of measurement and weighting scheme) for high-$z$ sources and a mean |RRM|$=58 \pm 4\,$rad m$^{-2}$ for the low-$z$ sources. Both high and low-$z$ sources have a median rest-frame polarization fraction $\Pi_{\rm e}\simeq3.3\,$per cent. Using bootstrap, KS and AD tests we detect no significant difference between high and low redshift sources. 

While some previous works found indications for higher RMs at high-$z$, indicating denser more highly magnetized environments at earlier times, we detect no significant difference in observed or intrinsic RMs or rest-frame polarization fractions. To properly answer the question, a larger sample and further subdivision of the sources by things like source types, spectral indices, absorbers or known cluster sources, etc is necessary. Our sample of 20 $z>3$ sources is too small to break down further and get accurate statistics. The uncertainty demonstrated by such a small sample of high redshift sources is further evidence of why future large surveys such as POSSUM are so important.

\section{Acknowledgments}
The Dunlap Institute is funded through an endowment established by the David Dunlap family and the University of Toronto. T.V. and B.M.G. acknowledge the support of the Natural Sciences and Engineering Research Council of Canada (NSERC) through grant RGPIN-2015-05948, and of the Canada Research Chairs program. We thank the staff of the JVLA, which is operated by the National Radio Astronomy Observatory (NRAO). We would like to thank Niels Oppermann for his input and for the use of his code. 

\bibliographystyle{mnras}

\bibliography{bib1}

\begin{thebibliography}{}
\makeatletter
\relax
\def\mn@urlcharsother{\let\do\@makeother \do\$\do\&\do\#\do\^\do\_\do\%\do\~}
\def\mn@doi{\begingroup\mn@urlcharsother \@ifnextchar [ {\mn@doi@}
  {\mn@doi@[]}}
\def\mn@doi@[#1]#2{\def\@tempa{#1}\ifx\@tempa\@empty \href
  {http://dx.doi.org/#2} {doi:#2}\else \href {http://dx.doi.org/#2} {#1}\fi
  \endgroup}
\def\mn@eprint#1#2{\mn@eprint@#1:#2::\@nil}
\def\mn@eprint@arXiv#1{\href {http://arxiv.org/abs/#1} {{\tt arXiv:#1}}}
\def\mn@eprint@dblp#1{\href {http://dblp.uni-trier.de/rec/bibtex/#1.xml}
  {dblp:#1}}
\def\mn@eprint@#1:#2:#3:#4\@nil{\def\@tempa {#1}\def\@tempb {#2}\def\@tempc
  {#3}\ifx \@tempc \@empty \let \@tempc \@tempb \let \@tempb \@tempa \fi \ifx
  \@tempb \@empty \def\@tempb {arXiv}\fi \@ifundefined
  {mn@eprint@\@tempb}{\@tempb:\@tempc}{\expandafter \expandafter \csname
  mn@eprint@\@tempb\endcsname \expandafter{\@tempc}}}

\bibitem[\protect\citeauthoryear{{Anderson}, {Gaensler}  \& {Feain}}{{Anderson}
  et~al.}{2016}]{Anderson16}
{Anderson} C.~S.,  {Gaensler} B.~M.,   {Feain} I.~J.,  2016, \mn@doi [\apj]
  {10.3847/0004-637X/825/1/59}, \href
  {http://adsabs.harvard.edu/abs/2016ApJ...825...59A} {825, 59}

\bibitem[\protect\citeauthoryear{{Athreya}, {Kapahi}, {McCarthy}  \& {van
  Breugel}}{{Athreya} et~al.}{1998}]{Athreya98}
{Athreya} R.~M.,  {Kapahi} V.~K.,  {McCarthy} P.~J.,   {van Breugel} W.,  1998,
  \aap, \href {http://adsabs.harvard.edu/abs/1998A%26A...329..809A} {329, 809}

\bibitem[\protect\citeauthoryear{{Becker}, {White}  \& {Edwards}}{{Becker}
  et~al.}{1991}]{Becker91}
{Becker} R.~H.,  {White} R.~L.,   {Edwards} A.~L.,  1991, \mn@doi [\apjs]
  {10.1086/191529}, \href {http://adsabs.harvard.edu/abs/1991ApJS...75....1B}
  {75, 1}

\bibitem[\protect\citeauthoryear{{Becker}, {White}  \& {Helfand}}{{Becker}
  et~al.}{1995}]{Becker95}
{Becker} R.~H.,  {White} R.~L.,   {Helfand} D.~J.,  1995, \mn@doi [\apj]
  {10.1086/176166}, \href {http://adsabs.harvard.edu/abs/1995ApJ...450..559B}
  {450, 559}

\bibitem[\protect\citeauthoryear{{Bernet}, {Miniati}  \& {Lilly}}{{Bernet}
  et~al.}{2012}]{Bernet12}
{Bernet} M.~L.,  {Miniati} F.,   {Lilly} S.~J.,  2012, \mn@doi [\apj]
  {10.1088/0004-637X/761/2/144}, \href
  {http://adsabs.harvard.edu/abs/2012ApJ...761..144B} {761, 144}

\bibitem[\protect\citeauthoryear{{Brentjens} \& {de Bruyn}}{{Brentjens} \& {de
  Bruyn}}{2005}]{Brentjens05}
{Brentjens} M.~A.,  {de Bruyn} A.~G.,  2005, \mn@doi [\aap]
  {10.1051/0004-6361:20052990}, \href
  {http://adsabs.harvard.edu/abs/2005A%26A...441.1217B} {441, 1217}

\bibitem[\protect\citeauthoryear{{Broderick}, {De Breuck}, {Hunstead}  \&
  {Seymour}}{{Broderick} et~al.}{2007}]{Broderick07a}
{Broderick} J.~W.,  {De Breuck} C.,  {Hunstead} R.~W.,   {Seymour} N.,  2007,
  \mn@doi [\mnras] {10.1111/j.1365-2966.2006.11375.x}, \href
  {http://adsabs.harvard.edu/abs/2007MNRAS.375.1059B} {375, 1059}

\bibitem[\protect\citeauthoryear{{Burn}}{{Burn}}{1966}]{Burn66}
{Burn} B.~J.,  1966, \mn@doi [\mnras] {10.1093/mnras/133.1.67}, \href
  {http://adsabs.harvard.edu/abs/1966MNRAS.133...67B} {133, 67}

\bibitem[\protect\citeauthoryear{{Carilli} \& {Taylor}}{{Carilli} \&
  {Taylor}}{2002}]{Carilli02}
{Carilli} C.~L.,  {Taylor} G.~B.,  2002, \mn@doi [\araa]
  {10.1146/annurev.astro.40.060401.093852}, \href
  {http://adsabs.harvard.edu/abs/2002ARA%26A..40..319C} {40, 319}

\bibitem[\protect\citeauthoryear{{Carilli}, {Owen}  \& {Harris}}{{Carilli}
  et~al.}{1994}]{Carilli94}
{Carilli} C.~L.,  {Owen} F.~N.,   {Harris} D.~E.,  1994, \mn@doi [\aj]
  {10.1086/116870}, \href {http://adsabs.harvard.edu/abs/1994AJ....107..480C}
  {107, 480}

\bibitem[\protect\citeauthoryear{{Carilli}, {R{\"o}ttgering}, {van Ojik},
  {Miley}, {Breugel}  \& {W.~J.~M.~van}}{{Carilli} et~al.}{1997}]{Carilli97}
{Carilli} C.~L.,  {R{\"o}ttgering} H.~J.~A.,  {van Ojik} R.,  {Miley} G.~K.,
  {Breugel}  {W.~J.~M.~van} 1997, \mn@doi [\apjs] {10.1086/312973}, \href
  {http://adsabs.harvard.edu/abs/1997ApJS..109....1C} {109, 1}

\bibitem[\protect\citeauthoryear{{Condon}, {Cotton}, {Greisen}, {Yin},
  {Perley}, {Taylor}  \& {Broderick}}{{Condon} et~al.}{1998}]{Condon98}
{Condon} J.~J.,  {Cotton} W.~D.,  {Greisen} E.~W.,  {Yin} Q.~F.,  {Perley}
  R.~A.,  {Taylor} G.~B.,   {Broderick} J.~J.,  1998, \mn@doi [\aj]
  {10.1086/300337}, \href {http://adsabs.harvard.edu/abs/1998AJ....115.1693C}
  {115, 1693}

\bibitem[\protect\citeauthoryear{{Conway}, {Haves}, {Kronberg}, {Stannard},
  {Vallee}  \& {Wardle}}{{Conway} et~al.}{1974}]{Conway74}
{Conway} R.~G.,  {Haves} P.,  {Kronberg} P.~P.,  {Stannard} D.,  {Vallee}
  J.~P.,   {Wardle} J.~F.~C.,  1974, \mn@doi [\mnras]
  {10.1093/mnras/168.1.137}, \href
  {http://adsabs.harvard.edu/abs/1974MNRAS.168..137C} {168}

\bibitem[\protect\citeauthoryear{{Darling}}{{Darling}}{1957}]{Darling57}
{Darling} D.~A.,  1957, \mn@doi [Ann. Math. Statist.]
  {10.1214/aoms/1177706788}, 28, 823

\bibitem[\protect\citeauthoryear{{Douglas}, {Bash}, {Bozyan}, {Torrence}  \&
  {Wolfe}}{{Douglas} et~al.}{1996}]{Douglas96}
{Douglas} J.~N.,  {Bash} F.~N.,  {Bozyan} F.~A.,  {Torrence} G.~W.,   {Wolfe}
  C.,  1996, \mn@doi [\aj] {10.1086/117932}, \href
  {http://adsabs.harvard.edu/abs/1996AJ....111.1945D} {111, 1945}

\bibitem[\protect\citeauthoryear{{Efron}}{{Efron}}{1979}]{Efron79}
{Efron} B.,  1979, \mn@doi [Ann. Statist.] {10.1214/aos/1176344552}, 7, 1

\bibitem[\protect\citeauthoryear{{Farnes}, {Gaensler}  \& {Carretti}}{{Farnes}
  et~al.}{2014a}]{Farnes14}
{Farnes} J.~S.,  {Gaensler} B.~M.,   {Carretti} E.,  2014a, \mn@doi [\apjs]
  {10.1088/0067-0049/212/1/15}, \href
  {http://adsabs.harvard.edu/abs/2014ApJS..212...15F} {212, 15}

\bibitem[\protect\citeauthoryear{{Farnes}, {O'Sullivan}, {Corrigan}  \&
  {Gaensler}}{{Farnes} et~al.}{2014b}]{Farnes141}
{Farnes} J.~S.,  {O'Sullivan} S.~P.,  {Corrigan} M.~E.,   {Gaensler} B.~M.,
  2014b, \mn@doi [\apj] {10.1088/0004-637X/795/1/63}, \href
  {http://adsabs.harvard.edu/abs/2014ApJ...795...63F} {795, 63}

\bibitem[\protect\citeauthoryear{{Farnes}, {Rudnick}, {Gaensler}, {Haverkorn},
  {O'Sullivan}  \& {Curran}}{{Farnes} et~al.}{2017}]{Farnes17a}
{Farnes} J.~S.,  {Rudnick} L.,  {Gaensler} B.~M.,  {Haverkorn} M.,
  {O'Sullivan} S.~P.,   {Curran} S.~J.,  2017, \mn@doi [\apj]
  {10.3847/1538-4357/aa7060}, \href
  {http://adsabs.harvard.edu/abs/2017ApJ...841...67F} {841, 67}

\bibitem[\protect\citeauthoryear{{Farnsworth}, {Rudnick}  \&
  {Brown}}{{Farnsworth} et~al.}{2011}]{Farnsworth11}
{Farnsworth} D.,  {Rudnick} L.,   {Brown} S.,  2011, \mn@doi [\aj]
  {10.1088/0004-6256/141/6/191}, \href
  {http://adsabs.harvard.edu/abs/2011AJ....141..191F} {141, 191}

\bibitem[\protect\citeauthoryear{{Flesch}}{{Flesch}}{2015}]{Flesch15}
{Flesch} E.~W.,  2015, \mn@doi [\pasa] {10.1017/pasa.2015.10}, \href
  {http://cdsads.u-strasbg.fr/abs/2015PASA...32...10F} {32, 10}

\bibitem[\protect\citeauthoryear{{Flesch}}{{Flesch}}{2016}]{Flesch16}
{Flesch} E.~W.,  2016, \mn@doi [\pasa] {10.1017/pasa.2016.44}, \href
  {http://adsabs.harvard.edu/abs/2016PASA...33...52F} {33, e052}

\bibitem[\protect\citeauthoryear{{Gaensler}, {Landecker}, {Taylor}  \& {POSSUM
  Collaboration}}{{Gaensler} et~al.}{2010}]{Gaensler10}
{Gaensler} B.~M.,  {Landecker} T.~L.,  {Taylor} A.~R.,   {POSSUM Collaboration}
  2010, in American Astronomical Society Meeting Abstracts \#215. p.~515

\bibitem[\protect\citeauthoryear{{Goodlet} \& {Kaiser}}{{Goodlet} \&
  {Kaiser}}{2005}]{Goodlet05}
{Goodlet} J.~A.,  {Kaiser} C.~R.,  2005, \mn@doi [\mnras]
  {10.1111/j.1365-2966.2005.09000.x}, \href
  {http://adsabs.harvard.edu/abs/2005MNRAS.359.1456G} {359, 1456}

\bibitem[\protect\citeauthoryear{{Govoni} \& {Feretti}}{{Govoni} \&
  {Feretti}}{2004}]{Govoni04}
{Govoni} F.,  {Feretti} L.,  2004, \mn@doi [International Journal of Modern
  Physics D] {10.1142/S0218271804005080}, \href
  {http://adsabs.harvard.edu/abs/2004IJMPD..13.1549G} {13, 1549}

\bibitem[\protect\citeauthoryear{{Gregory}, {Scott}, {Douglas}  \&
  {Condon}}{{Gregory} et~al.}{1996}]{Gregory96}
{Gregory} P.~C.,  {Scott} W.~K.,  {Douglas} K.,   {Condon} J.~J.,  1996,
  \mn@doi [\apjs] {10.1086/192282}, \href
  {http://adsabs.harvard.edu/abs/1996ApJS..103..427G} {103, 427}

\bibitem[\protect\citeauthoryear{{Hales}, {Gaensler}, {Norris}  \&
  {Middelberg}}{{Hales} et~al.}{2012}]{Hales12a}
{Hales} C.~A.,  {Gaensler} B.~M.,  {Norris} R.~P.,   {Middelberg} E.,  2012,
  \mn@doi [\mnras] {10.1111/j.1365-2966.2012.21372.x}, \href
  {http://adsabs.harvard.edu/abs/2012MNRAS.424.2160H} {424, 2160}

\bibitem[\protect\citeauthoryear{{Hammond}, {Robishaw}  \&
  {Gaensler}}{{Hammond} et~al.}{2012}]{Hammond12}
{Hammond} A.~M.,  {Robishaw} T.,   {Gaensler} B.~M.,  2012, preprint, \href
  {http://adsabs.harvard.edu/abs/2012arXiv1209.1438H} {} (\mn@eprint {arXiv}
  {1209.1438v3})

\bibitem[\protect\citeauthoryear{{Heald}, {Braun}  \& {Edmonds}}{{Heald}
  et~al.}{2009}]{Heald09}
{Heald} G.,  {Braun} R.,   {Edmonds} R.,  2009, \mn@doi [\aap]
  {10.1051/0004-6361/200912240}, \href
  {http://adsabs.harvard.edu/abs/2009A%26A...503..409H} {503, 409}

\bibitem[\protect\citeauthoryear{{Helou}, {Madore}, {Schmitz}, {Bicay}, {Wu}
  \& {Bennett}}{{Helou} et~al.}{1991}]{Helou91}
{Helou} G.,  {Madore} B.~F.,  {Schmitz} M.,  {Bicay} M.~D.,  {Wu} X.,
  {Bennett} J.,  1991, \mn@doi [Astrophysics and Space Science Library]
  {10.1007/978-94-011-3250-3_10}, \href
  {http://adsabs.harvard.edu/abs/1991ASSL..171...89H} {171, 89}

\bibitem[\protect\citeauthoryear{{Helou}, {Madore}, {Schmitz}, {Wu}, {Corwin},
  {Lague}, {Bennett}  \& {Sun}}{{Helou} et~al.}{1995}]{Helou95}
{Helou} G.,  {Madore} B.~F.,  {Schmitz} M.,  {Wu} X.,  {Corwin} Jr. H.~G.,
  {Lague} C.,  {Bennett} J.,   {Sun} H.,  1995, \mn@doi [Astrophysics and Space
  Science Library] {10.1007/978-94-011-0397-8_10}, \href
  {http://adsabs.harvard.edu/abs/1995ASSL..203...95H} {203, 95}

\bibitem[\protect\citeauthoryear{{Homan}, {Ojha}, {Wardle}, {Roberts}, {Aller},
  {Aller}  \& {Hughes}}{{Homan} et~al.}{2002}]{Homan02}
{Homan} D.~C.,  {Ojha} R.,  {Wardle} J.~F.~C.,  {Roberts} D.~H.,  {Aller}
  M.~F.,  {Aller} H.~D.,   {Hughes} P.~A.,  2002, \mn@doi [\apj]
  {10.1086/338701}, \href {http://adsabs.harvard.edu/abs/2002ApJ...568...99H}
  {568, 99}

\bibitem[\protect\citeauthoryear{{Hovatta}, {Lister}, {Aller}, {Aller},
  {Homan}, {Kovalev}, {Pushkarev}  \& {Savolainen}}{{Hovatta}
  et~al.}{2012}]{Hovatta12}
{Hovatta} T.,  {Lister} M.~L.,  {Aller} M.~F.,  {Aller} H.~D.,  {Homan} D.~C.,
  {Kovalev} Y.~Y.,  {Pushkarev} A.~B.,   {Savolainen} T.,  2012, \mn@doi [\aj]
  {10.1088/0004-6256/144/4/105}, \href
  {http://adsabs.harvard.edu/abs/2012AJ....144..105H} {144, 105}

\bibitem[\protect\citeauthoryear{{Kimball} \& {Ivezi{\'c}}}{{Kimball} \&
  {Ivezi{\'c}}}{2008}]{Kimball08}
{Kimball} A.~E.,  {Ivezi{\'c}} {\v Z}.,  2008, \mn@doi [\aj]
  {10.1088/0004-6256/136/2/684}, \href
  {http://adsabs.harvard.edu/abs/2008AJ....136..684K} {136, 684}

\bibitem[\protect\citeauthoryear{{Kimball} \& {Ivezic}}{{Kimball} \&
  {Ivezic}}{2014}]{Kimball14}
{Kimball} A.,  {Ivezic} Z.,  2014, preprint, \href
  {http://adsabs.harvard.edu/abs/2014arXiv1401.1535K} {} (\mn@eprint {arXiv}
  {1401.1535})

\bibitem[\protect\citeauthoryear{{Klein}, {Mack}, {Gregorini}  \&
  {Vigotti}}{{Klein} et~al.}{2003}]{Klein03}
{Klein} U.,  {Mack} K.-H.,  {Gregorini} L.,   {Vigotti} M.,  2003, \mn@doi
  [\aap] {10.1051/0004-6361:20030825}, \href
  {http://adsabs.harvard.edu/abs/2003A%26A...406..579K} {406, 579}

\bibitem[\protect\citeauthoryear{{Kronberg} \& {Simard-Normandin}}{{Kronberg}
  \& {Simard-Normandin}}{1976}]{Kronberg76}
{Kronberg} P.~P.,  {Simard-Normandin} M.,  1976, \mn@doi [\nat]
  {10.1038/263653a0}, \href {http://adsabs.harvard.edu/abs/1976Natur.263..653K}
  {263, 653}

\bibitem[\protect\citeauthoryear{{Kronberg}, {Dufton}, {Li}  \&
  {Colgate}}{{Kronberg} et~al.}{2001}]{Kronberg01}
{Kronberg} P.~P.,  {Dufton} Q.~W.,  {Li} H.,   {Colgate} S.~A.,  2001, \mn@doi
  [\apj] {10.1086/322767}, \href
  {http://adsabs.harvard.edu/abs/2001ApJ...560..178K} {560, 178}

\bibitem[\protect\citeauthoryear{{Kronberg}, {Bernet}, {Miniati}, {Lilly},
  {Short}  \& {Higdon}}{{Kronberg} et~al.}{2008}]{Kronberg08}
{Kronberg} P.~P.,  {Bernet} M.~L.,  {Miniati} F.,  {Lilly} S.~J.,  {Short}
  M.~B.,   {Higdon} D.~M.,  2008, \mn@doi [\apj] {10.1086/527281}, \href
  {http://adsabs.harvard.edu/abs/2008ApJ...676...70K} {676, 70}

\bibitem[\protect\citeauthoryear{{Labadi}, {Masuadi}  \& {Zarepour}}{{Labadi}
  et~al.}{2014}]{Labadi14}
{Labadi} L.~A.,  {Masuadi} E.,   {Zarepour} M.,  2014, preprint, \href
  {http://adsabs.harvard.edu/abs/2014arXiv1411.3427L} {} (\mn@eprint {arXiv}
  {1411.3427})

\bibitem[\protect\citeauthoryear{{Lamee}, {Rudnick}, {Farnes}, {Carretti},
  {Gaensler}, {Haverkorn}  \& {Poppi}}{{Lamee} et~al.}{2016}]{Lamee16}
{Lamee} M.,  {Rudnick} L.,  {Farnes} J.~S.,  {Carretti} E.,  {Gaensler} B.~M.,
  {Haverkorn} M.,   {Poppi} S.,  2016, \mn@doi [\apj]
  {10.3847/0004-637X/829/1/5}, \href
  {http://adsabs.harvard.edu/abs/2016ApJ...829....5L} {829, 5}

\bibitem[\protect\citeauthoryear{{Landman}, {Roussel-Dupre}  \&
  {Tanigawa}}{{Landman} et~al.}{1982}]{Landman82}
{Landman} D.~A.,  {Roussel-Dupre} R.,   {Tanigawa} G.,  1982, \mn@doi [\apj]
  {10.1086/160383}, \href {http://adsabs.harvard.edu/abs/1982ApJ...261..732L}
  {261, 732}

\bibitem[\protect\citeauthoryear{{Liu}, {Jiang}, {Gu}  \& {Gurvits}}{{Liu}
  et~al.}{2017}]{Liu17}
{Liu} Y.,  {Jiang} D.~R.,  {Gu} M.,   {Gurvits} L.~I.,  2017, \mn@doi [\mnras]
  {10.1093/mnras/stx617}, \href
  {http://adsabs.harvard.edu/abs/2017MNRAS.468.2699L} {468, 2699}

\bibitem[\protect\citeauthoryear{{Mantovani}, {Mack}, {Montenegro-Montes},
  {Rossetti}  \& {Kraus}}{{Mantovani} et~al.}{2009}]{Mantovani09}
{Mantovani} F.,  {Mack} K.-H.,  {Montenegro-Montes} F.~M.,  {Rossetti} A.,
  {Kraus} A.,  2009, \mn@doi [\aap] {10.1051/0004-6361/200911815}, \href
  {http://adsabs.harvard.edu/abs/2009A%26A...502...61M} {502, 61}

\bibitem[\protect\citeauthoryear{{Mao} et~al.,}{{Mao} et~al.}{2014}]{Mao14}
{Mao} S.~A.,  et~al., 2014, preprint, \href
  {http://adsabs.harvard.edu/abs/2014arXiv1401.1875M} {} (\mn@eprint {arXiv}
  {1401.1875})

\bibitem[\protect\citeauthoryear{{Mestel} \& {Paris}}{{Mestel} \&
  {Paris}}{1984}]{Mestel84}
{Mestel} L.,  {Paris} R.~B.,  1984, \aap, \href
  {http://adsabs.harvard.edu/abs/1984A%26A...136...98M} {136, 98}

\bibitem[\protect\citeauthoryear{{Miley} \& {De Breuck}}{{Miley} \& {De
  Breuck}}{2008}]{Miley08}
{Miley} G.,  {De Breuck} C.,  2008, \mn@doi [\aapr]
  {10.1007/s00159-007-0008-z}, \href
  {http://adsabs.harvard.edu/abs/2008A%26ARv..15...67M} {15, 67}

\bibitem[\protect\citeauthoryear{{Murphy} et~al.,}{{Murphy}
  et~al.}{2010}]{Murphy10}
{Murphy} T.,  et~al., 2010, \mn@doi [\mnras]
  {10.1111/j.1365-2966.2009.15961.x}, \href
  {http://adsabs.harvard.edu/abs/2010MNRAS.402.2403M} {402, 2403}

\bibitem[\protect\citeauthoryear{{O'Sullivan}, {Gabuzda}  \&
  {Gurvits}}{{O'Sullivan} et~al.}{2011}]{O'Sullivan11}
{O'Sullivan} S.~P.,  {Gabuzda} D.~C.,   {Gurvits} L.~I.,  2011, \mn@doi
  [\mnras] {10.1111/j.1365-2966.2011.18915.x}, \href
  {http://adsabs.harvard.edu/abs/2011MNRAS.415.3049O} {415, 3049}

\bibitem[\protect\citeauthoryear{{O'Sullivan} et~al.,}{{O'Sullivan}
  et~al.}{2012}]{O'Sullivan12}
{O'Sullivan} S.~P.,  et~al., 2012, \mn@doi [\mnras]
  {10.1111/j.1365-2966.2012.20554.x}, \href
  {http://adsabs.harvard.edu/abs/2012MNRAS.421.3300O} {421, 3300}

\bibitem[\protect\citeauthoryear{{O'Sullivan}, {Purcell}, {Anderson}, {Farnes},
  {Sun}  \& {Gaensler}}{{O'Sullivan} et~al.}{2017}]{O'Sullivan17}
{O'Sullivan} S.~P.,  {Purcell} C.~R.,  {Anderson} C.~S.,  {Farnes} J.~S.,
  {Sun} X.~H.,   {Gaensler} B.~M.,  2017, \mn@doi [\mnras]
  {10.1093/mnras/stx1133}, \href
  {http://adsabs.harvard.edu/abs/2017MNRAS.469.4034O} {469, 4034}

\bibitem[\protect\citeauthoryear{{Offringa} \& {Smirnov}}{{Offringa} \&
  {Smirnov}}{2017}]{Offringa17}
{Offringa} A.~R.,  {Smirnov} O.,  2017, preprint, \href
  {http://adsabs.harvard.edu/abs/2017arXiv170606786O} {} (\mn@eprint {arXiv}
  {1706.06786})

\bibitem[\protect\citeauthoryear{{Offringa} et~al.,}{{Offringa}
  et~al.}{2014}]{Offringa14}
{Offringa} A.~R.,  et~al., 2014, \mn@doi [\mnras] {10.1093/mnras/stu1368},
  \href {http://adsabs.harvard.edu/abs/2014MNRAS.444..606O} {444, 606}

\bibitem[\protect\citeauthoryear{{Oppermann} et~al.,}{{Oppermann}
  et~al.}{2012}]{Oppermann12}
{Oppermann} N.,  et~al., 2012, \mn@doi [\aap] {10.1051/0004-6361/201118526},
  \href {http://adsabs.harvard.edu/abs/2012A%26A...542A..93O} {542, A93}

\bibitem[\protect\citeauthoryear{{Oppermann} et~al.,}{{Oppermann}
  et~al.}{2015}]{Oppermann15}
{Oppermann} N.,  et~al., 2015, \mn@doi [\aap] {10.1051/0004-6361/201423995},
  \href {http://adsabs.harvard.edu/abs/2015A%26A...575A.118O} {575, A118}

\bibitem[\protect\citeauthoryear{{Pentericci}, {Van Reeven}, {Carilli},
  {R{\"o}ttgering}  \& {Miley}}{{Pentericci} et~al.}{2000}]{Pentericci00}
{Pentericci} L.,  {Van Reeven} W.,  {Carilli} C.~L.,  {R{\"o}ttgering}
  H.~J.~A.,   {Miley} G.~K.,  2000, \mn@doi [\aaps] {10.1051/aas:2000104},
  \href {http://adsabs.harvard.edu/abs/2000A%26AS..145..121P} {145, 121}

\bibitem[\protect\citeauthoryear{{Planck Collaboration} et~al.,}{{Planck
  Collaboration} et~al.}{2014}]{Planck14a}
{Planck Collaboration} et~al., 2014, \mn@doi [\aap]
  {10.1051/0004-6361/201321591}, \href
  {http://adsabs.harvard.edu/abs/2014A%26A...571A..16P} {571, A16}

\bibitem[\protect\citeauthoryear{{Pratley} \& {Johnston-Hollitt}}{{Pratley} \&
  {Johnston-Hollitt}}{2016}]{Pratley16}
{Pratley} L.,  {Johnston-Hollitt} M.,  2016, \mn@doi [\mnras]
  {10.1093/mnras/stw1377}, \href
  {http://adsabs.harvard.edu/abs/2016MNRAS.462.3483P} {462, 3483}

\bibitem[\protect\citeauthoryear{{Pshirkov}, {Tinyakov}  \& {Urban}}{{Pshirkov}
  et~al.}{2015}]{Pshirkov15}
{Pshirkov} M.~S.,  {Tinyakov} P.~G.,   {Urban} F.~R.,  2015, \mn@doi [\mnras]
  {10.1093/mnras/stv1273}, \href
  {http://adsabs.harvard.edu/abs/2015MNRAS.452.2851P} {452, 2851}

\bibitem[\protect\citeauthoryear{{Rees}}{{Rees}}{1987}]{Rees87}
{Rees} M.~J.,  1987, \qjras, \href
  {http://adsabs.harvard.edu/abs/1987QJRAS..28..197R} {28, 197}

\bibitem[\protect\citeauthoryear{{Rees} \& {Reinhardt}}{{Rees} \&
  {Reinhardt}}{1972}]{Rees72}
{Rees} M.~J.,  {Reinhardt} M.,  1972, \aap, \href
  {http://adsabs.harvard.edu/abs/1972A%26A....19..189R} {19, 189}

\bibitem[\protect\citeauthoryear{{Rengelink}, {Tang}, {de Bruyn}, {Miley},
  {Bremer}, {Roettgering}  \& {Bremer}}{{Rengelink} et~al.}{1997}]{Rengelink97}
{Rengelink} R.~B.,  {Tang} Y.,  {de Bruyn} A.~G.,  {Miley} G.~K.,  {Bremer}
  M.~N.,  {Roettgering} H.~J.~A.,   {Bremer} M.~A.~R.,  1997, \mn@doi [\aaps]
  {10.1051/aas:1997358}, \href
  {http://adsabs.harvard.edu/abs/1997A%26AS..124..259R} {124}

\bibitem[\protect\citeauthoryear{{Rossetti}, {Dallacasa}, {Fanti}, {Fanti}  \&
  {Mack}}{{Rossetti} et~al.}{2008}]{Rossetti08}
{Rossetti} A.,  {Dallacasa} D.,  {Fanti} C.,  {Fanti} R.,   {Mack} K.-H.,
  2008, \mn@doi [\aap] {10.1051/0004-6361:20079047}, \href
  {http://adsabs.harvard.edu/abs/2008A%26A...487..865R} {487, 865}

\bibitem[\protect\citeauthoryear{{Scholz} \& {Stephens}}{{Scholz} \&
  {Stephens}}{1987}]{Scholz87}
{Scholz} F.~W.,  {Stephens} M.~A.,  1987, Journal of the American Statistical
  Association, 82, 918

\bibitem[\protect\citeauthoryear{{Simmons} \& {Stewart}}{{Simmons} \&
  {Stewart}}{1985}]{Simmons85}
{Simmons} J.~F.~L.,  {Stewart} B.~G.,  1985, \aap, \href
  {http://adsabs.harvard.edu/abs/1985A%26A...142..100S} {142, 100}

\bibitem[\protect\citeauthoryear{{Sun} et~al.,}{{Sun} et~al.}{2015}]{Sun15}
{Sun} X.~H.,  et~al., 2015, \mn@doi [\apj] {10.1088/0004-637X/811/1/40}, \href
  {http://adsabs.harvard.edu/abs/2015ApJ...811...40S} {811, 40}

\bibitem[\protect\citeauthoryear{{Taylor}, {Stil}  \& {Sunstrum}}{{Taylor}
  et~al.}{2009}]{Taylor09}
{Taylor} A.~R.,  {Stil} J.~M.,   {Sunstrum} C.,  2009, \mn@doi [\apj]
  {10.1088/0004-637X/702/2/1230}, \href
  {http://adsabs.harvard.edu/abs/2009ApJ...702.1230T} {702, 1230}

\bibitem[\protect\citeauthoryear{{Tribble}}{{Tribble}}{1991}]{Tribble91}
{Tribble} P.~C.,  1991, \mn@doi [\mnras] {10.1093/mnras/250.4.726}, \href
  {http://adsabs.harvard.edu/abs/1991MNRAS.250..726T} {250, 726}

\bibitem[\protect\citeauthoryear{{Urry} \& {Padovani}}{{Urry} \&
  {Padovani}}{1995}]{Urry95}
{Urry} C.~M.,  {Padovani} P.,  1995, \mn@doi [\pasp] {10.1086/133630}, \href
  {http://adsabs.harvard.edu/abs/1995PASP..107..803U} {107, 803}

\bibitem[\protect\citeauthoryear{{Vacca} et~al.,}{{Vacca}
  et~al.}{2016}]{Vacca16}
{Vacca} V.,  et~al., 2016, \mn@doi [\aap] {10.1051/0004-6361/201527291}, \href
  {http://adsabs.harvard.edu/abs/2016A%26A...591A..13V} {591, A13}

\bibitem[\protect\citeauthoryear{{Vaillancourt}}{{Vaillancourt}}{2006}]{Vaillancourt06}
{Vaillancourt} J.~E.,  2006, \mn@doi [\pasp] {10.1086/507472}, \href
  {http://adsabs.harvard.edu/abs/2006PASP..118.1340V} {118, 1340}

\bibitem[\protect\citeauthoryear{{Welter}, {Perry}  \& {Kronberg}}{{Welter}
  et~al.}{1984}]{Welter84}
{Welter} G.~L.,  {Perry} J.~J.,   {Kronberg} P.~P.,  1984, \mn@doi [\apj]
  {10.1086/161862}, \href {http://adsabs.harvard.edu/abs/1984ApJ...279...19W}
  {279, 19}

\bibitem[\protect\citeauthoryear{{Wenger} et~al.,}{{Wenger}
  et~al.}{2000}]{Wenger00}
{Wenger} M.,  et~al., 2000, \mn@doi [\aaps] {10.1051/aas:2000332}, \href
  {http://adsabs.harvard.edu/abs/2000A%26AS..143....9W} {143, 9}

\bibitem[\protect\citeauthoryear{{Widrow}, {Ryu}, {Schleicher}, {Subramanian},
  {Tsagas}  \& {Treumann}}{{Widrow} et~al.}{2012}]{Widrow12}
{Widrow} L.~M.,  {Ryu} D.,  {Schleicher} D.~R.~G.,  {Subramanian} K.,  {Tsagas}
  C.~G.,   {Treumann} R.~A.,  2012, \mn@doi [\ssr] {10.1007/s11214-011-9833-5},
  \href {http://adsabs.harvard.edu/abs/2012SSRv..166...37W} {166, 37}

\bibitem[\protect\citeauthoryear{{York} et~al.,}{{York} et~al.}{2000}]{York00}
{York} D.~G.,  et~al., 2000, \mn@doi [\aj] {10.1086/301513}, \href
  {http://adsabs.harvard.edu/abs/2000AJ....120.1579Y} {120, 1579}

\bibitem[\protect\citeauthoryear{{You}, {Han}  \& {Chen}}{{You}
  et~al.}{2003}]{You03}
{You} X.~P.,  {Han} J.~L.,   {Chen} Y.,  2003, Acta Astronomica Sinica, \href
  {http://adsabs.harvard.edu/abs/2003AcASn..44S.155Y} {44, 155}

\makeatother
\end{thebibliography}

\bsp

\appendix 

\section{Debiasing}
\label{sec:debias}

The polarized flux density is computed from the Stokes $Q$ and $U$ flux densities such that
\begin{equation}
P_0=\sqrt{Q^2+U^2}.
\label{eq:pol1}
\end{equation}
This yields a Rician, rather than Gaussian, noise distribution for the polarized images. The noise in a polarized intensity image has a non-zero mean and has higher probability of positive peaks above a given detection threshold than Gaussian noise. Therefore, the measured polarized flux density needs to be corrected for noise bias in order to obtain an estimate of the true polarized intensity $P$,
\begin{equation}
P=\sqrt{Q^2+U^2-(f\sigma_{QU})^2},
\label{eq:pol2}
\end{equation}
where $f$ is the debias factor and $\sigma_{QU}$ is the average noise of $Q$ and $U$. One can take a maximum likelihood approach to find $f$ \citep{Simmons85,Vaillancourt06,Hales12a}. The probability distribution function for  $P$ and $P_0$ is
\begin{equation}
F(P_0|P)=\frac{P_0}{\sigma_{QU}^2}J_0\left ( \frac{P_0 P}{\sigma_{QU}^2} \right ) \exp \left [ - \frac{P_0^2+P^2}{2\sigma_{QU}^2} \right ],
\label{eq:debiasp}
 \end{equation}
 where $J_0$ is a zero order Bessel function. The maximum likelihood estimator of $P$ is defined as the value of $P$ which maximizes $F(P_0|P)$ for a given $P_0$. This is equivalent to solving for $P$ using
 \begin{equation}
 P_0J_1 \left ( \frac{P_0 P}{\sigma_{QU}^2} \right ) -  PJ_0 \left ( \frac{P_0 P}{\sigma_{QU}^2} \right ) = 0,
 \label{eq:plike}
 \end{equation}
where $J_1$ is the first order Bessel function. This yields the debias factor $f$ for the given $P$ as
\begin{equation}
f= \sqrt{P_0^2-P^2}/\sigma_{QU}.
 \label{eq:plike2}
 \end{equation}
This value is generally found to be approximately one when the source has a S/N$\ge3$. 

Rather than numerically find $f$ for each channel of each source, we found one value of $f$ for each source using the median $\sigma_{QU}$ and median $P_0$ from all the channels. New values of $P_i$ were then calculated for each $i$th channel using the debias factor. Throughout the paper $P$ refers to the debiased value calculated using eq.~(\ref{eq:pol1}).

\section{Depolarization models}
\label{sec:append1}
Inhomogeneous Faraday screens cause Faraday rotation and depolarization
of the signal coming from background radio sources.

Fluctuations on scales smaller than the spatial resolution of radio
observations cause a depolarization by increasing the observing wavelength
of the signal. \citet{Burn66} assumes that these
fluctuations happen on a single characteristic scale and, in this case,
the depolarization can be approximated by the law,
\begin{equation}
P=P_0\exp(-c\lambda^4),
\label{eq:burn}
\end{equation}
where $c$ is quantity describing the unresolved rotation measure
fluctuations and $P_0$ is the intrinsic percentage of polarization.

This law does not give an appropriate description of the depolarization at
long wavelengths since the observed polarization at these wavelengths is
higher than the value predicted by this law. Assuming that the
fluctuations are not associated with a single scale but rather happening on
a range of scales, \citet{Tribble91} finds that the polarization at long
wavelengths is indeed larger and that the depolarization can be described
with a power law
\begin{equation}
P=A/\lambda^2,
\end{equation}
where $A$ is a constant depending on the spatial resolution of the
observations and the rotation measure dispersion.

\citet{Rossetti08} argue that, while at short wavelengths a
depolarization of the signal is observed, at longer wavelengths the
polarization rather stays constant. This behaviour is better described if a
Faraday screen that only partially covers the source is considered,
\begin{equation}
P=P_0(f_c\exp(-c\lambda^4) +(1-f_c)),
\end{equation}
where $f_c$ is the fraction of the source covered by the Faraday screen
\citep{Rossetti08, Mantovani09}. Though this depolarization model turns out be unphysical it may also reflect multiple components or more complex Faraday behaviour \citep[for more discussion on this see appendix A of][]{Farnes14}.

\section{KS $\&$ AD formulae}
\label{sec:ksad}

For the KS test, the maximum distance between the CDFs of the two samples KS$_{n_1,n_2}$ is defined as 
\begin{equation}
{\rm KS}_{n_1,n_2}=\sup{|F_1(x)-F_2(x)|}_x,
\label{eq:ks21} 
\end{equation}
where $F_1(x)$ is the CDF of sample 1 and $F_2(x)$ is the CDF of sample 2, with $x$ in our case being $|{\rm RRM}|$, $|{\rm RRM}_z|$, or $\Pi_{\rm e}$ . The null hypothesis (that the two samples are from the same population) can be rejected at level $\alpha$ if
\begin{equation}
{\rm KS}_{n_1,n_2}>c(\alpha)\sqrt{\frac{n_1+n_2}{n_1n_2}},
\label{eq:ks22} 
\end{equation}
where $c(\alpha)$ is approximated as
\begin{equation}
c(\alpha)= \sqrt{-\frac{1}{2}\log{ \left ( \frac{\alpha}{2}\right )} }.
\label{eq:ks23} 
\end{equation}
\citep{Darling57}. This gives a p$^*$ value of
\begin{equation}
{\rm p}^* \simeq 2 \, {\rm exp} \left [ -2 \left ({\rm KS}_{n_1,n_2}\sqrt{\frac{n_1n_2}{n_1+n_2}} \right )^2 \right ].
\label{eq:ks24} 
\end{equation}

For the AD test, the formula for calculating AD is
\begin{equation}
AD= \frac{1}{n_1n_2}\sum_{i=1}^{n_1+n_2} (N_i Z_{(n_1+n_2-n_2i)})^2\frac{1}{iZ_{(n_1+n_2-i)}},
\label{eq:ad1}
\end{equation}
where $Z_{(n_1+n_2)}$ represents the combined and ordered $X_1$ and $X_2$ of sizes $n_1$ and $n_2$, respectively, and $N_i$ represents the number of samples (or sources) in $X_2$ that are equal 
to or smaller than the $i$th observation in $Z_{(n_1+n_2)}$ \citep{Darling57,Scholz87}.

\section{Derived and fitted parameters for new sources}
\label{sec:aptables}
\FloatBarrier
%\begin{center}
%\onecolumn
\clearpage

\begin{table*}
 \setlength{\tabcolsep}{4.5pt}

\caption{Stokes I fit parameters. The parameters $\alpha_1$, $k$, $\nu_{\rm peak}$, and $\alpha_2$ are from eq.(\ref{eq:stksi1}) and (\ref{eq:stksi2}). The $\Delta$s are the $1\sigma$ uncertainties. The frequency $\nu_{\rm obs}$ is the observed frequency, and $\lambda_{\rm obs}$ the observed wavelength, for each source for a rest-frame frequency of $15\,$GHz, or wavelength of $2\,$cm. $I_{{\rm e}}$ is the fitted rest-frame ($15\,$GHz) Stokes $I$ flux density, with $L_{\rm e}$ being the rest-frame luminosity in units of W m$^{-2}$ Hz$^{-1}$. This table is an excerpt, with the full table available online.}
\label{tab:stksif}

%\begin{tabu}{lrrrrrrrrrrrcc}
\begin{tabular}{lSSSSSSSSSSSSS}
%\begin{tabu}{cr@{.}lr@{.}lr@{.}lr@{.}lr@{.}lr@{.}l@{.}l}
\hline 
 \multicolumn{1}{c}{Name}  &\multicolumn{1}{c}{ $\alpha_1$} & \multicolumn{1}{c}{$\Delta {\alpha_1}$} & \multicolumn{1}{c}{$k$ }& \multicolumn{1}{c}{$\Delta k$} & \multicolumn{1}{c}{$\alpha_2$ }& \multicolumn{1}{c}{$\Delta {\alpha_2}$ }& \multicolumn{1}{c}{$\nu_{\rm peak}$ }& \multicolumn{1}{c}{$\Delta {\nu_{\rm peak}}$ }&\multicolumn{1}{c}{$\nu_{\rm obs}$} &\multicolumn{1}{c}{$\lambda_{\rm obs}$} &\multicolumn{1}{c}{$I_{{\rm e}}$} &\multicolumn{1}{c}{$\Delta {I_{{\rm e}}}$ }& \multicolumn{1}{c}{$\log_{10}[L_{\rm e}] $}\\
 & & & \multicolumn{1}{c}{ [mJy]} & \multicolumn{1}{c}{[mJy]} & & & \multicolumn{1}{c}{[GHz]} & \multicolumn{1}{c}{[GHz]} & \multicolumn{1}{c}{[GHz]} &\multicolumn{1}{c}{[cm]} &\multicolumn{1}{c}{[mJy]} & \multicolumn{1}{c}{[mJy]} & \\
 \hline
J001115+144603 & -0.42 & 0.03 & 35.8 & 0.3 & {--} & {--} & {--} & {--} & 2.51 & 11.90 & 24.3 & 0.2 & 27.01  \\
J003126+150738 & 0.4 & 0.1 & 45.4 & 0.1 & {--} & {--} & {--} & {--} & 2.83 & 10.60 & 66.6 & 0.4 & 27.35  \\
J021042$-$001818 & -0.37 & 0.07 & 13.8 & 0.2 & {--} & {--} & {--} & {--} & 2.62 & 11.50 & 9.6 & 0.2 & 26.57  \\
J081333+350812 & -0.93 & 0.07 & 56.1 & 0.3 & {--} & {--} & {--} & {--} & 2.52 & 11.90 & 23.7 & 0.1 & 27.00  \\
J083644+005451 & -0.5 & 0.4 & 1.47 & 0.04 & {--} & {--} & {--} & {--} & 2.22 & 13.50 & 0.99 & 0.04 & 25.72  \\
J083946+511202 & -0.19 & 0.03 & 65.7 & 0.3 & {--} & {--} & {--} & {--} & 2.78 & 10.80 & 54.0 & 0.4 & 27.28  \\
J085111+142338 & -0.50 & 0.06 & 13.2 & 0.2 & {--} & {--} & {--} & {--} & 2.90 & 10.40 & 7.8 & 0.1 & 26.40  \\
J085853+345826 & -0.94 & 0.07 & 26.2 & 0.2 & {--} & {--} & {--} & {--} & 6.42 & 4.67 & 4.56 & 0.03 & 25.31  \\
J090600+574730 & -0.70 & 0.08 & 29.1 & 0.1 & {--} & {--} & {--} & {--} & 6.40 & 4.69 & 7.94 & 0.03 & 25.56  \\
J091316+591920 & -0.99 & 0.06 & 29.7 & 0.1 & {--} & {--} & {--} & {--} & 2.45 & 12.20 & 12.24 & 0.05 & 26.73  \\
J091824+063653 & -0.14 & 0.02 & 53.3 & 0.2 & {--} & {--} & {--} & {--} & 2.91 & 10.30 & 45.7 & 0.3 & 27.16  \\
J100424+122924 & 0.54 & 0.04 & 5.06 & 0.06 & {--} & {--} & {--} & {--} & 2.72 & 11.00 & 8.67 & 0.10 & 26.50  \\
J100645+462716 & -0.40 & 0.04 & 14.3 & 0.2 & {--} & {--} & {--} & {--} & 2.81 & 10.70 & 9.4 & 0.1 & 26.51  \\
J102551+192314 & -0.8 & 0.1 & 54.5 & 0.3 & {--} & {--} & {--} & {--} & 6.92 & 4.34 & 11.82 & 0.07 & 25.61  \\
J102623+254259 & -0.60 & 0.07 & 285.8 & 0.7 & {--} & {--} & {--} & {--} & 2.39 & 12.50 & 169.8 & 0.4 & 27.89  \\
J103601+500831 & -0.83 & 0.03 & 15.6 & 0.3 & {--} & {--} & {--} & {--} & 2.73 & 11.00 & 6.8 & 0.1 & 26.39  \\
J104624+590524a & -1.52 & 0.09 & 2.75 & 0.06 & {--} & {--} & {--} & {--} & 3.24 & 9.26 & 0.460 & 0.010 & 25.07  \\
J104624+590524b & -1.44 & 0.06 & 32.8 & 0.4 & {--} & {--} & {--} & {--} & 3.24 & 9.26 & 6.03 & 0.07 & 26.19  \\
J105320$-$001650 & -0.62 & 0.04 & 16.4 & 0.3 & {--} & {--} & {--} & {--} & 2.83 & 10.60 & 8.6 & 0.1 & 26.46  \\
J130738+150752 & -0.61 & 0.03 & 12.4 & 0.2 & {--} & {--} & {--} & {--} & 2.95 & 10.20 & 6.44 & 0.08 & 26.30  \\
J130940+573311 & -0.54 & 0.04 & 20.1 & 0.3 & {--} & {--} & {--} & {--} & 2.84 & 10.60 & 11.5 & 0.1 & 26.58  \\
J132512+112330 & -0.34 & 0.03 & 80.0 & 0.5 & {--} & {--} & {--} & {--} & 2.77 & 10.80 & 56.5 & 0.4 & 27.30  \\
J133342+491625 & -0.80 & 0.02 & 46.7 & 0.5 & {--} & {--} & {--} & {--} & 6.27 & 4.78 & 10.75 & 0.10 & 25.72  \\
J135135+284015 & -1.23 & 0.03 & 5.7 & 0.2 & {--} & {--} & {--} & {--} & 2.62 & 11.50 & 1.73 & 0.05 & 25.83  \\
J142738+331242 & -0.9 & 0.5 & 2.21 & 0.08 & {--} & {--} & {--} & {--} & 2.11 & 14.20 & 1.12 & 0.05 & 25.81  \\
J142952+544717 & -0.6 & 0.8 & 4.4 & 0.2 & {--} & {--} & {--} & {--} & 2.08 & 14.40 & 2.7 & 0.2 & 26.21  \\
J151002+570243 & -0.39 & 0.05 & 365.4 & 0.4 & {--} & {--} & {--} & {--} & 2.83 & 10.60 & 243.3 & 0.3 & 27.91  \\
J155633+351757 & -0.20 & 0.07 & 30.1 & 0.6 & {--} & {--} & {--} & {--} & 2.65 & 11.30 & 25.5 & 0.6 & 26.99  \\
J161105+084437 & -0.4 & 0.2 & 11.4 & 0.3 & 0.62 & 0.04 & 2.1 & 0.1 & 2.71 & 11.10 & 19.6 & 0.6 & 26.86  \\
J165913+210116 & -0.69 & 0.07 & 34.6 & 0.2 & {--} & {--} & {--} & {--} & 2.55 & 11.80 & 18.02 & 0.09 & 26.87  \\
J221356$-$002457 & -1.12 & 0.06 & 0 & 1 & {--} & {--} & {--} & {--} & 7.28 & 4.12 & 19.8 & 0.1 & 25.76  \\
J222032+002535 & -1.04 & 0.02 & 134.0 & 0.3 & {--} & {--} & {--} & {--} & 2.88 & 10.40 & 44.61 & 0.09 & 27.16  \\
J222235+001536 & -0.26 & 0.02 & 66.9 & 0.3 & {--} & {--} & {--} & {--} & 6.36 & 4.72 & 41.5 & 0.2 & 26.29  \\
J222843+011032 & -0.1 & 0.1 & 0.30 & 0.01 & {--} & {--} & {--} & {--} & 2.16 & 13.90 & 0.27 & 0.05 & 25.17  \\
J224924+004750 & -0.68 & 0.02 & 18.5 & 0.2 & {--} & {--} & {--} & {--} & 2.74 & 11.00 & 9.31 & 0.10 & 26.52  \\
J231443$-$090637 & 0.5 & 0.1 & 2.23 & 0.05 & {--} & {--} & {--} & {--} & 6.55 & 4.58 & 5.5 & 0.1 & 25.36  \\
J232604+001333 & -0.90 & 0.02 & 13.9 & 0.2 & {--} & {--} & {--} & {--} & 7.51 & 4.00 & 2.27 & 0.03 & 24.76  \\
J235018$-$000658 & -1.00 & 0.02 & 335.1 & 0.6 & {--} & {--} & {--} & {--} & 6.34 & 4.73 & 52.83 & 0.08 & 26.39  \\
\hline
%\end{tabu}
\end{tabular}
\normalsize
\end{table*}

%\onecolumn
%\clearpage

\begin{table*}
 \setlength{\tabcolsep}{4.5pt}

\caption{Polarized fraction fit parameters. The variable $f$ is the debias factor described in eq.(~\ref{eq:plike2}).The parameters $\beta$ and $c_1$ are for the power-law model of eq.(\ref{eq:pfrac1}). The parameters $c_2$, $c_3$, and $c_4$ are for the Gaussian model of eq.(\ref{eq:pfrac2}) and the offset Gaussian of eq.(\ref{eq:pfrac3}), with the parameter $c_5$ for the offset Gaussian of eq.(\ref{eq:pfrac3}). The $\Delta$s are the $1\sigma$ uncertainties. $\Pi_{\rm e}$ and  $P_{{\rm e}}$ are the rest-frame (15 GHz) polarized fraction and polarized intensity. This table is an excerpt, with the full table available online.}
\label{tab:stkspf}

%\begin{tabu}{lrrrrrrrrrrrrrrr}
\begin{tabular}{lSSSSSSSSSSSSSSS}
%\begin{tabu}{cr@{.}lr@{.}lr@{.}lr@{.}lr@{.}lr@{.}l@{.}l}
\hline 
 \multicolumn{1}{c}{Name}&\multicolumn{1}{c}{$f$}  &\multicolumn{1}{c}{ $\beta$} & \multicolumn{1}{c}{$\Delta {\beta}$} & \multicolumn{1}{c}{$c_1$ }& \multicolumn{1}{c}{$\Delta {c_1}$} & \multicolumn{1}{c}{$c_2$ }& \multicolumn{1}{c}{$\Delta {c_2}$ }& \multicolumn{1}{c}{$c_3$}& \multicolumn{1}{c}{$\Delta {c_3}$ }&\multicolumn{1}{c}{$c_4$} &\multicolumn{1}{c}{$\Delta {c_4}$} &\multicolumn{1}{c}{$c_5$ }&\multicolumn{1}{c}{$\Delta {c_5}$ } &\multicolumn{1}{r}{$\Pi_{\rm e}$}&\multicolumn{1}{c}{$P_{{\rm e}}$}\\
 % \rowfont{\tiny}
 & & & &  &  & & &  &  &  & &  & & \multicolumn{1}{r}{[$\, \%$\,]}& \multicolumn{1}{r}{[mJy]}\\
 \hline
J001115+144603 & 1.01 & {--} & {--} & {--} & {--} & 2.4 & 0.2 & 12 & 1 & 4.1 & 0.2 & 0.24 & 0.01 & 2.60 & 0.64  \\
J003126+150738 & 1.04 & {--} & {--} & {--} & {--} & 0.47 & 0.02 & 15.4 & 0.9 & 5.2 & 0.3 & {--} & {--} & 0.31 & 0.21  \\
J021042$-$001818 & 1.02 & {--} & {--} & {--} & {--} & 4.6 & 0.4 & 2.5 & 0.2 & 12 & 1 & {--} & {--} & 3.40 & 0.33  \\
J081333+350812 & 1.09 & {--} & {--} & {--} & {--} & 7.8 & 0.4 & 0.64 & 0.07 & 11.3 & 0.9 & {--} & {--} & 4.70 & 1.10  \\
J083644+005451 & 1.40 & -0.198 & -0.008 & 8.0 & 0.9 & {--} & {--} & {--} & {--} & {--} & {--} & {--} & {--} & 4.80 & 0.05  \\
J083946+511202 & 1.03 & {--} & {--} & {--} & {--} & 1.6 & 0.1 & 11 & 1 & 7.5 & 0.7 & {--} & {--} & 1.60 & 0.84  \\
J085111+142338 & 1.20 & -0.169 & -0.018 & 3.4 & 0.2 & {--} & {--} & {--} & {--} & {--} & {--} & {--} & {--} & 2.30 & 0.18  \\
J085853+345826 & 1.28 & 0.21 & 0.02 & 0.53 & 0.03 & {--} & {--} & {--} & {--} & {--} & {--} & {--} & {--} & 0.73 & 0.03  \\
J090600+574730 & 1.16 & {--} & {--} & {--} & {--} & 6.6 & 0.7 & 12 & 1 & 6.9 & 0.3 & {--} & {--} & 3.50 & 0.28  \\
J091316+591920 & 1.11 & -0.377 & -0.015 & 2.0 & 0.1 & {--} & {--} & {--} & {--} & {--} & {--} & {--} & {--} & 0.76 & 0.09  \\
J091824+063653 & 1.03 & -0.44 & -0.04 & 3.7 & 0.3 & {--} & {--} & {--} & {--} & {--} & {--} & {--} & {--} & 1.30 & 0.60  \\
J100424+122924 & 1.01 & 0.38 & 0.04 & 0.30 & 0.04 & {--} & {--} & {--} & {--} & {--} & {--} & {--} & {--} & 0.73 & 0.06  \\
J100645+462716 & 1.00 & -0.320 & -0.030 & 1.4 & 0.1 & {--} & {--} & {--} & {--} & {--} & {--} & {--} & {--} & 0.66 & 0.06  \\
J102551+192314 & 1.04 & {--} & {--} & {--} & {--} & 5.0 & 0.2 & 1.6 & 0.1 & 11 & 1 & {--} & {--} & 4.90 & 0.57  \\
J102623+254259 & 1.00 & {--} & {--} & {--} & {--} & 8.2 & 0.6 & 11 & 1 & 9.5 & 0.8 & {--} & {--} & 8.10 & 14.00  \\
J103601+500831 & 1.24 & -0.420 & -0.027 & 5.4 & 0.3 & {--} & {--} & {--} & {--} & {--} & {--} & {--} & {--} & 2.00 & 0.13  \\
J104624+590524a & 1.23 & {--} & {--} & {--} & {--} & 12.8 & 0.7 & 2.5 & 0.2 & 9.4 & 0.7 & {--} & {--} & 9.90 & 0.05  \\
J104624+590524b & 1.04 & -0.301 & -0.028 & 13.7 & 0.8 & {--} & {--} & {--} & {--} & {--} & {--} & {--} & {--} & 7.00 & 0.42  \\
J105320$-$001650 & 1.42 & 0.24 & 0.02 & 0.44 & 0.03 & {--} & {--} & {--} & {--} & {--} & {--} & {--} & {--} & 0.79 & 0.07  \\
J130738+150752 & 1.02 & {--} & {--} & {--} & {--} & 5.2 & 0.3 & 3.6 & 0.4 & 9.7 & 0.6 & {--} & {--} & 4.10 & 0.27  \\
J130940+573311 & 1.00 & 0.52 & 0.03 & 0.15 & 0.01 & {--} & {--} & {--} & {--} & {--} & {--} & {--} & {--} & 0.50 & 0.06  \\
J132512+112330 & 1.03 & 0.38 & 0.04 & 0.31 & 0.02 & {--} & {--} & {--} & {--} & {--} & {--} & {--} & {--} & 0.76 & 0.43  \\
J133342+491625 & 1.05 & -0.118 & -0.009 & 7.3 & 0.6 & {--} & {--} & {--} & {--} & {--} & {--} & {--} & {--} & 6.00 & 0.65  \\
J135135+284015 & 1.00 & -0.266 & -0.012 & 15.4 & 0.9 & {--} & {--} & {--} & {--} & {--} & {--} & {--} & {--} & 8.10 & 0.14  \\
J142738+331242 & 1.41 & {--} & {--} & {--} & {--} & 11 & 1 & 1.89 & 0.08 & 10 & 1 & 3.2 & 0.2 & 7.90 & 0.09  \\
J142952+544717 & 1.08 & -0.313 & -0.020 & 9.4 & 0.8 & {--} & {--} & {--} & {--} & {--} & {--} & {--} & {--} & 4.10 & 0.11  \\
J151002+570243 & 1.00 & {--} & {--} & {--} & {--} & 4.6 & 0.4 & 8.8 & 0.5 & 2.9 & 0.3 & {--} & {--} & 3.80 & 9.20  \\
J155633+351757 & 1.03 & -0.333 & -0.025 & 20 & 1 & {--} & {--} & {--} & {--} & {--} & {--} & {--} & {--} & 9.10 & 2.30  \\
J161105+084437 & 1.36 & {--} & {--} & {--} & {--} & 2.9 & 0.2 & 17.3 & 0.9 & 4.6 & 0.2 & 2.4 & 0.2 & 3.60 & 0.70  \\
J165913+210116 & 1.20 & {--} & {--} & {--} & {--} & 2.2 & 0.2 & 2.1 & 0.2 & 4.9 & 0.5 & 1.7 & 0.1 & 2.00 & 0.37  \\
J221356$-$002457 & 1.03 & {--} & {--} & {--} & {--} & 1.08 & 0.08 & 29 & 4 & 18 & 2 & {--} & {--} & 0.40 & 0.08  \\
J222032+002535 & 1.07 & -0.163 & -0.021 & 10.3 & 0.9 & {--} & {--} & {--} & {--} & {--} & {--} & {--} & {--} & 7.00 & 3.10  \\
J222235+001536 & 1.04 & {--} & {--} & {--} & {--} & 6.4 & 0.3 & 18 & 2 & 15.4 & 1.0 & {--} & {--} & 4.30 & 1.80  \\
J222843+011032 & 1.00 & -0.465 & -0.031 & 61 & 5 & {--} & {--} & {--} & {--} & {--} & {--} & {--} & {--} & 18.00 & 0.05  \\
J224924+004750 & 1.01 & {--} & {--} & {--} & {--} & 9.6 & 0.8 & 12 & 1 & 11 & 1 & {--} & {--} & 9.50 & 0.89  \\
J231443$-$090637 & 1.10 & 0.32 & 0.02 & 1.33 & 0.06 & {--} & {--} & {--} & {--} & {--} & {--} & {--} & {--} & 2.20 & 0.12  \\
J232604+001333 & 1.26 & 0.42 & 0.02 & 0.81 & 0.09 & {--} & {--} & {--} & {--} & {--} & {--} & {--} & {--} & 1.50 & 0.03  \\
J235018$-$000658 & 1.02 & {--} & {--} & {--} & {--} & 2.7 & 0.2 & 1.6 & 0.2 & 10 & 1 & 1.3 & 0.2 & 3.90 & 2.10  \\
\hline
\end{tabular}
\normalsize
\end{table*}

%\onecolumn
%\clearpage

\begin{table*}
 \setlength{\tabcolsep}{4.pt}
\begin{scriptsize}
\caption{RM-synthesis fit parameters. The flag column indicates if a peak in the Faraday dispersion function was detected (1) or not (0). The full width at half maximum of the RMSF is given by $\Phi$.The RM gives the position of the peak, $A$, while $\Delta {\rm RM}$ and $\Delta A$ are the $1\sigma$ uncertainties. This table is an excerpt, with the full table available online.}
\label{tab:rmsynf}

%\begin{tabu}{lrcrrrrrrrrrrrr}
\begin{tabular}{lccSSSSSSSSSSS}
%\begin{tabu}{cr@{.}lr@{.}lr@{.}lr@{.}lr@{.}lr@{.}l@{.}l}
\hline 
& & &\multicolumn{5}{c}{no-wt} & \multicolumn{5}{c}{sd-wt}\\
 \multicolumn{1}{c}{Name} & \multicolumn{1}{c}{RM$_{\chi}$} & \multicolumn{1}{c}{flag}  &\multicolumn{1}{c}{ $\Phi$} & \multicolumn{1}{c}{RM}& \multicolumn{1}{c}{$\Delta {\rm RM}$}& \multicolumn{1}{c}{$A$}& \multicolumn{1}{c}{$\Delta {A}$ }&\multicolumn{1}{c}{S/N} & \multicolumn{1}{c}{RM}& \multicolumn{1}{c}{$\Delta {\rm RM}$ }&\multicolumn{1}{c}{$A$} &\multicolumn{1}{c}{$\Delta {A}$} &\multicolumn{1}{c}{S/N} \\
\rowfont{\scriptsize}
& \multicolumn{1}{c}{[rad m$^{-2}$]} &&  \multicolumn{1}{c}{[rad m$^{-2}$]}&\multicolumn{1}{c}{[rad m$^{-2}$]}  &\multicolumn{1}{c}{[rad m$^{-2}$]} &\multicolumn{1}{c}{[$\%$]}&\multicolumn{1}{c}{[$\%$]} &  &\multicolumn{1}{c}{[rad m$^{-2}$]} &\multicolumn{1}{c}{[rad m$^{-2}$]}  &\multicolumn{1}{c}{[$\%$]}&\multicolumn{1}{c}{[$\%$]} & \\
 \hline
 
\scriptsize J001115+144603 & -11 & 1 & 76 & -11.9 & 0.6 & 1.51 & 0.02 & 28 & -6.5 & 0.5 & 1.85 & 0.03 & 32  \\
\scriptsize J003126+150738 & 22 & 1 & 135 & 15 & 2 & 0.260 & 0.007 & 17 & 10 & 2 & 0.236 & 0.007 & 16  \\
\scriptsize J021042$-$001818 & -2 & 1 & 82 & 5.1 & 0.7 & 2.65 & 0.05 & 25 & -4.5 & 0.7 & 2.80 & 0.05 & 25  \\
\scriptsize J081333+350812 & 12 & 1 & 224 & 12.6 & 0.6 & 5.05 & 0.03 & 78 & 11.9 & 0.6 & 5.12 & 0.03 & 77  \\
\scriptsize J083644+005451 & 3 & 0 & 56 & -8682 & 4 & 3.0 & 0.5 & 6 & -8688 & 5 & 2.7 & 0.5 & 6  \\
\scriptsize J083946+511202 & 12 & 1 & 228 & 10.5 & 0.8 & 1.45 & 0.01 & 63 & 13.5 & 0.8 & 1.47 & 0.01 & 63  \\
\scriptsize J085111+142338 & 39 & 0 & 213 & -60 & 30 & 1.2 & 0.3 & 4 & -3 & 20 & 1.4 & 0.3 & 4  \\
\scriptsize J085853+345826 & -49 & 0 & 239 & 8480 & 40 & 0.4 & 0.1 & 3 & 8720 & 40 & 0.4 & 0.1 & 3  \\
\scriptsize J090600+574730 & -5 & 1 & 219 & -5.5 & 0.6 & 5.87 & 0.03 & 84 & -5.2 & 0.6 & 5.90 & 0.03 & 83  \\
\scriptsize J091316+591920 & 33 & 0 & 232 & -7810 & 30 & 0.34 & 0.09 & 4 & -7810 & 30 & 0.33 & 0.09 & 4  \\
\scriptsize J091824+063653 & 38 & 1 & 218 & 256.2 & 0.9 & 1.045 & 0.009 & 50 & 259.0 & 0.9 & 1.051 & 0.009 & 51  \\
\scriptsize J100424+122924 & 204 & 0 & 383 & 150 & 50 & 0.4 & 0.1 & 3 & 150 & 50 & 0.4 & 0.1 & 4  \\
\scriptsize J100645+462716 & -37 & 0 & 225 & -2860 & 30 & 0.4 & 0.1 & 4 & -2860 & 30 & 0.4 & 0.1 & 3  \\
\scriptsize J102551+192314 & 13 & 1 & 224 & 17.3 & 0.6 & 3.30 & 0.02 & 76 & 16.0 & 0.7 & 3.33 & 0.02 & 75  \\
\scriptsize J102623+254259 & 10 & 1 & 390 & 9.92 & 0.06 & 8.11 & 0.01 & 1363 & 9.90 & 0.06 & 8.10 & 0.01 & 1343  \\
\scriptsize J103601+500831 & -5 & 0 & 222 & 2310 & 30 & 1.4 & 0.4 & 4 & 2330 & 30 & 1.3 & 0.4 & 3  \\
\scriptsize J104624+590524a & -3 & 1 & 241 & -12 & 6 & 7.3 & 0.4 & 9 & -7 & 5 & 7.7 & 0.4 & 10  \\
\scriptsize J104624+590524b & 6 & 1 & 241 & 8.3 & 0.5 & 6.43 & 0.03 & 111 & 8.3 & 0.5 & 6.40 & 0.03 & 110  \\
\scriptsize J105320$-$001650 & -161 & 0 & 241 & -1160 & 30 & 0.4 & 0.1 & 4 & -6690 & 30 & 0.4 & 0.1 & 4  \\
\scriptsize J130738+150752 & 11 & 1 & 474 & 6 & 4 & 3.91 & 0.06 & 29 & 11 & 3 & 3.95 & 0.06 & 30  \\
\scriptsize J130940+573311 & 5 & 0 & 251 & 9370 & 40 & 0.27 & 0.09 & 3 & 30 & 50 & 0.24 & 0.09 & 3  \\
\scriptsize J132512+112330 & 102 & 1 & 223 & 55 & 2 & 0.63 & 0.01 & 25 & 78 & 2 & 0.64 & 0.01 & 25  \\
\scriptsize J133342+491625 & 12 & 1 & 225 & 12.6 & 0.6 & 5.50 & 0.03 & 77 & 12.7 & 0.6 & 5.47 & 0.03 & 77  \\
\scriptsize J135135+284015 & 41 & 0 & 242 & -1980 & 40 & 4 & 1 & 3 & 2690 & 40 & 4 & 1 & 3  \\
\scriptsize J142738+331242 & 14 & 0 & 53 & 75 & 5 & 2.5 & 0.5 & 5 & 4802 & 5 & 2.4 & 0.5 & 5  \\
\scriptsize J142952+544717 & 7 & 0 & 53 & -6891 & 6 & 2.4 & 0.5 & 5 & -6880 & 6 & 2.0 & 0.5 & 4  \\
\scriptsize J151002+570243 & -12 & 1 & 215 & -74.3 & 0.3 & 2.904 & 0.008 & 159 & -82.9 & 0.3 & 2.595 & 0.008 & 145  \\
\scriptsize J155633+351757 & 6 & 1 & 73 & 7.8 & 0.7 & 7.2 & 0.1 & 24 & 7.4 & 0.6 & 7.6 & 0.1 & 26  \\
\scriptsize J161105+084437 & 1 & 0 & 42 & 4601 & 3 & 1.5 & 0.2 & 6 & 4547 & 4 & 1.3 & 0.2 & 6  \\
\scriptsize J165913+210116 & 6 & 1 & 40 & 99 & 1 & 1.20 & 0.06 & 9 & 96 & 1 & 1.30 & 0.08 & 11  \\
\scriptsize J221356$-$002457 & -44 & 1 & 218 & -35 & 2 & 0.48 & 0.01 & 19 & -34 & 3 & 0.44 & 0.01 & 17  \\
\scriptsize J222032+002535 & -13 & 1 & 43 & -12.90 & 0.06 & 6.58 & 0.01 & 259 & -12.83 & 0.05 & 6.53 & 0.02 & 182  \\
\scriptsize J222235+001536 & -17 & 1 & 43 & -16.78 & 0.06 & 5.84 & 0.01 & 200 & -16.92 & 0.06 & 5.37 & 0.01 & 164  \\
\scriptsize J222843+011032 & 16 & 0 & 59 & -7123 & 9 & 12 & 3 & 3 & -7123 & 9 & 11 & 3 & 3  \\
\scriptsize J224924+004750 & 2 & 1 & 81 & 4.5 & 0.2 & 8.16 & 0.05 & 79 & 4.1 & 0.2 & 8.85 & 0.05 & 79  \\
\scriptsize J231443$-$090637 & 0 & 0 & 73 & 7 & 8 & 2.1 & 0.5 & 4 & 8910 & 10 & 1.4 & 0.4 & 3  \\
\scriptsize J232604+001333 & 18 & 0 & 76 & -8174 & 5 & 1.5 & 0.2 & 8 & 2280 & 10 & 0.7 & 0.2 & 3  \\
\scriptsize J235018$-$000658 & 5 & 1 & 43 & 3.45 & 0.07 & 2.215 & 0.007 & 142 & 5.93 & 0.08 & 2.291 & 0.009 & 112  \\

\hline
\end{tabular}
\normalsize
\end{scriptsize}
\end{table*}

%\onecolumn
%\clearpage

\begin{table*}
 \setlength{\tabcolsep}{4.pt}

\caption{QU-fitting parameters. The model column designates which model was the best fit to the data, with ``D'' being a delta function, or thin component, and ``G" being a modified Gaussian, or thick component. The number of each in the model designates how many thin and thick components there are. If a source has multiple components each component is listed on a separate row, with a number following the model name designating which component (1, 2, or 3).  For a ``D" component $p$ is the modulus, $\phi_0$ is the position of the delta function, and $\psi_0$ is the angle. For a ``G" component $p$ is the Gaussian peak, $\phi_0$ is the position of the peak, $\psi_0$ is the angle, $\sigma_{\phi}$ is the width, and $N$ determines its deviation from Normality. The $\Delta$s are the $1\sigma$ uncertainties obtained from the MCMC fitting. The $F_{\rm max}$ columns indicate the peak after summing all of the model components. This table is an excerpt, with the full table available online.}
\label{tab:quff}

%\begin{tabu}{lrrrrrrrrrrrrr}
\begin{tabular}{lcSSSSSSSSSSSS}
%\begin{tabu}{cr@{.}lr@{.}lr@{.}lr@{.}lr@{.}lr@{.}l@{.}l}
\hline 
 \multicolumn{1}{c}{Name}  &\multicolumn{1}{c}{ model} &\multicolumn{1}{c}{ $p$} &\multicolumn{1}{c}{ $\Delta p$}&\multicolumn{1}{c}{ $\psi_0$}  &\multicolumn{1}{c}{ $\Delta {\psi_0}$}  &\multicolumn{1}{c}{ $\phi_0$}  &\multicolumn{1}{c}{ $\Delta {\phi_0}$}  &\multicolumn{1}{c}{ $\sigma_{\phi}$}  &\multicolumn{1}{c}{ $\Delta {\sigma_{\phi}}$}  &\multicolumn{1}{c}{ $N$}  &\multicolumn{1}{c}{ $\Delta {N}$}&\multicolumn{1}{c}{$F_{\rm max}$}  &\multicolumn{1}{c}{ $F_{\rm max}$}\\    
\rowfont{\tiny}
&&\multicolumn{1}{c}{[$\%$]}&\multicolumn{1}{c}{[$\%$]}&\multicolumn{1}{c}{[rad]}&\multicolumn{1}{c}{[rad]}&\multicolumn{1}{c}{[rad m$^{-2}$]}&\multicolumn{1}{c}{[rad m$^{-2}$]}&\multicolumn{1}{c}{[rad m$^{-2}$]}&\multicolumn{1}{c}{[rad m$^{-2}$]}&&& \multicolumn{1}{c}{[$\%$]}& \multicolumn{1}{c}{[mJy]}\\
 \hline
\scriptsize J001115+144603 & \scriptsize GG1 & 0.97 & 0.06 & 1.5 & 0.3 & 27.0 & 2.0 & 7.0 & 8.0 & 5.7 & 0.4 & 0.11 & 0.03  \\
\scriptsize J001115+144603 & \scriptsize GG2 & 1.6 & 0.2 & 1.6 & 0.2 & -7.0 & 2.0 & 5.0 & 8.0 & 2.0 & 1.0 & 0.11 & 0.03  \\
\scriptsize J003126+150738 & \scriptsize DD1 & 0.2 & 0.1 & 0.6 & 0.4 & -185 & 5 & {--} & {--} & {--} & {--} & 0.30 & 0.20  \\
\scriptsize J003126+150738 & \scriptsize DD2 & 0.3 & 0.1 & 0.5 & 0.2 & 7 & 3 & {--} & {--} & {--} & {--} & 0.30 & 0.20  \\
\scriptsize J021042$-$001818 & \scriptsize DG1 & 0.9 & 0.3 & 0.7 & 0.3 & 100 & 300 & {--} & {--} & {--} & {--} & 0.93 & 0.09  \\
\scriptsize J021042$-$001818 & \scriptsize DG2 & 3.511 & 0.006 & 1.5 & 0.1 & -9 & 1 & 9 & 10 & 5 & 1 & 0.93 & 0.09  \\
\scriptsize J081333+350812 & \scriptsize DDD1 & 4.4 & 0.3 & 0.72 & 0.05 & 6.5 & 0.5 & {--} & {--} & {--} & {--} & 4.40 & 0.65  \\
\scriptsize J081333+350812 & \scriptsize DDD2 & 0.3 & 0.2 & 0.8 & 0.4 & -340 & 20 & {--} & {--} & {--} & {--} & 4.40 & 0.65  \\
\scriptsize J081333+350812 & \scriptsize DDD3 & 2.6 & 0.4 & 1.1 & 0.1 & 38 & 1 & {--} & {--} & {--} & {--} & 4.40 & 0.65  \\
\scriptsize J083644+005451 & \scriptsize D & 2 & 1 & 0.7 & 0.7 & 10 & 60 & {--} & {--} & {--} & {--} & 2.50 & 0.03  \\
\scriptsize J083946+511202 & \scriptsize DDD1 & 0.3 & 0.1 & 1.0 & 0.3 & -48 & 2 & {--} & {--} & {--} & {--} & 1.30 & 0.68  \\
\scriptsize J083946+511202 & \scriptsize DDD2 & 1.3 & 0.2 & -0.4 & 0.1 & 2 & 1 & {--} & {--} & {--} & {--} & 1.30 & 0.68  \\
\scriptsize J083946+511202 & \scriptsize DDD3 & 0.3 & 0.2 & -1.2 & 0.4 & 103 & 5 & {--} & {--} & {--} & {--} & 1.30 & 0.68  \\
\scriptsize J085111+142338 & \scriptsize D & 1.3 & 0.5 & -1.6 & 0.4 & 30 & 20 & {--} & {--} & {--} & {--} & 1.30 & 0.10  \\
\scriptsize J085853+345826 & \scriptsize D & 0.3 & 0.2 & 0.3 & 0.7 & -900 & 70 & {--} & {--} & {--} & {--} & 0.33 & 0.03  \\
\scriptsize J090600+574730 & \scriptsize DG1 & 6.0 & 0.4 & -1.20 & 0.06 & 2 & 50 & {--} & {--} & {--} & {--} & 6.00 & 0.80  \\
\scriptsize J090600+574730 & \scriptsize DG2 & 1.188 & 0.003 & -1.6 & 0.4 & -220 & 7 & 70 & 10 & 13 & 1 & 6.00 & 0.80  \\
\scriptsize J091316+591920 & \scriptsize D & 0.3 & 0.2 & -0.3 & 0.8 & -1010 & 60 & {--} & {--} & {--} & {--} & 0.30 & 0.03  \\
\scriptsize J091824+063653 & \scriptsize G & 2.1 & 0.1 & -0.5 & 0.1 & 251.0 & 2.0 & 44.0 & 7.0 & 7.0 & 8.0 & 0.01 & 0.01  \\
\scriptsize J100424+122924 & \scriptsize D & 0.4 & 0.2 & -0.4 & 0.6 & 120 & 40 & {--} & {--} & {--} & {--} & 0.37 & 0.04  \\
\scriptsize J100645+462716 & \scriptsize D & 0.3 & 0.2 & -1.1 & 0.6 & 70 & 60 & {--} & {--} & {--} & {--} & 0.31 & 0.03  \\
\scriptsize J102551+192314 & \scriptsize DD1 & 3.6 & 0.3 & -1.06 & 0.09 & 15 & 1 & {--} & {--} & {--} & {--} & 3.60 & 0.86  \\
\scriptsize J102551+192314 & \scriptsize DD2 & 0.6 & 0.1 & -0.4 & 0.6 & -112 & 6 & {--} & {--} & {--} & {--} & 3.60 & 0.86  \\
\scriptsize J102623+254259 & \scriptsize DDD1 & 1.6 & 0.1 & 0.22 & 0.04 & -3.4 & 0.4 & {--} & {--} & {--} & {--} & 4.10 & 6.20  \\
\scriptsize J102623+254259 & \scriptsize DDD2 & 4.1 & 0.2 & 0.43 & 0.02 & -8.3 & 0.2 & {--} & {--} & {--} & {--} & 4.10 & 6.20  \\
\scriptsize J102623+254259 & \scriptsize DDD3 & 2.6 & 0.2 & -0.28 & 0.04 & 47.8 & 0.4 & {--} & {--} & {--} & {--} & 4.10 & 6.20  \\
\scriptsize J103601+500831 & \scriptsize D & 1.1 & 0.4 & 0.3 & 0.7 & -1100 & 50 & {--} & {--} & {--} & {--} & 1.10 & 0.07  \\
\scriptsize J104624+590524a & \scriptsize G & 11.0 & 2.0 & 1.4 & 0.3 & -7.0 & 3.0 & 30.0 & 30.0 & 12.0 & 10.0 & 0.11 & 0.00  \\
\scriptsize J104624+590524b & \scriptsize GG1 & 1.2 & 0.3 & 0.5 & 0.2 & -610.0 & 20.0 & 90.0 & 100.0 & 4.0 & 2.0 & 0.08 & 0.01  \\
\scriptsize J104624+590524b & \scriptsize GG2 & 11.0 & 0.8 & -1.37 & 0.06 & 6.5 & 0.5 & 39.0 & 3.0 & 8.0 & 3.0 & 0.08 & 0.01  \\
\scriptsize J105320$-$001650 & \scriptsize D & 0.3 & 0.2 & 0.4 & 0.7 & 270 & 40 & {--} & {--} & {--} & {--} & 0.34 & 0.03  \\
\scriptsize J130738+150752 & \scriptsize DD1 & 2.5 & 0.5 & -0.3 & 0.2 & -47 & 2 & {--} & {--} & {--} & {--} & 2.50 & 0.16  \\
\scriptsize J130738+150752 & \scriptsize DD2 & 1.8 & 0.4 & -1.5 & 0.3 & 94 & 3 & {--} & {--} & {--} & {--} & 2.50 & 0.16  \\
\scriptsize J130940+573311 & \scriptsize D & 0.2 & 0.1 & -0.8 & 0.7 & 30 & 50 & {--} & {--} & {--} & {--} & 0.23 & 0.03  \\
\scriptsize J132512+112330 & \scriptsize DD1 & 0.20 & 0.03 & -1.6 & 0.7 & 522 & 6 & {--} & {--} & {--} & {--} & 0.69 & 0.38  \\
\scriptsize J132512+112330 & \scriptsize DD2 & 0.69 & 0.06 & 0.1 & 0.2 & 65 & 1 & {--} & {--} & {--} & {--} & 0.69 & 0.38  \\
\scriptsize J133342+491625 & \scriptsize DD1 & 0.3 & 0.3 & -0.5 & 0.4 & 147 & 5 & {--} & {--} & {--} & {--} & 5.40 & 1.00  \\
\scriptsize J133342+491625 & \scriptsize DD2 & 5.4 & 0.4 & 0.22 & 0.04 & 15.2 & 0.4 & {--} & {--} & {--} & {--} & 5.40 & 1.00  \\
\scriptsize J135135+284015 & \scriptsize D & 3 & 2 & 0.5 & 0.8 & -1530 & 90 & {--} & {--} & {--} & {--} & 2.90 & 0.04  \\
\scriptsize J142738+331242 & \scriptsize D & 2 & 1 & 1.4 & 0.8 & -560 & 90 & {--} & {--} & {--} & {--} & 1.60 & 0.02  \\
\scriptsize J142952+544717 & \scriptsize D & 2 & 1 & -0.9 & 0.7 & -1010 & 90 & {--} & {--} & {--} & {--} & 2.00 & 0.07  \\
\scriptsize J151002+570243 & \scriptsize DDD1 & 2.1 & 0.1 & 1.6 & 0.1 & 18 & 1 & {--} & {--} & {--} & {--} & 2.20 & 5.40  \\
\scriptsize J151002+570243 & \scriptsize DDD2 & 2.2 & 0.2 & -0.33 & 0.08 & -127 & 1 & {--} & {--} & {--} & {--} & 2.20 & 5.40  \\
\scriptsize J151002+570243 & \scriptsize DDD3 & 0.54 & 0.04 & 1.00 & 0.08 & 128 & 2 & {--} & {--} & {--} & {--} & 2.20 & 5.40  \\
\scriptsize J155633+351757 & \scriptsize DD1 & 2 & 2 & 0.4 & 0.3 & 240 & 10 & {--} & {--} & {--} & {--} & 7.00 & 1.50  \\
\scriptsize J155633+351757 & \scriptsize DD2 & 7 & 2 & 0.8 & 0.3 & 7 & 6 & {--} & {--} & {--} & {--} & 7.00 & 1.50  \\
\scriptsize J161105+084437 & \scriptsize D & 1.4 & 0.8 & 1.5 & 0.7 & 1060 & 60 & {--} & {--} & {--} & {--} & 1.40 & 0.18  \\
\scriptsize J165913+210116 & \scriptsize DD1 & 1.1 & 0.4 & 0.3 & 0.5 & 290 & 30 & {--} & {--} & {--} & {--} & 1.20 & 0.15  \\
\scriptsize J165913+210116 & \scriptsize DD2 & 1.2 & 0.5 & 1.1 & 0.5 & 90 & 30 & {--} & {--} & {--} & {--} & 1.20 & 0.15  \\
\scriptsize J221356$-$002457 & \scriptsize DDD1 & 0.46 & 0.05 & 0.2 & 0.2 & -31 & 1 & {--} & {--} & {--} & {--} & 0.46 & 0.25  \\
\scriptsize J221356$-$002457 & \scriptsize DDD2 & 0.20 & 0.03 & 0.1 & 0.2 & 242 & 2 & {--} & {--} & {--} & {--} & 0.46 & 0.25  \\
\scriptsize J221356$-$002457 & \scriptsize DDD3 & 0.23 & 0.04 & -0.1 & 0.2 & -271 & 1 & {--} & {--} & {--} & {--} & 0.46 & 0.25  \\
\scriptsize J222032+002535 & \scriptsize DG1 & 6.6 & 0.2 & 1.32 & 0.04 & -10 & 30 & {--} & {--} & {--} & {--} & 6.60 & 3.50  \\
\scriptsize J222032+002535 & \scriptsize DG2 & 2.472 & 0.002 & -1.1 & 0.1 & 260 & 20 & 100 & 10 & 9 & 2 & 6.60 & 3.50  \\
\scriptsize J222235+001536 & \scriptsize DD1 & 4.7 & 0.2 & -0.63 & 0.09 & -23 & 1 & {--} & {--} & {--} & {--} & 4.70 & 2.40  \\
\scriptsize J222235+001536 & \scriptsize DD2 & 1.8 & 0.5 & -1.6 & 0.1 & 3 & 2 & {--} & {--} & {--} & {--} & 4.70 & 2.40  \\
\scriptsize J222843+011032 & \scriptsize D & 8 & 4 & 0.2 & 0.7 & -9 & 80 & {--} & {--} & {--} & {--} & 7.70 & 0.02  \\
\scriptsize J224924+004750 & \scriptsize DD1 & 8.5 & 0.5 & 1.12 & 0.09 & 5.3 & 0.5 & {--} & {--} & {--} & {--} & 8.50 & 0.77  \\
\scriptsize J224924+004750 & \scriptsize DD2 & 1.8 & 0.7 & 1.2 & 0.4 & 45 & 3 & {--} & {--} & {--} & {--} & 8.50 & 0.77  \\
\scriptsize J231443$-$090637 & \scriptsize D & 1.4 & 0.8 & 0.5 & 0.8 & 1250 & 70 & {--} & {--} & {--} & {--} & 1.40 & 0.05  \\
\scriptsize J232604+001333 & \scriptsize D & 0.7 & 0.7 & -1.3 & 0.7 & 400 & 80 & {--} & {--} & {--} & {--} & 0.65 & 0.04  \\
\scriptsize J235018$-$000658 & \scriptsize DG1 & 1.3 & 0.2 & 0.30 & 0.10 & 5 & 50 & {--} & {--} & {--} & {--} & 1.30 & 1.60  \\
\scriptsize J235018$-$000658 & \scriptsize DG2 & 2.785 & 0.002 & 0.49 & 0.06 & 17.2 & 0.6 & 19 & 3 & 7 & 1 & 1.30 & 1.60  \\
\hline
\end{tabular}
\normalsize
\end{table*}

%\onecolumn
%\clearpage
\FloatBarrier

%\twocolumn
%\onecolumn
\onecolumn

\section{List of archival sources}
\label{sec:appendA}
\FloatBarrier
%\onecolumn
\begin{centering}
\begin{longtable}{ccrrrrrrrr}
\caption{List of archival source properties. The values of $I_{\rm e}$, $L_{\rm e}$, and $\Pi_{\rm e}$ are the $15\,$GHz rest-frame values of the brightness, luminosity, and polarization fraction. The units of $L_{\rm e}$ are W m$^{-2}$ Hz$^{-1}$. The RM, GRM, and RRM are the reported rotation measure, Galactic RM, and residual RM, respectively. This table is an excerpt, with the full table available online.}\label{tab:archsrcs} \\

\hline
RA & Dec & \multicolumn{1}{c}{$z$} & \multicolumn{1}{c}{$I_{e}$ }& \multicolumn{1}{c}{$\alpha$}&\multicolumn{1}{c}{$\log_{10}[L_{e}]$}&\multicolumn{1}{c}{$\Pi_{e}$}&\multicolumn{1}{c}{RM} & \multicolumn{1}{c}{GRM} & \multicolumn{1}{c}{RRM}  \\
\hline
(J2000) & (J2000) &  & \multicolumn{1}{c}{[mJy]} & & &\multicolumn{1}{c}{[$\%$]} & \multicolumn{1}{c}{[rad m$^{-2}$]}& \multicolumn{1}{c}{[rad m$^{-2}$]}& \multicolumn{1}{c}{[rad m$^{-2}$]}\\
\hline
\endfirsthead
\caption*{List of archival source properties.}\\
\hline
RA & Dec & \multicolumn{1}{c}{$z$} & \multicolumn{1}{c}{$I_{e}$ }& \multicolumn{1}{c}{$\alpha$}&\multicolumn{1}{c}{$\log_{10}[L_{e}]$}&\multicolumn{1}{c}{$\Pi_{e}$}&\multicolumn{1}{c}{RM} & \multicolumn{1}{c}{GRM} & \multicolumn{1}{c}{RRM}  \\
\hline
(J2000) & (J2000) &  & \multicolumn{1}{c}{[mJy]} & & &\multicolumn{1}{c}{[$\%$]} & \multicolumn{1}{c}{[rad m$^{-2}$]}& \multicolumn{1}{c}{[rad m$^{-2}$]}& \multicolumn{1}{c}{[rad m$^{-2}$]}\\
\hline
\endhead
\hline 
\multicolumn{10}{r}{{Continued on next page}} \\
\endfoot

\hline \\
\endlastfoot

00:03:22.00 & $-$17:27:11.40 &$ 1.47 $&$ 884.10 $&$ -0.71 $&$ 27.70 $&$ 2.54 $&$ -26.9 $&$ -2.3 $&$ -25.0 $ \\
00:05:59.41 & +16:09:46.70 &$ 0.45 $&$ 204.10 $&$ -0.70 $&$ 26.00 $&$ 1.28 $&$ -33.4 $&$ -21.8 $&$ -12.0 $ \\
00:06:13.87 & $-$06:23:35.20 &$ 0.35 $&$ 1627.00 $&$ -0.09 $&$ 26.70 $&$ 3.21 $&$ -409.0 $&$ 2.9 $&$ -410.0 $ \\
00:06:22.60 & $-$00:04:25.10 &$ 1.04 $&$ 968.90 $&$ -0.83 $&$ 27.40 $&$ 1.59 $&$ 20.3 $&$ -2.6 $&$ 23.0 $ \\
00:13:31.09 & +40:51:36.00 &$ 0.26 $&$ 632.10 $&$ -0.35 $&$ 26.00 $&$ 0.98 $&$ -55.2 $&$ -74.6 $&$ 19.0 $ \\
00:15:59.98 & +39:00:27.20 &$ 1.72 $&$ 73.25 $&$ -0.94 $&$ 26.70 $&$ 5.15 $&$ -124.0 $&$ -116.0 $&$ -7.7 $ \\
00:18:51.38 & $-$12:42:33.50 &$ 1.59 $&$ 492.20 $&$ -1.02 $&$ 27.50 $&$ 3.13 $&$ 8.7 $&$ 2.3 $&$ 6.4 $ \\
00:20:25.32 & +15:40:52.70 &$ 2.02 $&$ 483.70 $&$ -1.19 $&$ 27.70 $&$ 8.28 $&$ -20.3 $&$ -16.7 $&$ -3.6 $ \\
00:24:30.12 & $-$29:28:48.90 &$ 0.41 $&$ 278.40 $&$ -1.16 $&$ 26.10 $&$ 4.69 $&$ 18.9 $&$ 3.5 $&$ 15.0 $ \\
00:25:26.15 & +39:19:35.70 &$ 1.95 $&$ 580.40 $&$ -0.19 $&$ 27.70 $&$ 3.50 $&$ -98.5 $&$ -104.0 $&$ 5.5 $ \\

\end{longtable}
\end{centering}

%\end{center}
%\clearpage
%\onecolumn
\twocolumn
\section{Source spectra and Faraday functions}
\label{sec:appendB}
%\FloatBarrier
All Figures in this section are available with the online version of this paper. 
 
\begin{figure*}
\includegraphics[scale=0.35]{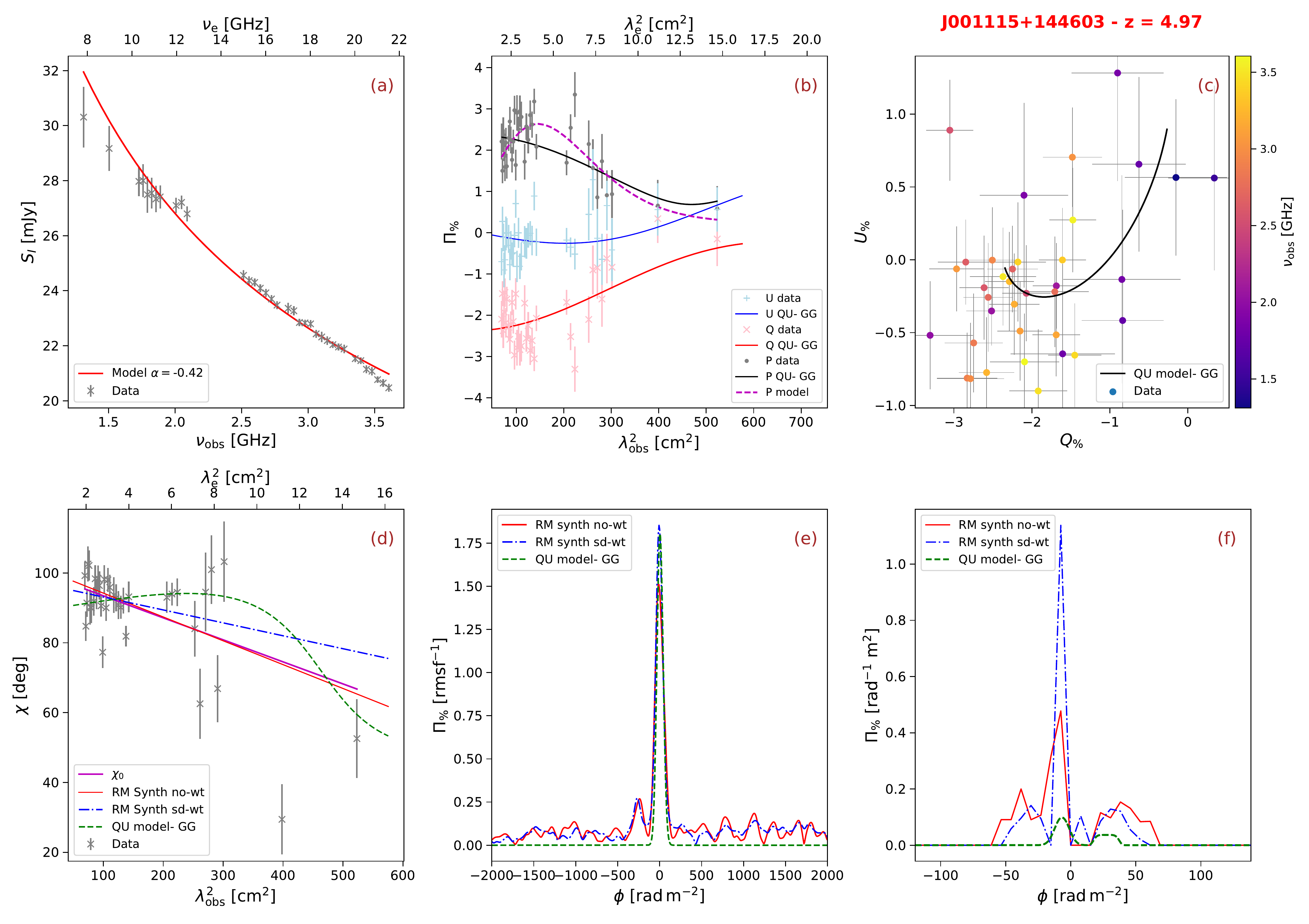}
\caption{As for Fig. 5. Source: J001115+144603}
\label{fig:spec1}
\end{figure*}

\begin{figure*}
\includegraphics[scale=0.35]{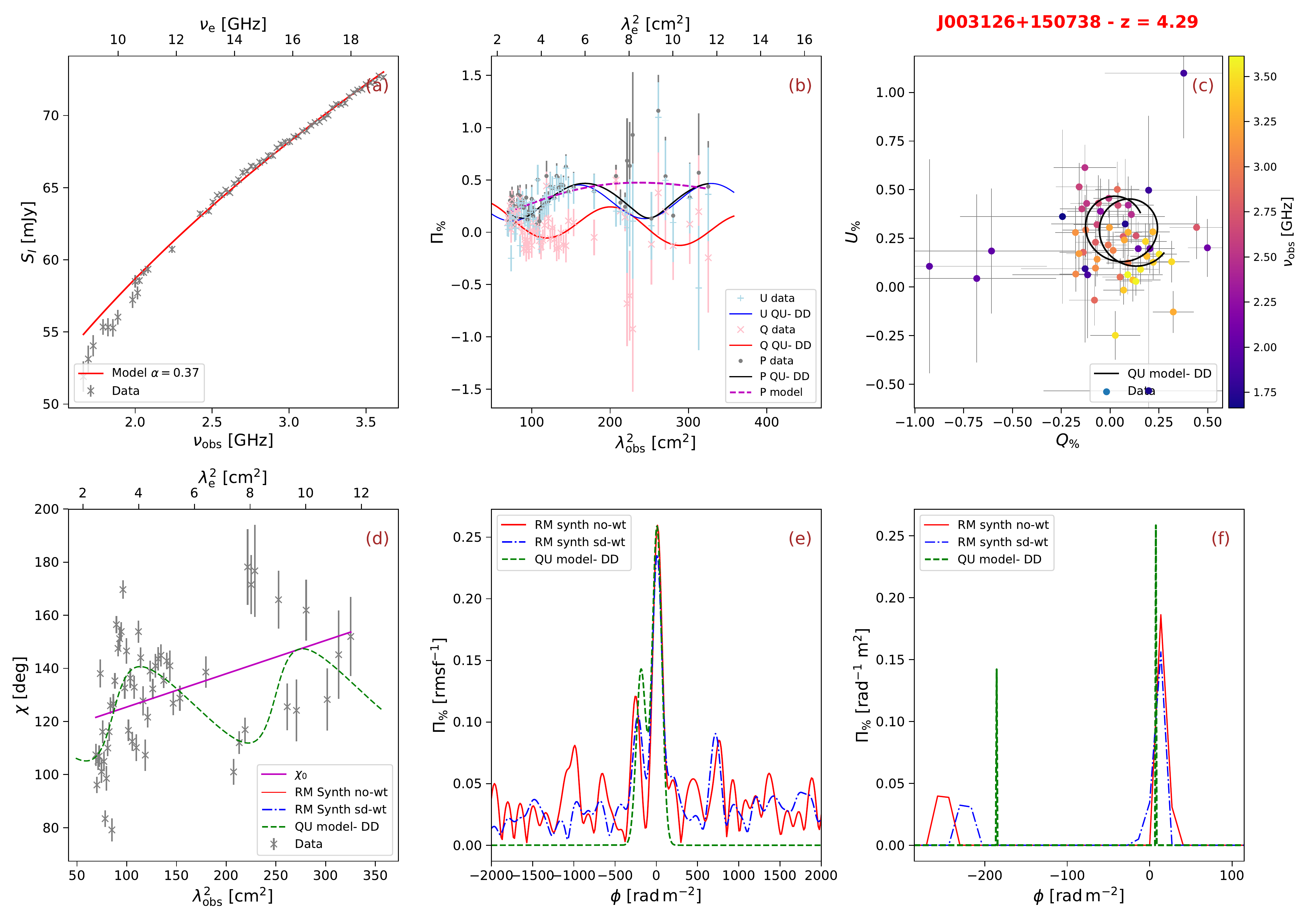}
\caption{As for Fig. 5. Source: J003126+150738}
\label{fig:spec2}
\end{figure*}

\begin{figure*}
\includegraphics[scale=0.35]{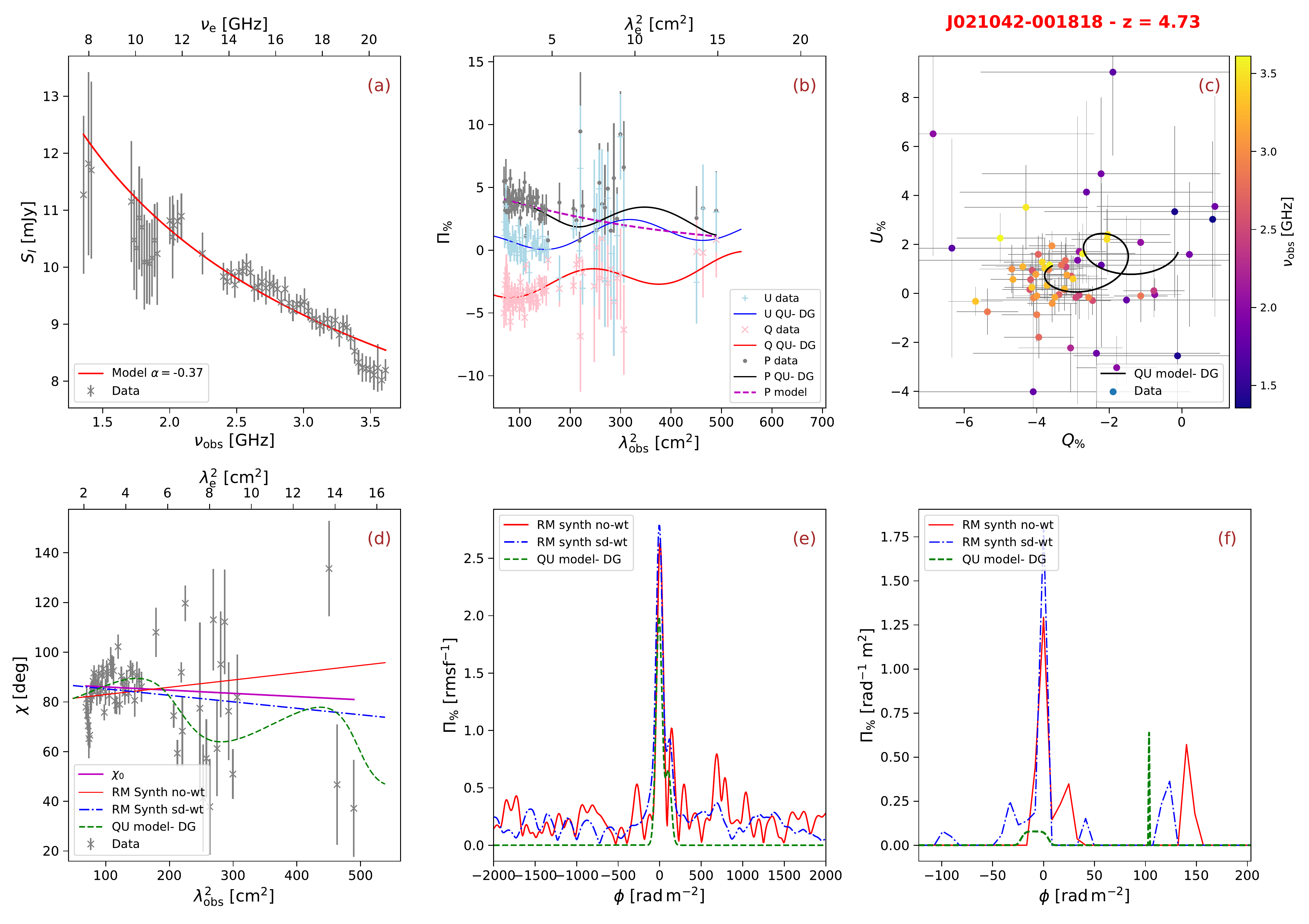}
\caption{As for Fig. 5. Source: J021042$-$001818}
\label{fig:spec3}
\end{figure*}

\begin{figure*}
\includegraphics[scale=0.35]{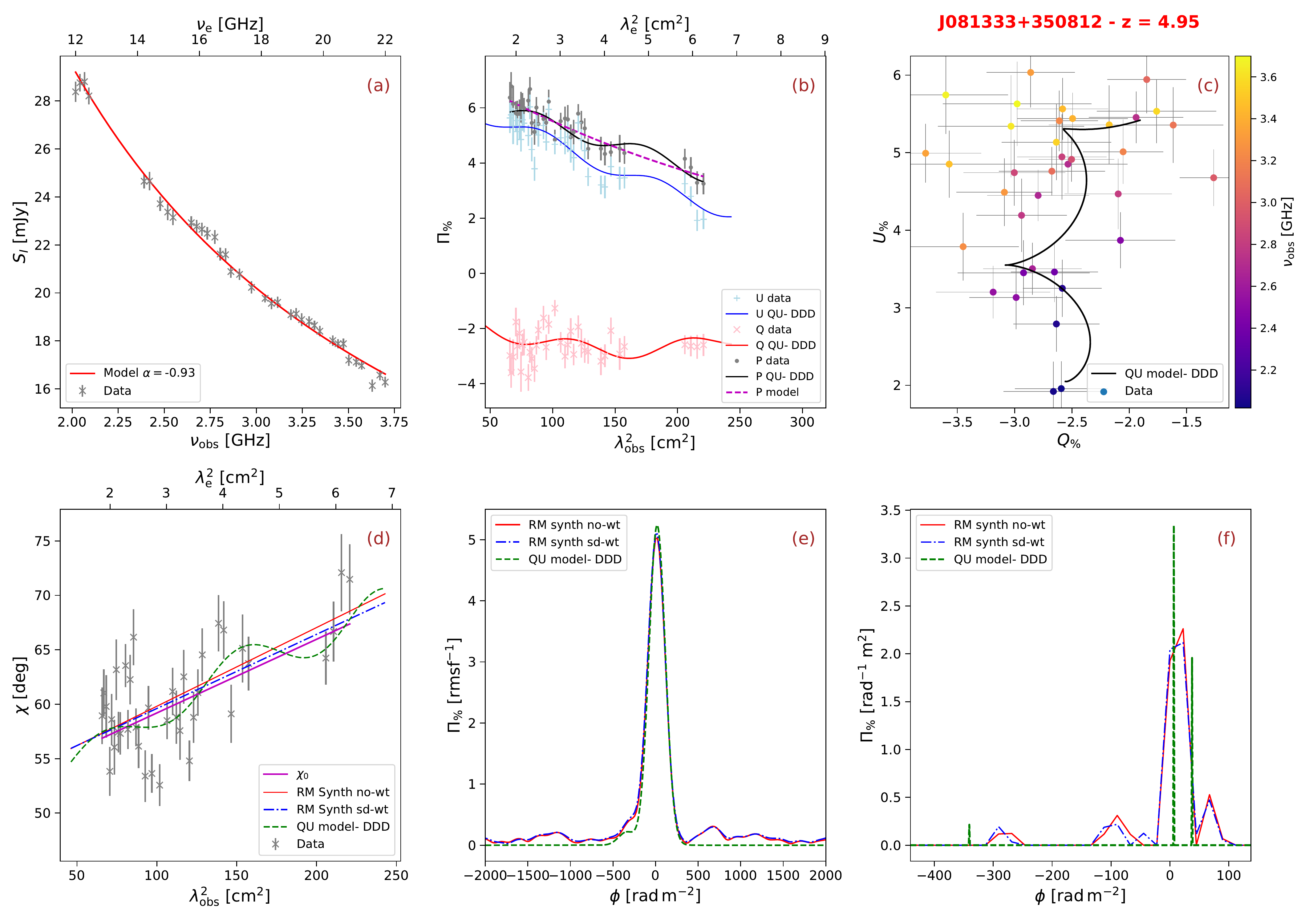}
\caption{As for Fig. 5. Source: J081333+350812}
\label{fig:spec4}
\end{figure*}

\begin{figure*}
\includegraphics[scale=0.35]{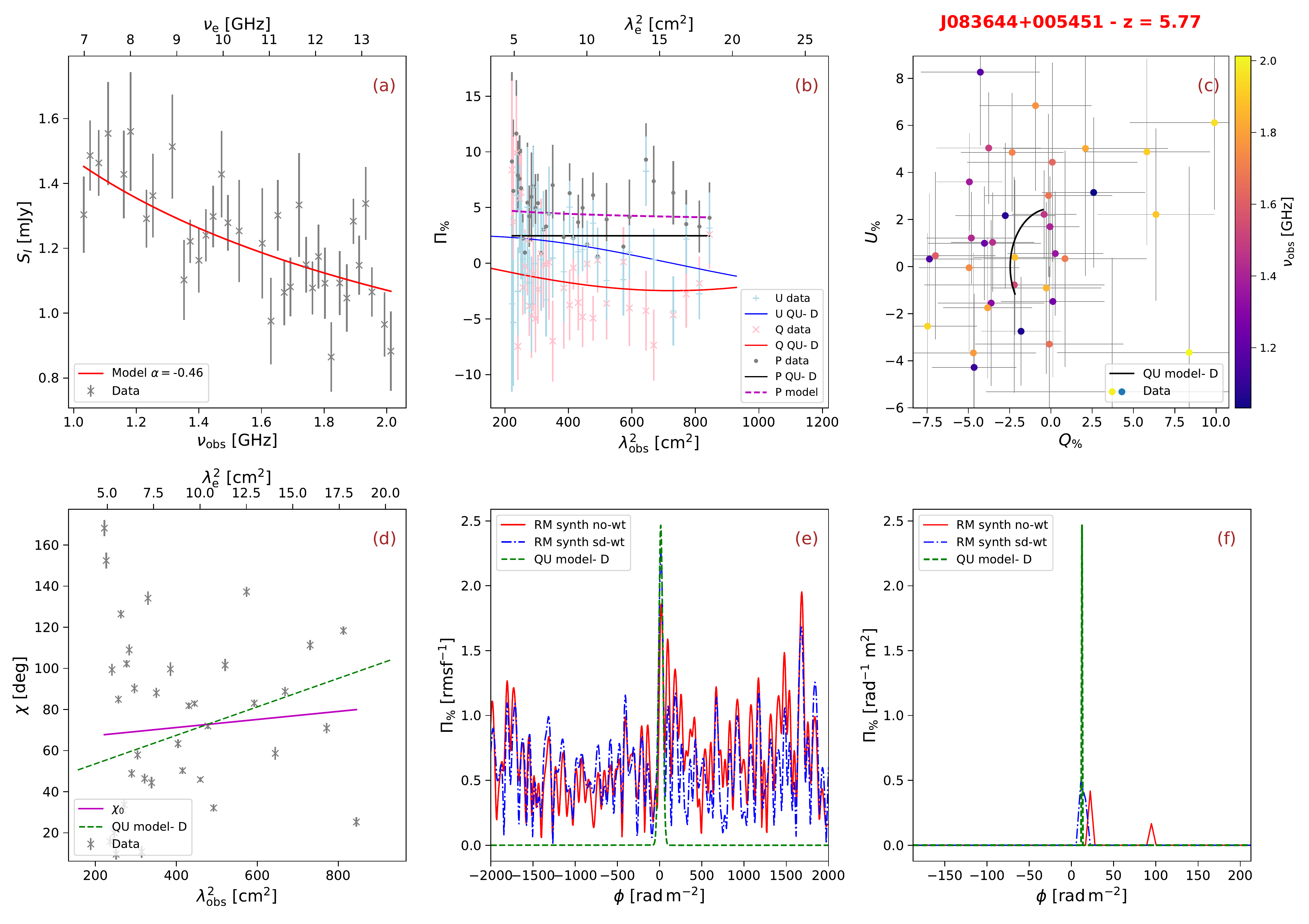}
\caption{As for Fig. 5. Source: J083644+005451}
\label{fig:spec5}
\end{figure*}

\begin{figure*}
\includegraphics[scale=0.35]{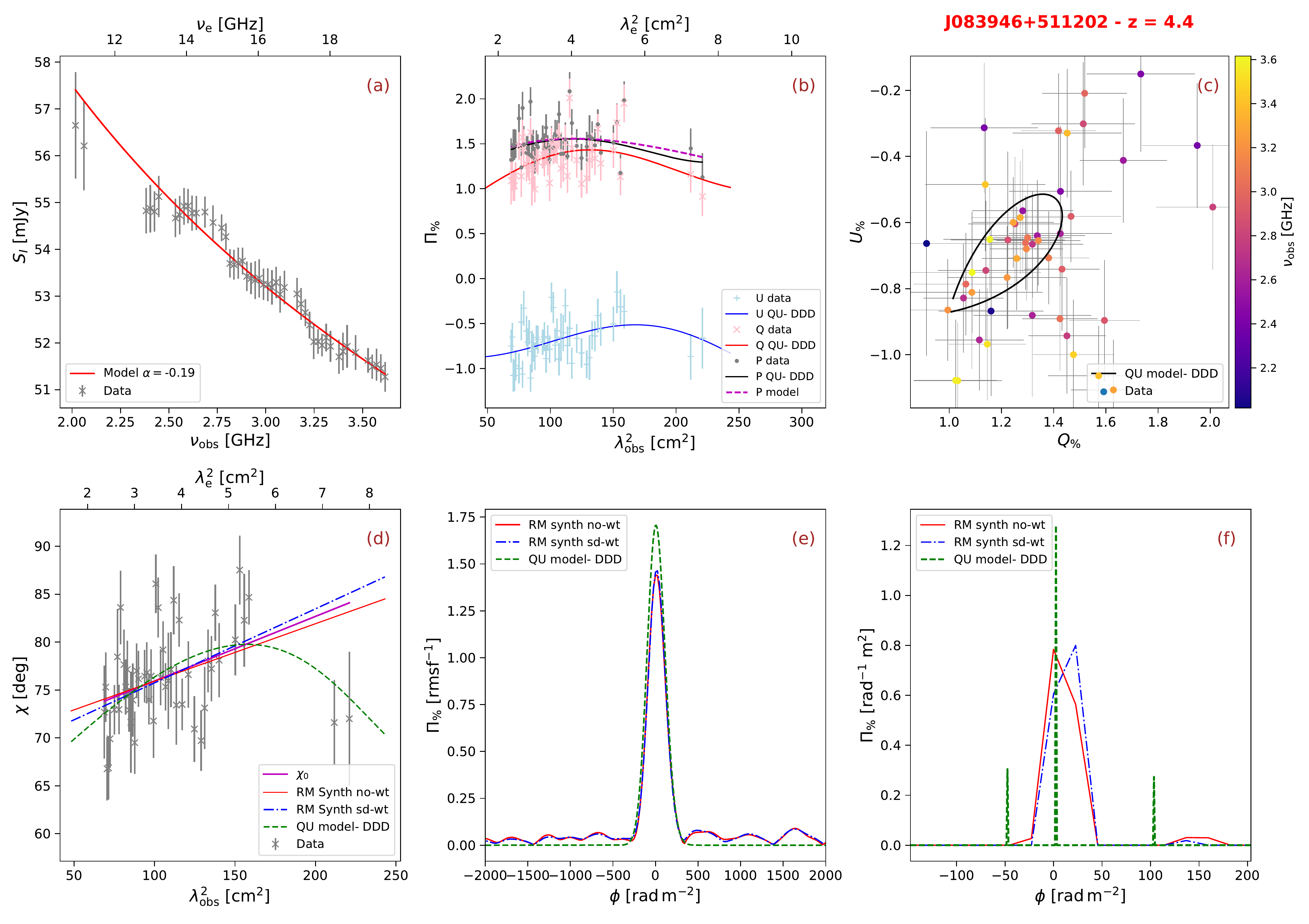}
\caption{As for Fig. 5. Source: J083946+511202}
\label{fig:spec6}
\end{figure*}

\begin{figure*}
\includegraphics[scale=0.35]{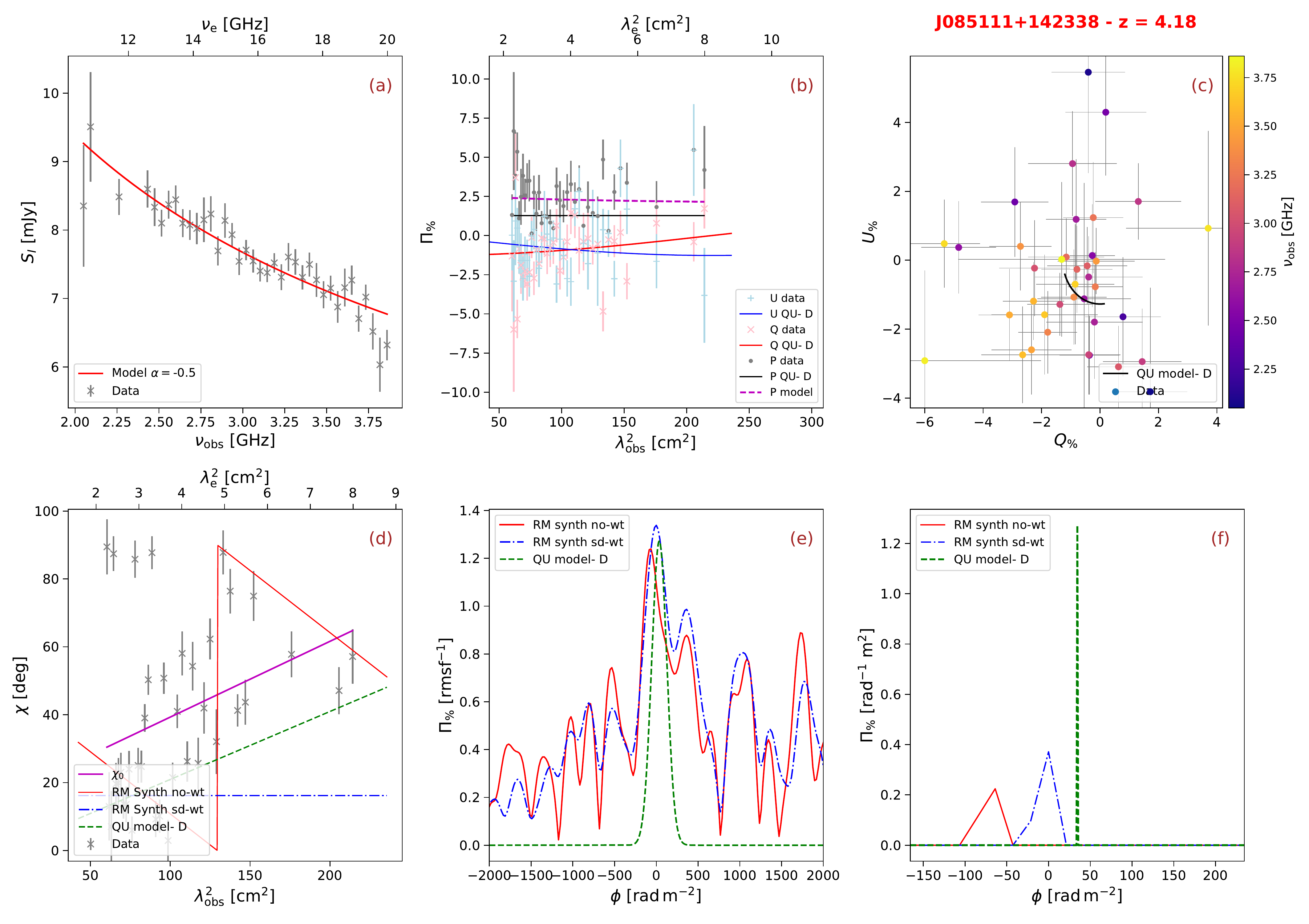}
\caption{As for Fig. 5. Source: J085111+142338}
\label{fig:spec7}
\end{figure*}

\begin{figure*}
\includegraphics[scale=0.35]{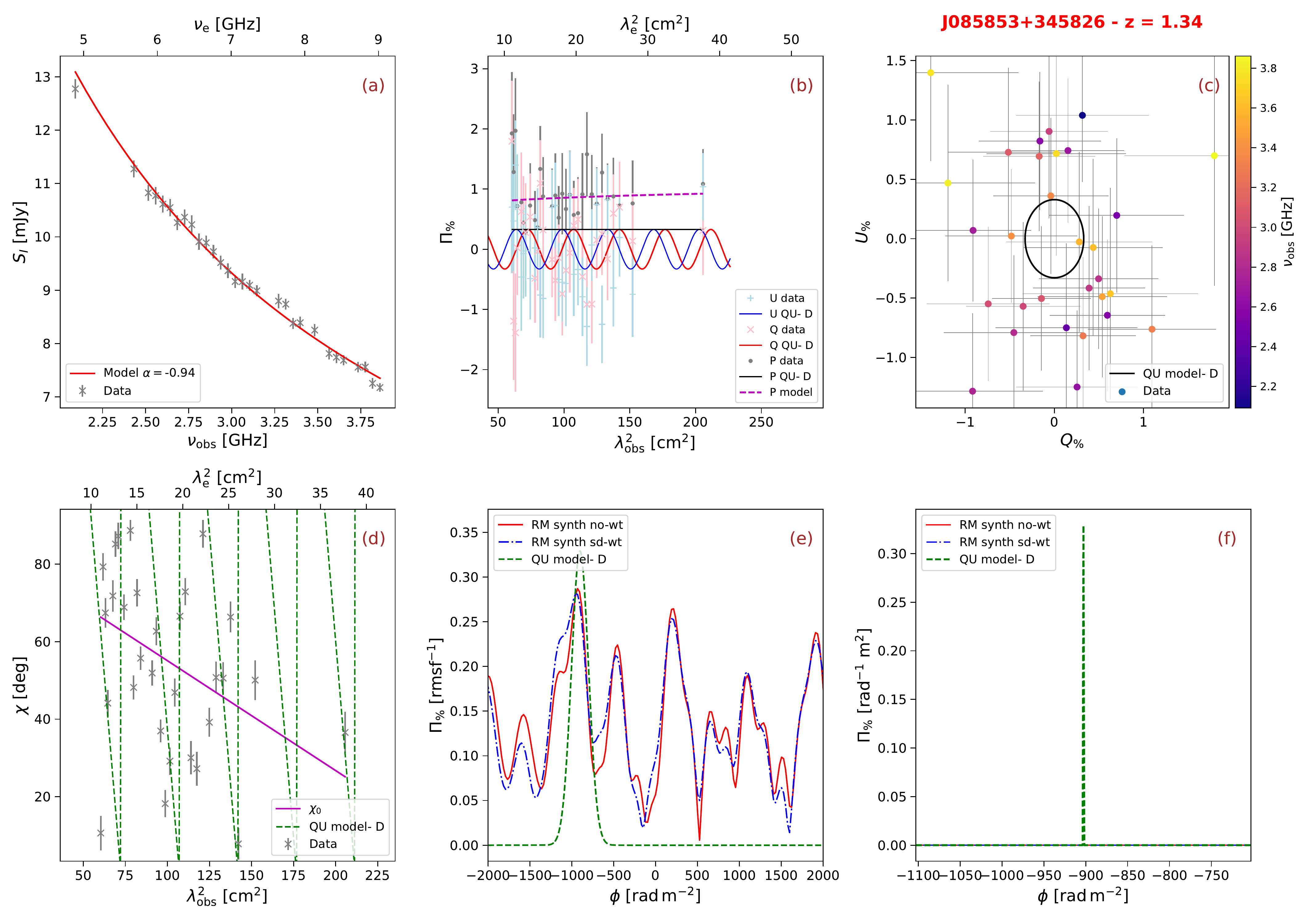}
\caption{As for Fig. 5. Source: J085853+345826}
\label{fig:spec8}
\end{figure*}

\begin{figure*}
\includegraphics[scale=0.35]{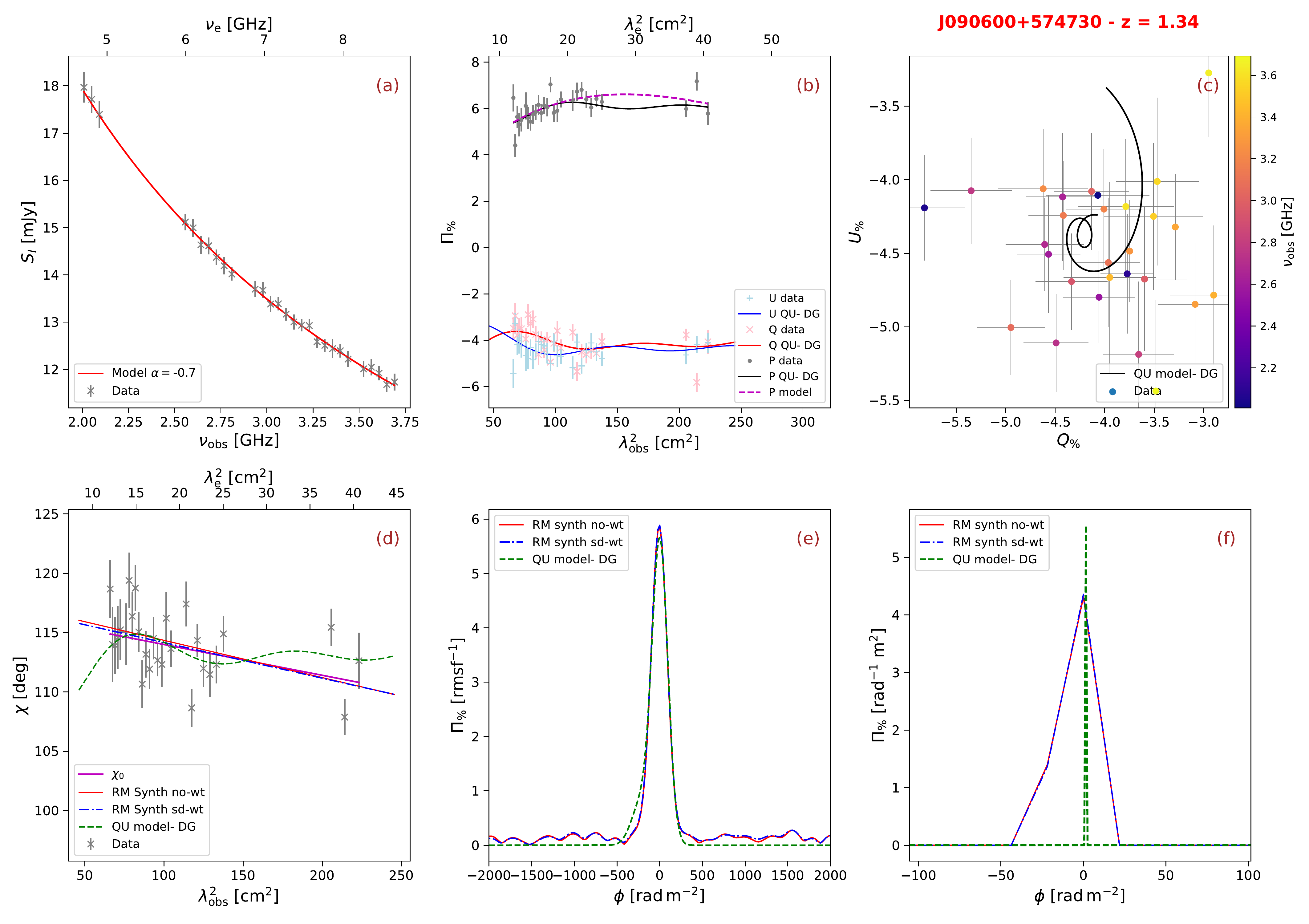}
\caption{As for Fig. 5. Source: J090600+574730}
\label{fig:spec9}
\end{figure*}

\begin{figure*}
\includegraphics[scale=0.35]{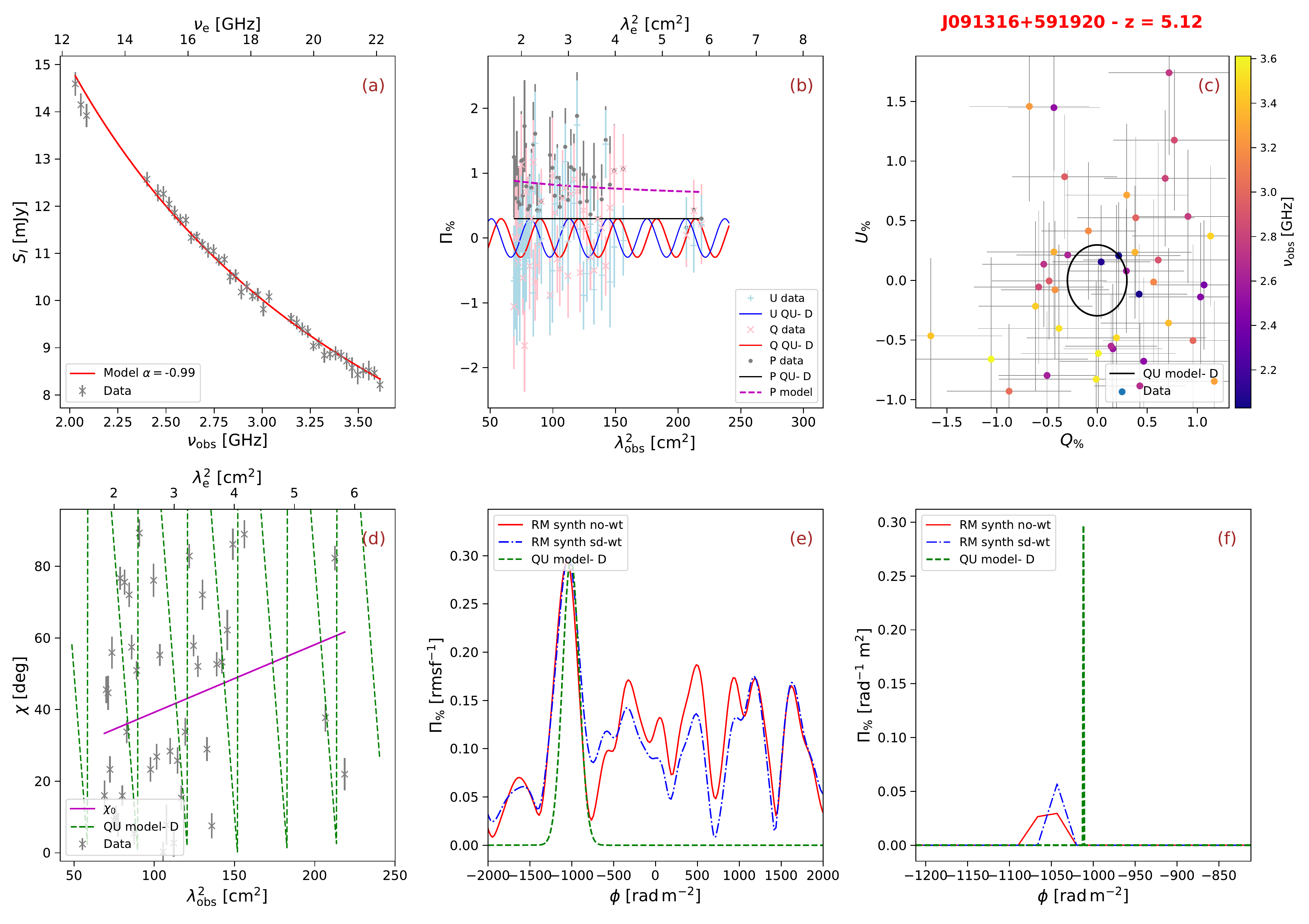}
\caption{As for Fig. 5. Source: J091316+591920}
\label{fig:spec10}
\end{figure*}

\begin{figure*}
\includegraphics[scale=0.35]{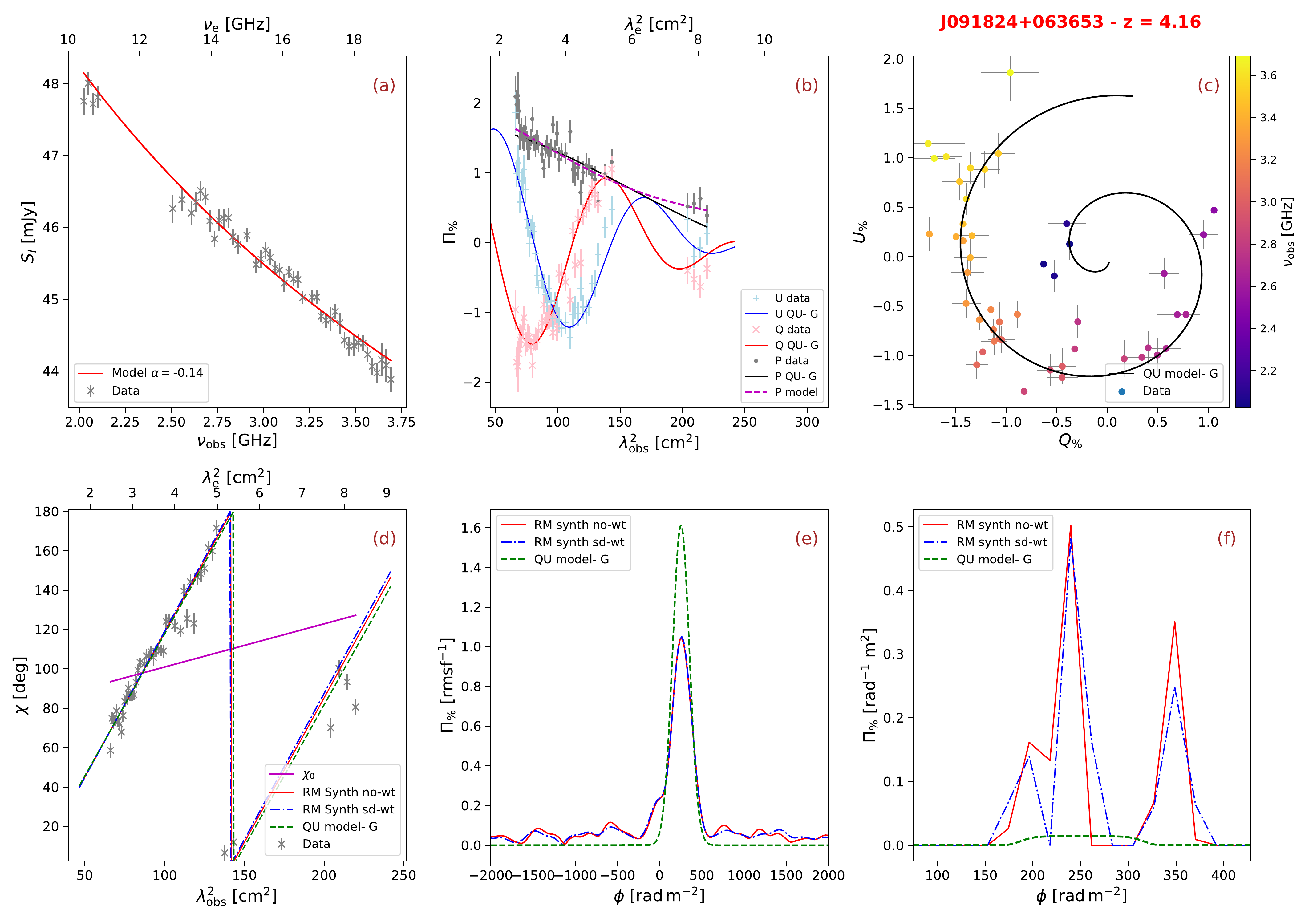}
\caption{As for Fig. 5. Source: J091824+063653}
\label{fig:spec11}
\end{figure*}

\begin{figure*}
\includegraphics[scale=0.35]{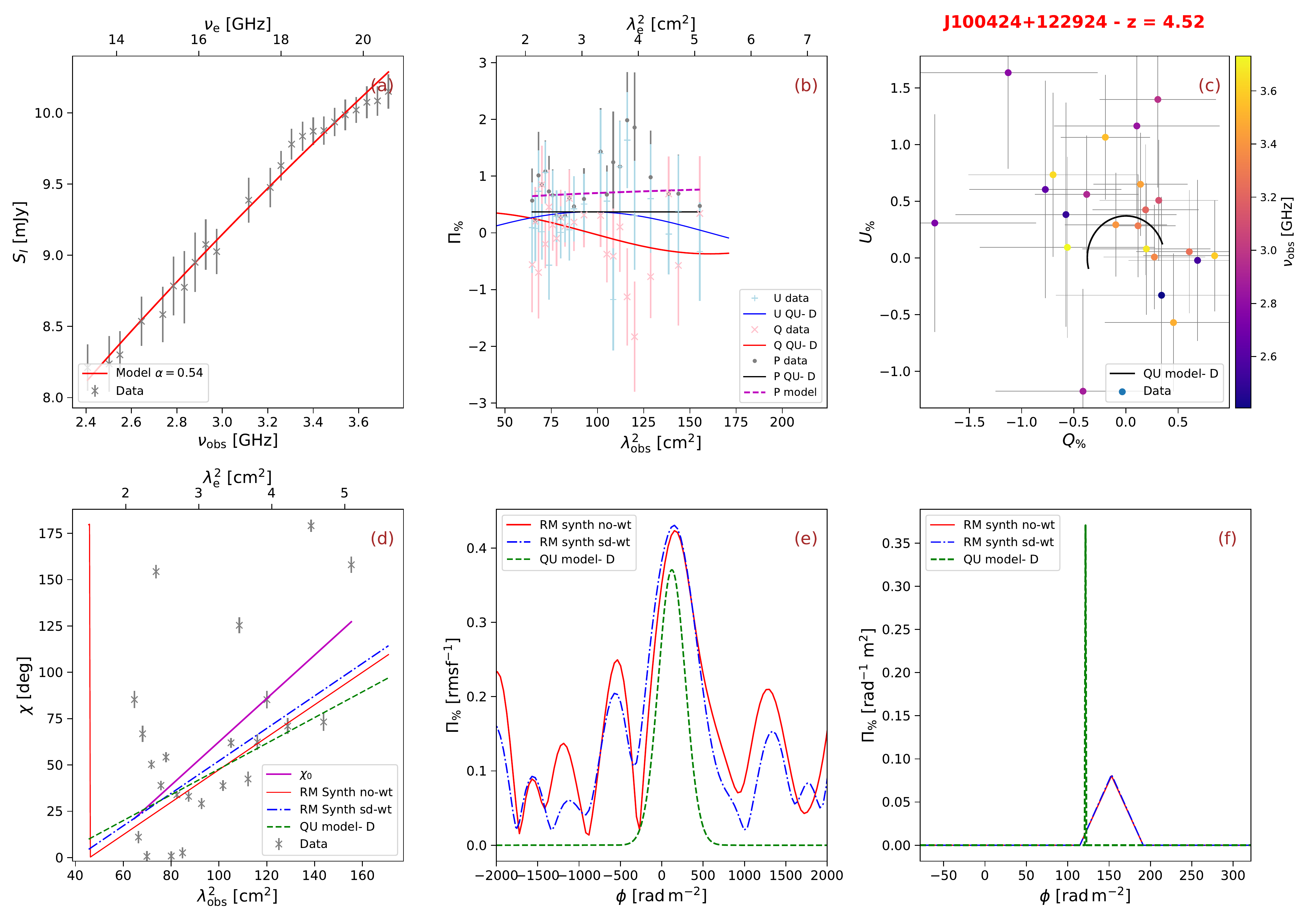}
\caption{As for Fig. 5. Source: J100424+122924}
\label{fig:spec12}
\end{figure*}

\begin{figure*}
\includegraphics[scale=0.35]{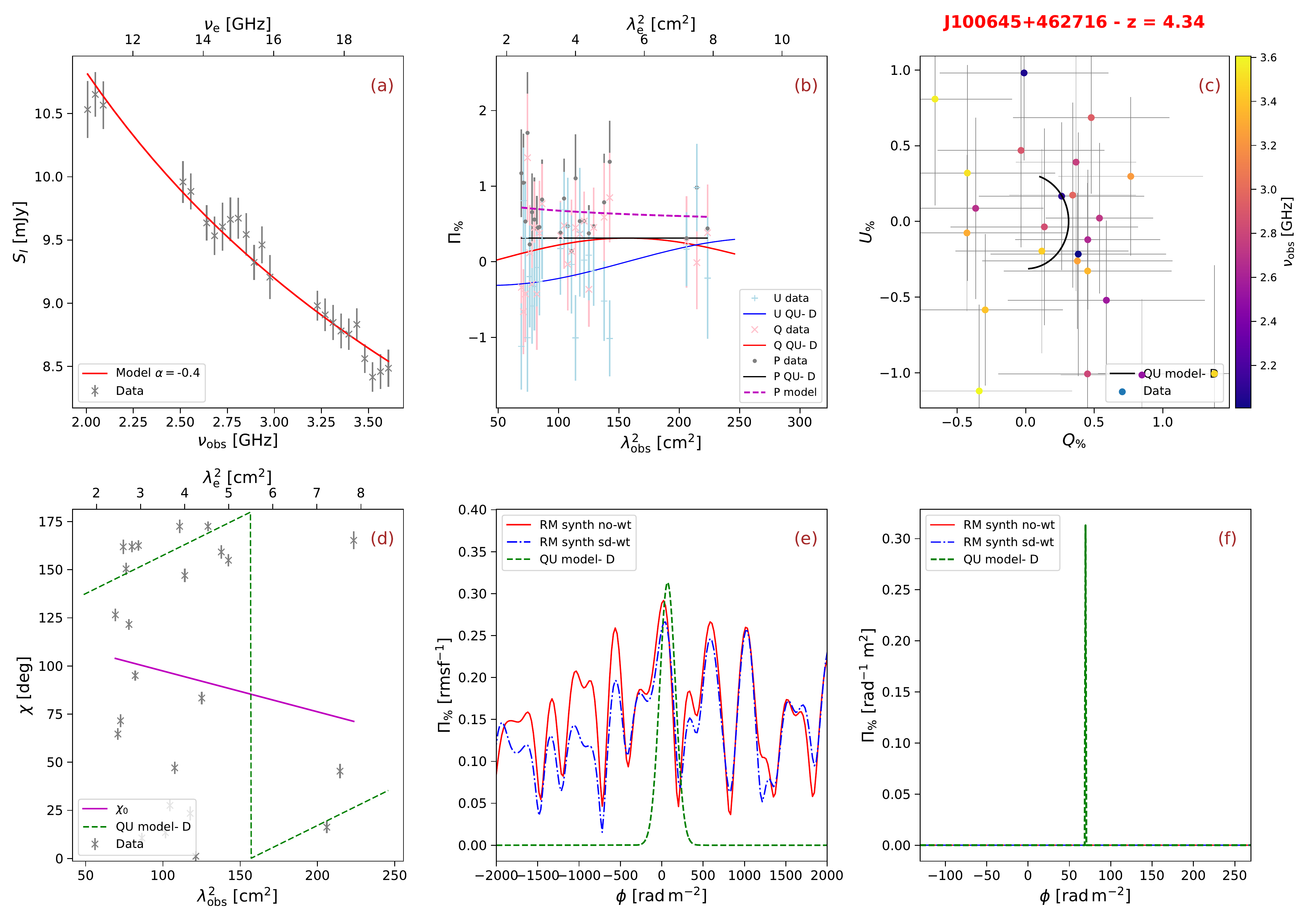}
\caption{As for Fig. 5. Source: J100645+462716}
\label{fig:spec13}
\end{figure*}

\begin{figure*}
\includegraphics[scale=0.35]{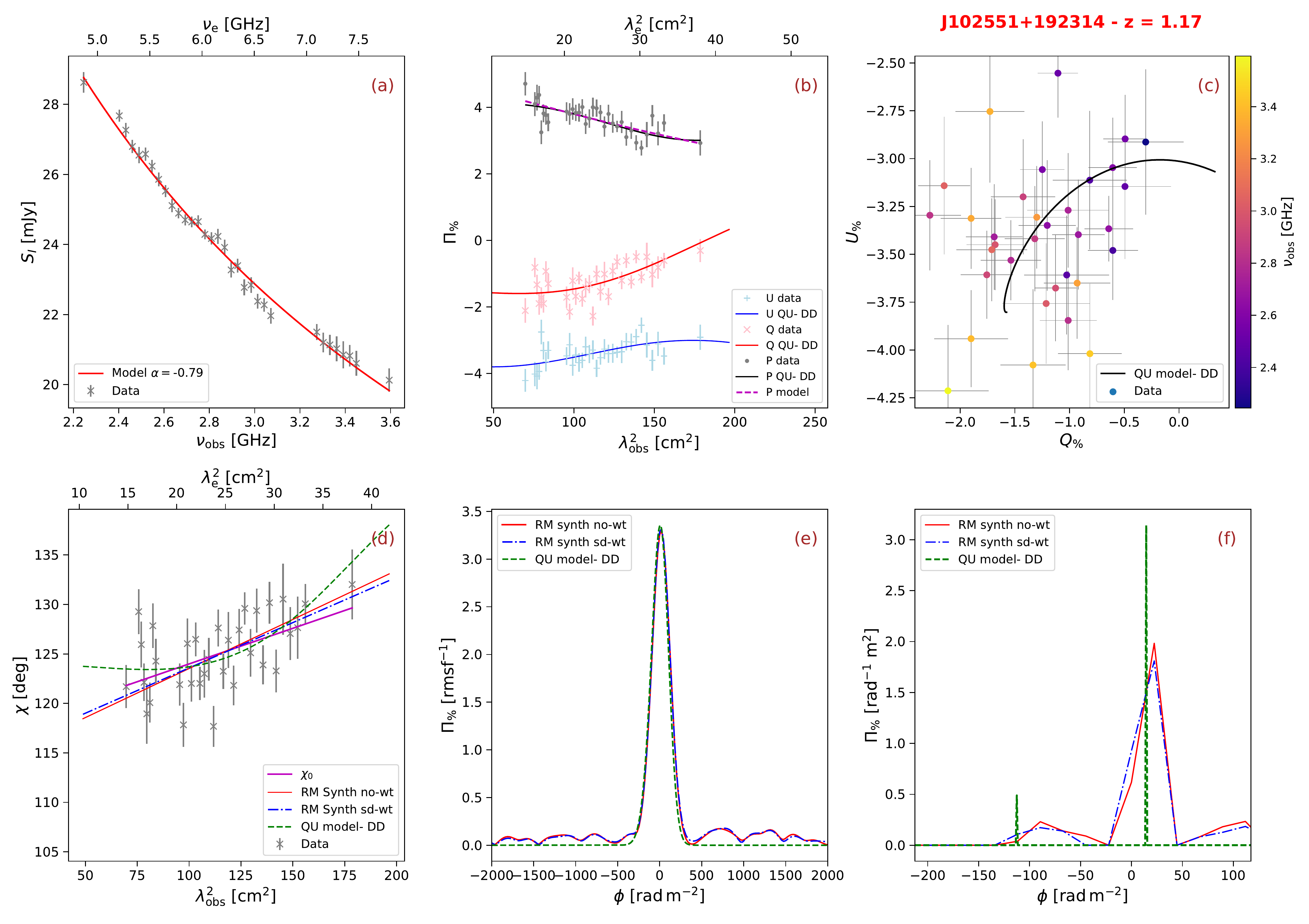}
\caption{As for Fig. 5. Source: J102551+192314}
\label{fig:spec14}
\end{figure*}

\begin{figure*}
\includegraphics[scale=0.35]{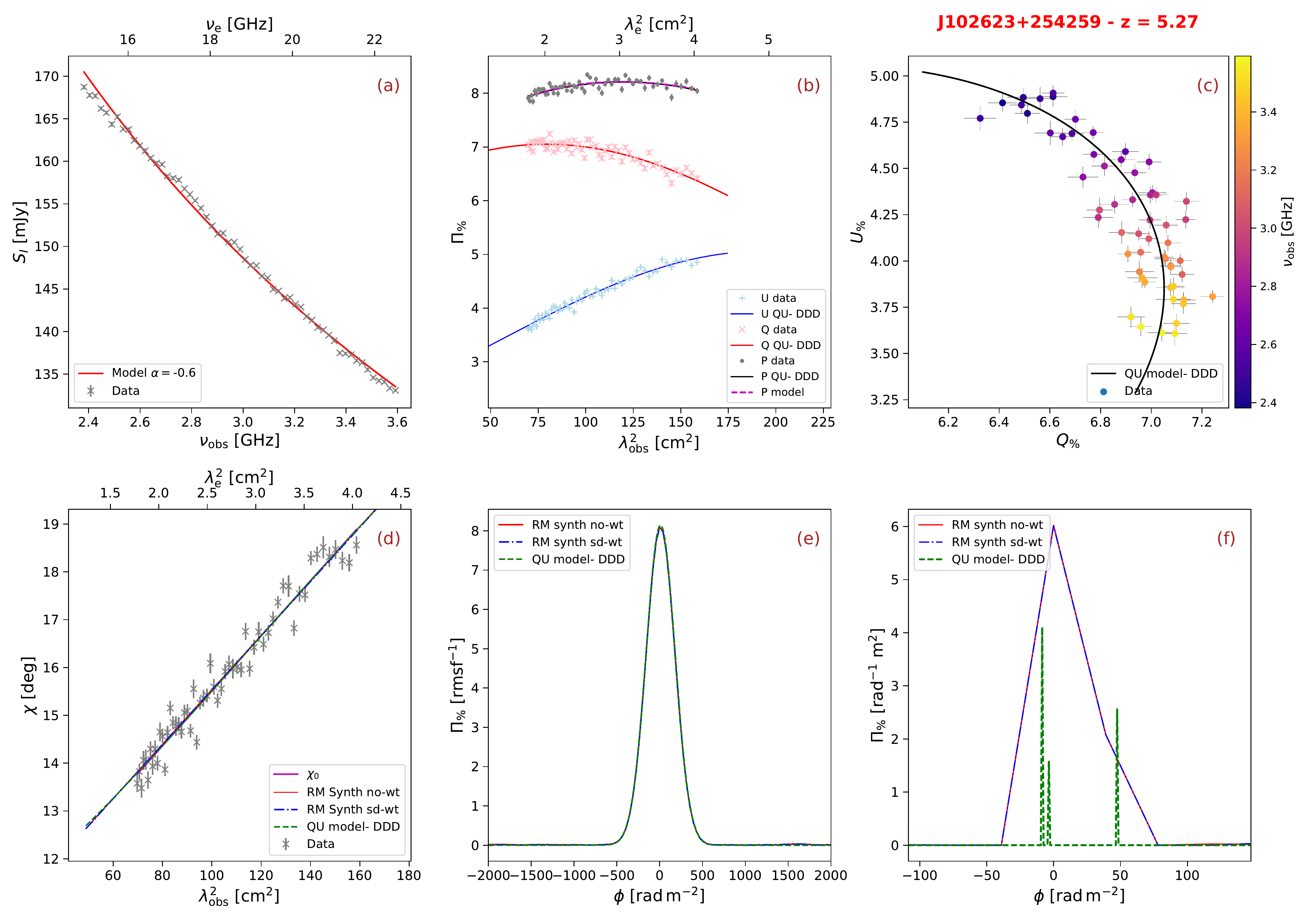}
\caption{As for Fig. 5. Source: J102623+254259}
\label{fig:spec15}
\end{figure*}

\begin{figure*}
\includegraphics[scale=0.35]{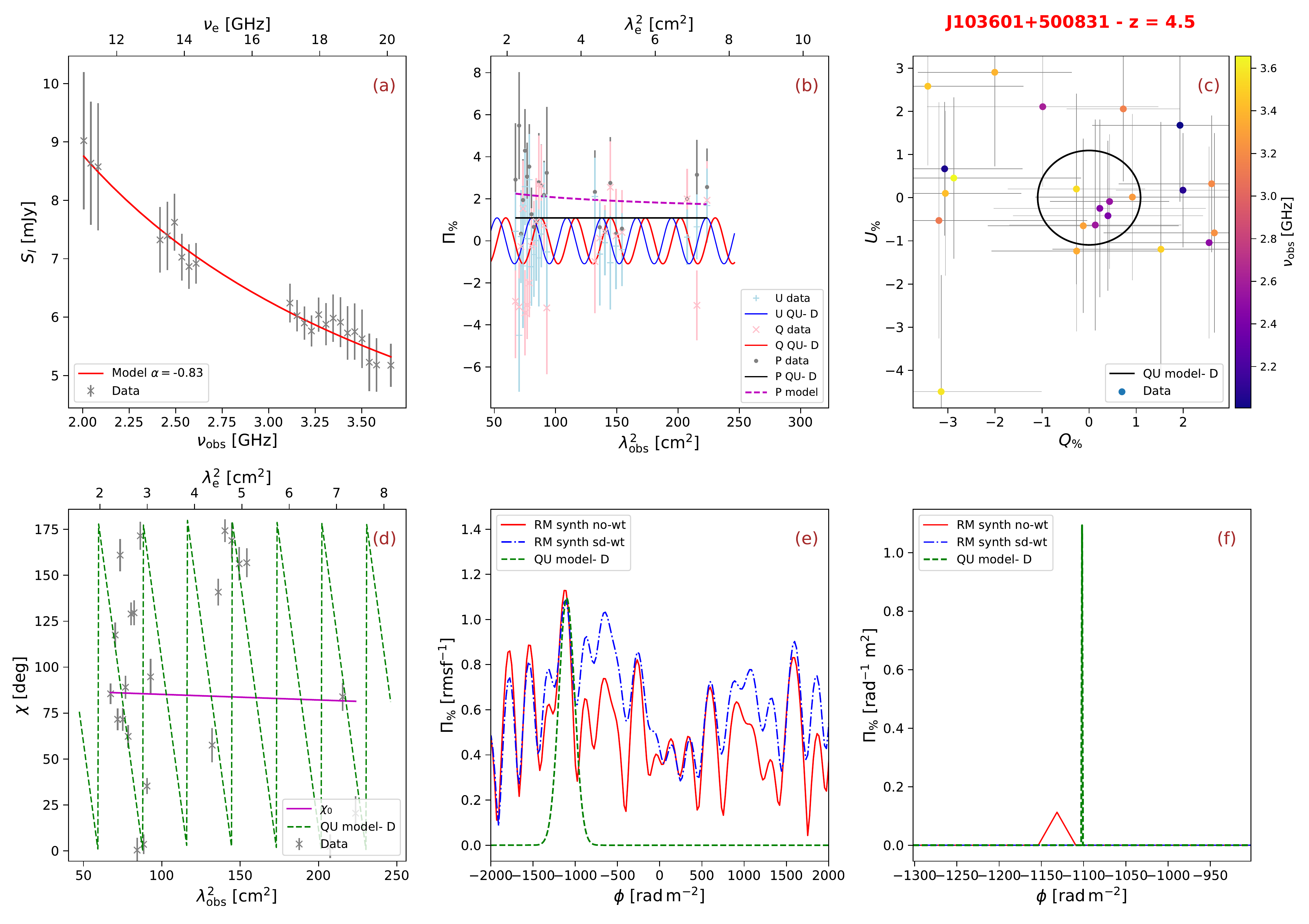}
\caption{As for Fig. 5. Source: J103601+500831}
\label{fig:spec16}
\end{figure*}

\begin{figure*}
\includegraphics[scale=0.35]{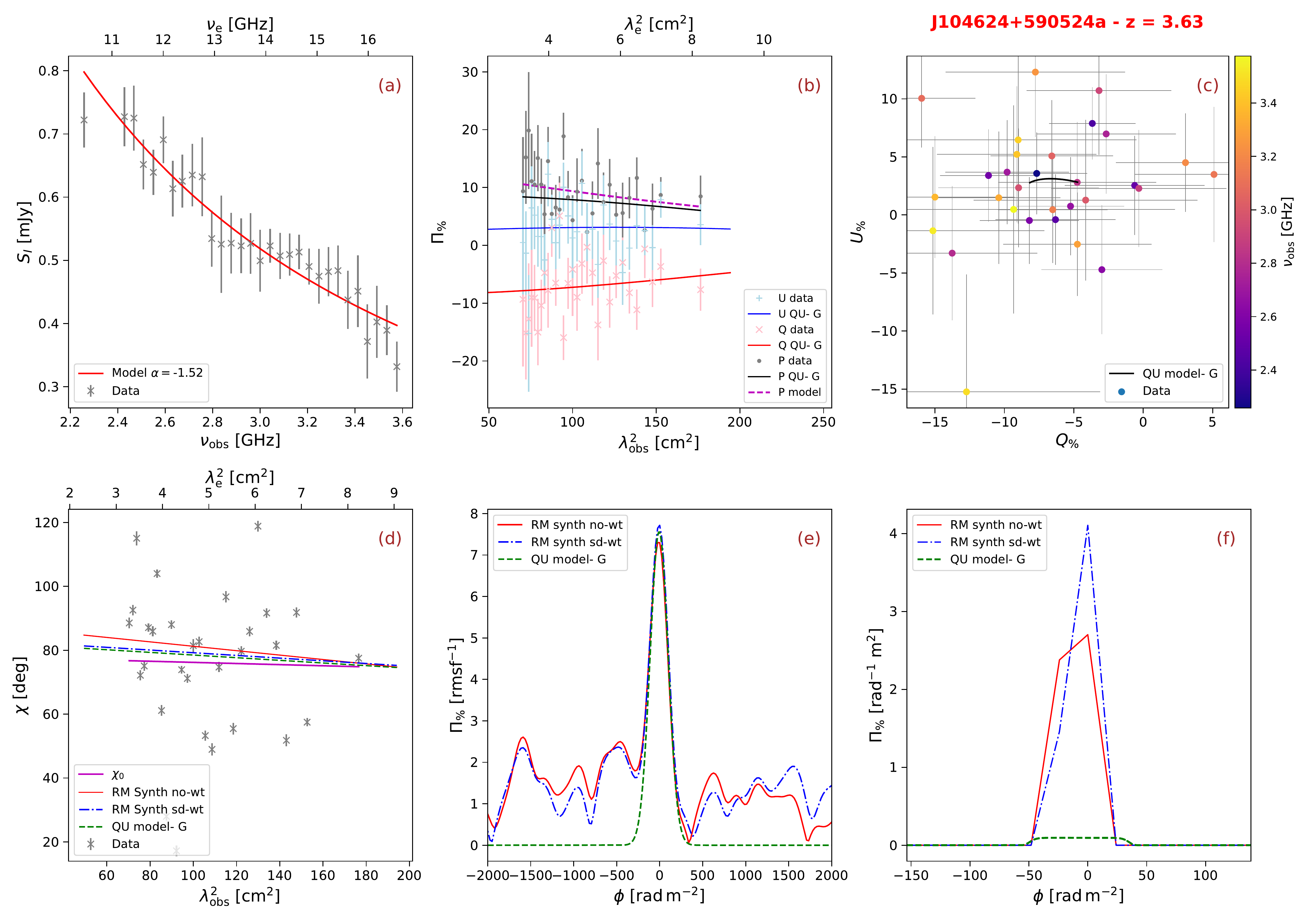}
\caption{As for Fig. 5. Source: J104624+590524a}
\label{fig:spec17}
\end{figure*}

\begin{figure*}
\includegraphics[scale=0.35]{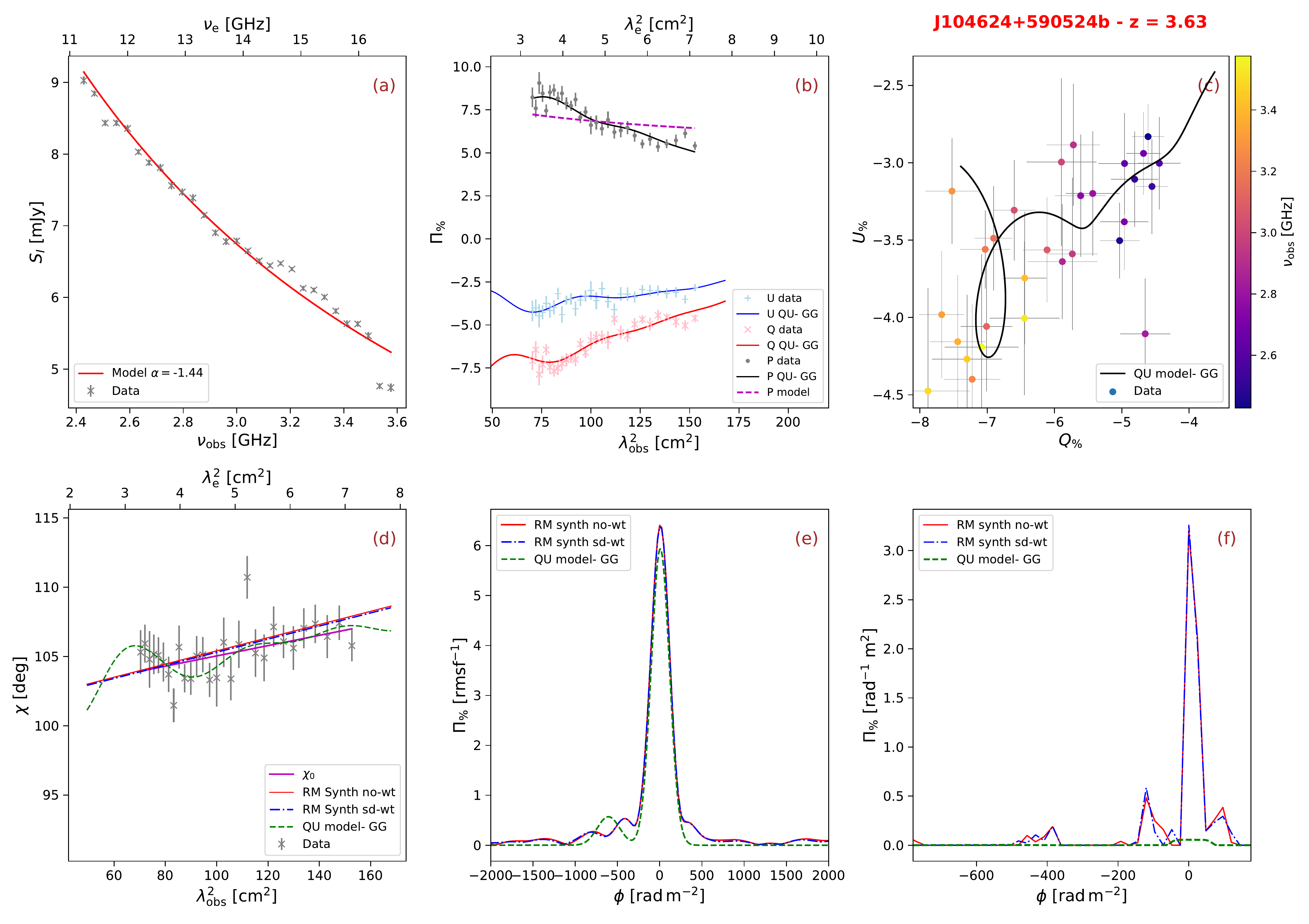}
\caption{As for Fig. 5. Source: J104624+590524b}
\label{fig:spec18}
\end{figure*}

\begin{figure*}
\includegraphics[scale=0.35]{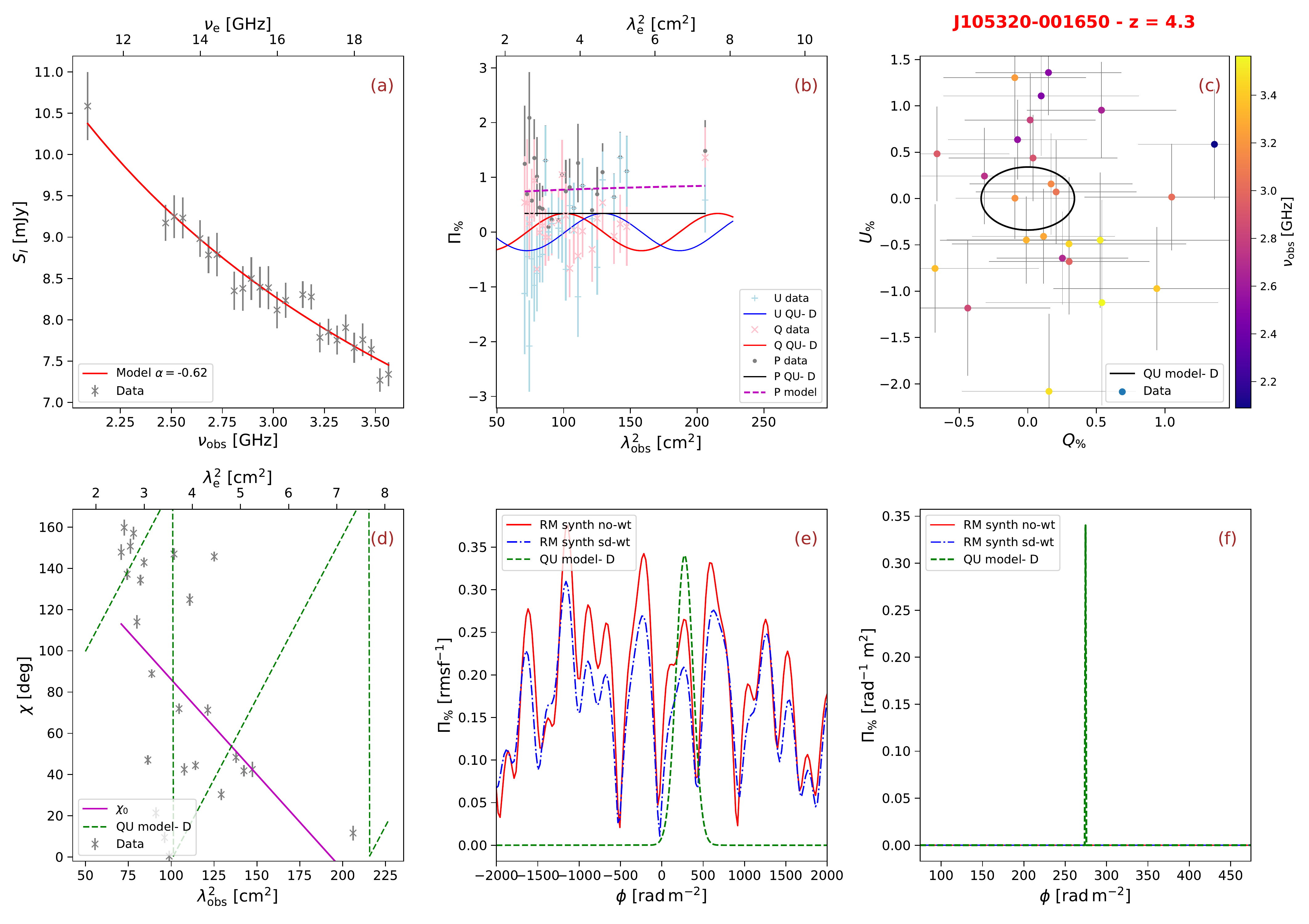}
\caption{As for Fig. 5. Source: J105320$-$001650}
\label{fig:spec19}
\end{figure*}

\begin{figure*}
\includegraphics[scale=0.35]{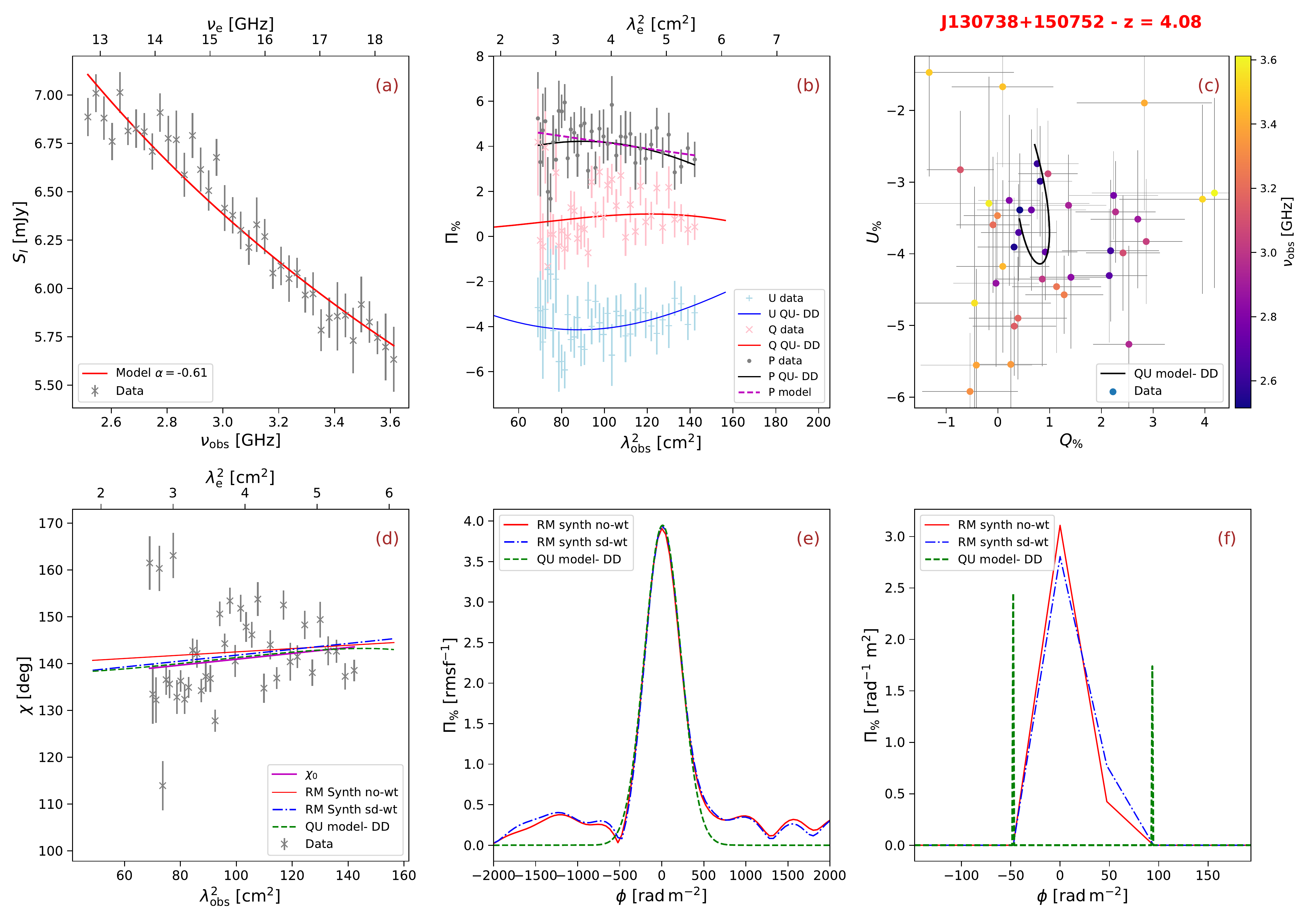}
\caption{As for Fig. 5. Source: J130738+150752}
\label{fig:spec20}
\end{figure*}

\begin{figure*}
\includegraphics[scale=0.35]{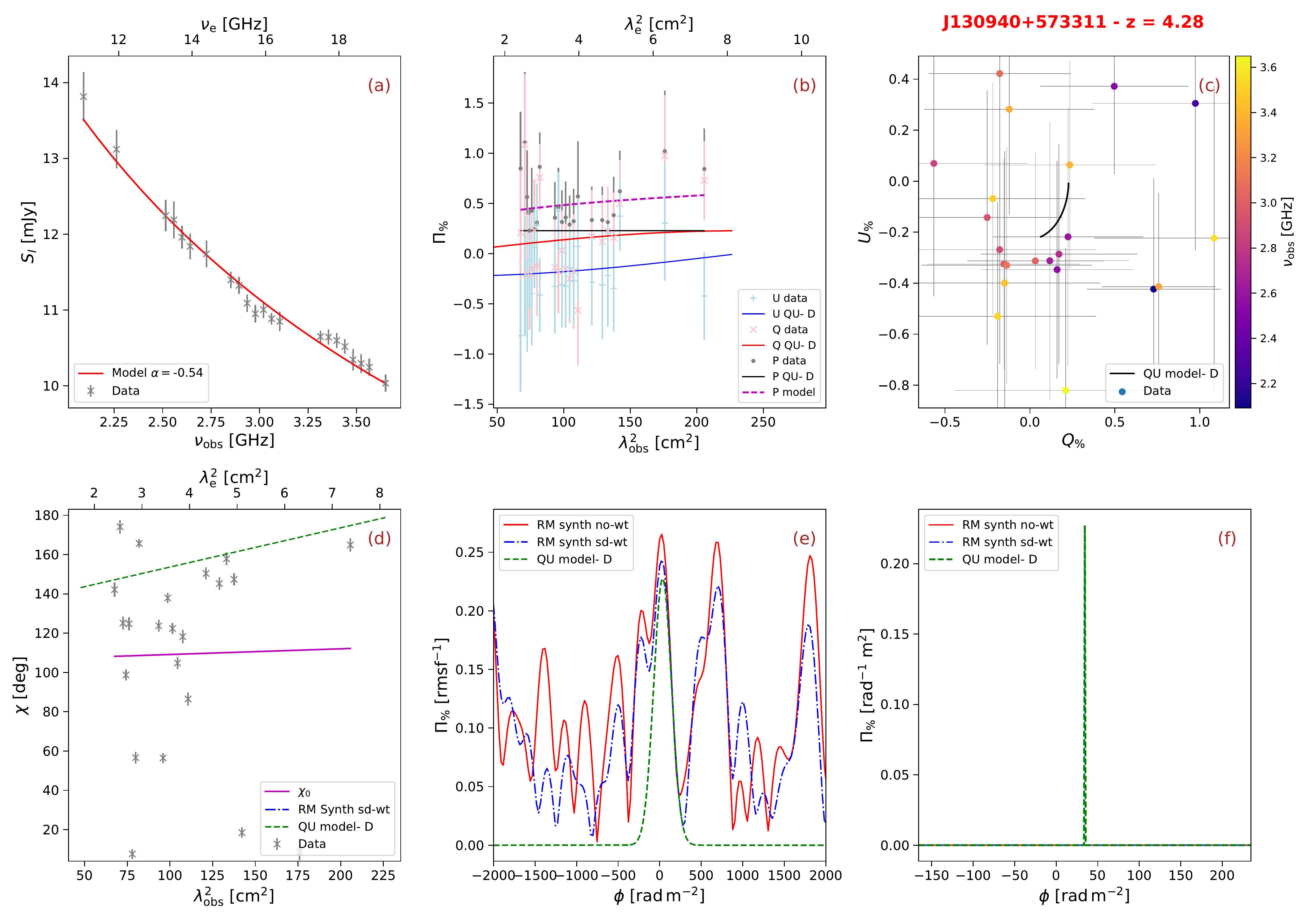}
\caption{As for Fig. 5. Source: J130940+573311}
\label{fig:spec21}
\end{figure*}

\begin{figure*}
\includegraphics[scale=0.35]{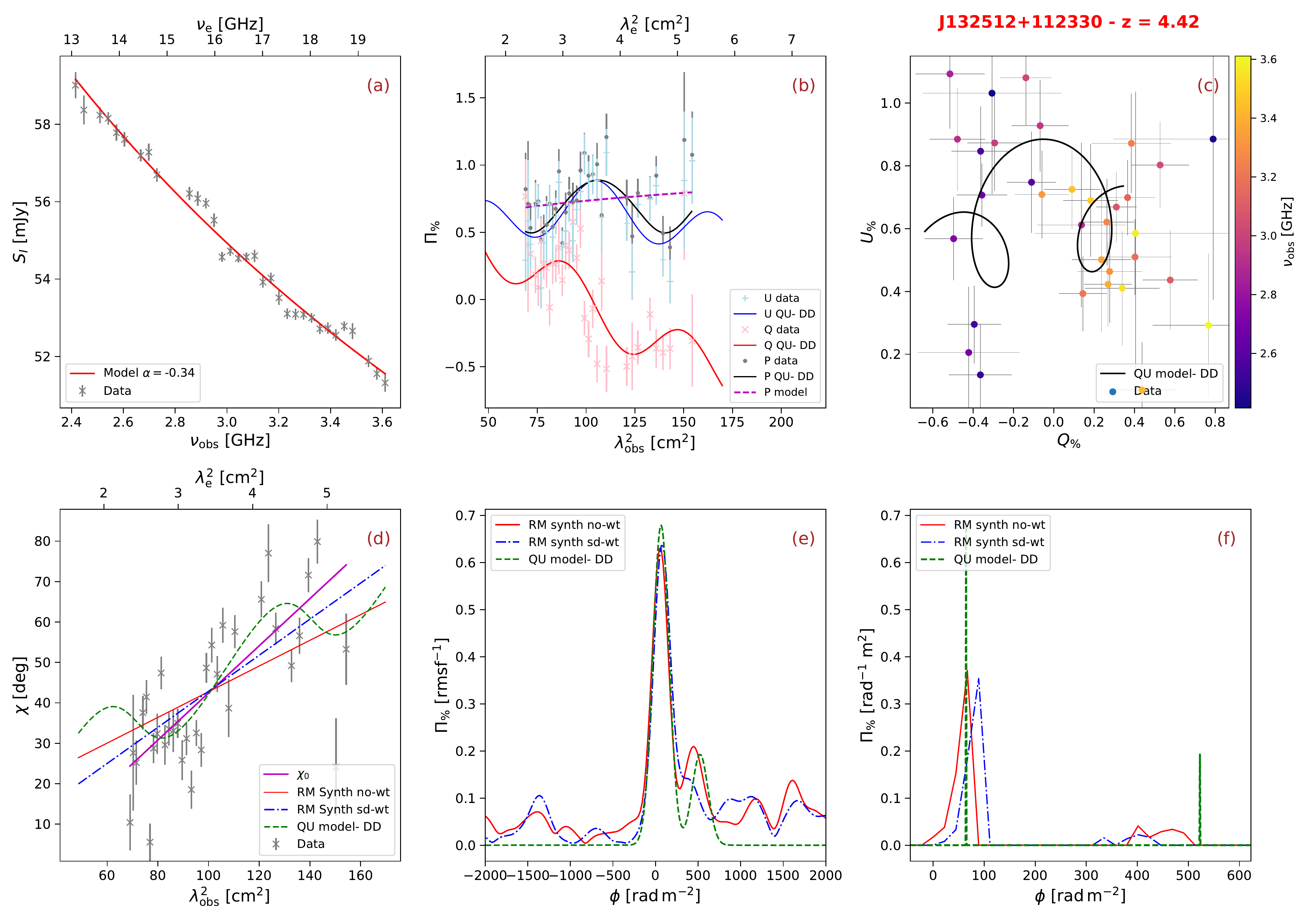}
\caption{As for Fig. 5. Source: J132512+112330}
\label{fig:spec22}
\end{figure*}

\begin{figure*}
\includegraphics[scale=0.35]{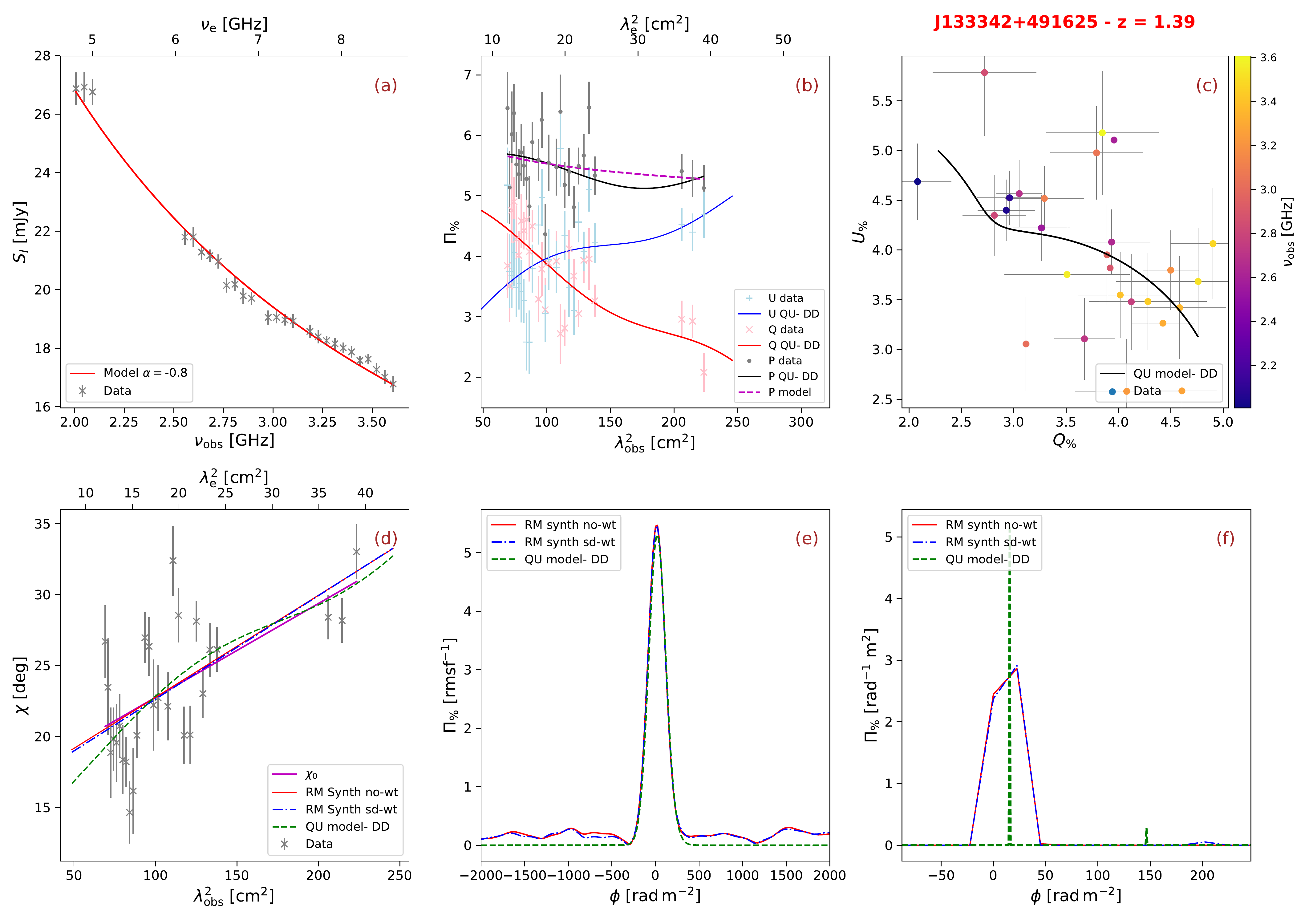}
\caption{As for Fig. 5. Source: J133342+491625}
\label{fig:spec23}
\end{figure*}

\begin{figure*}
\includegraphics[scale=0.35]{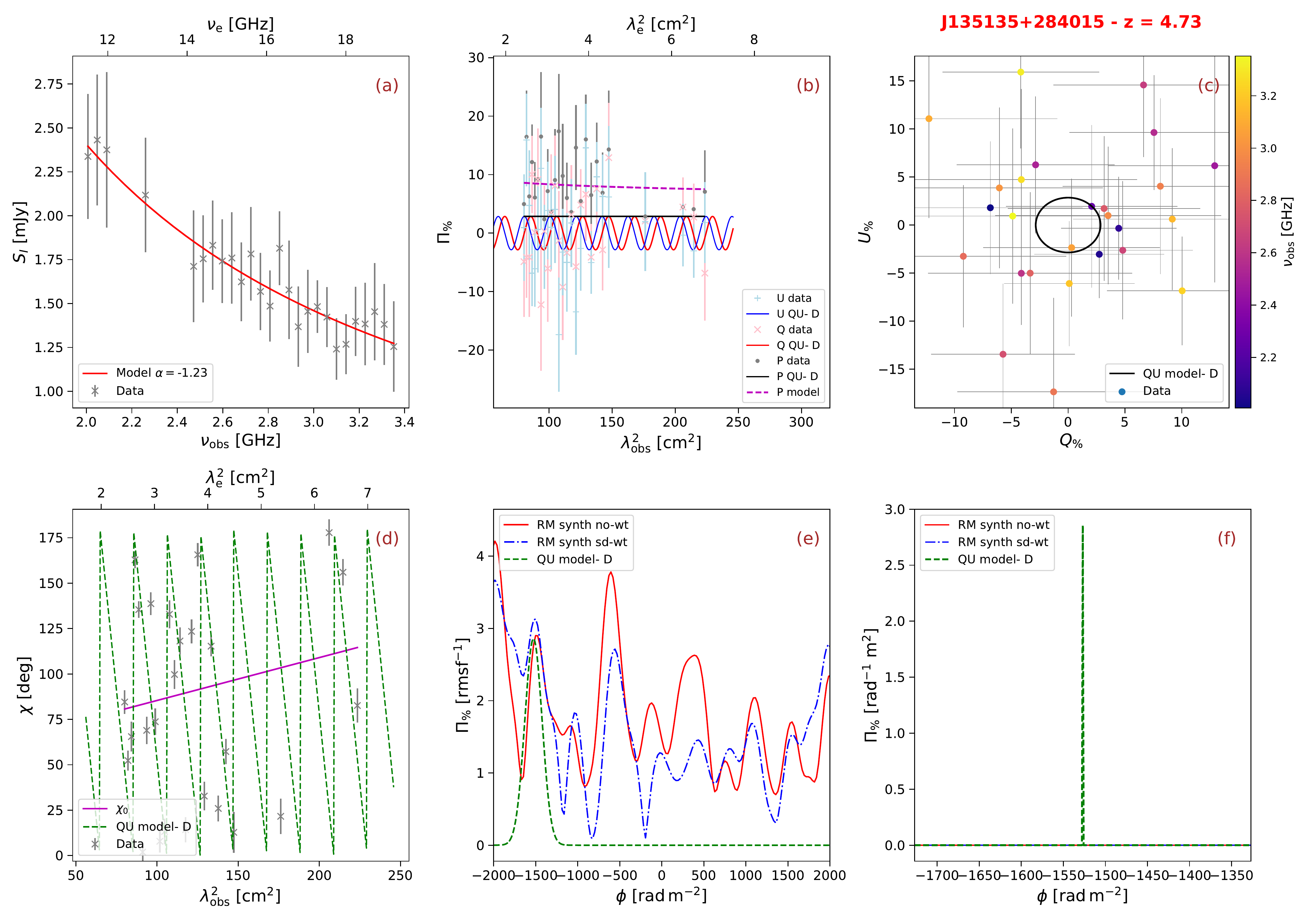}
\caption{As for Fig. 5. Source: J135135+284015}
\label{fig:spec24}
\end{figure*}

\begin{figure*}
\includegraphics[scale=0.35]{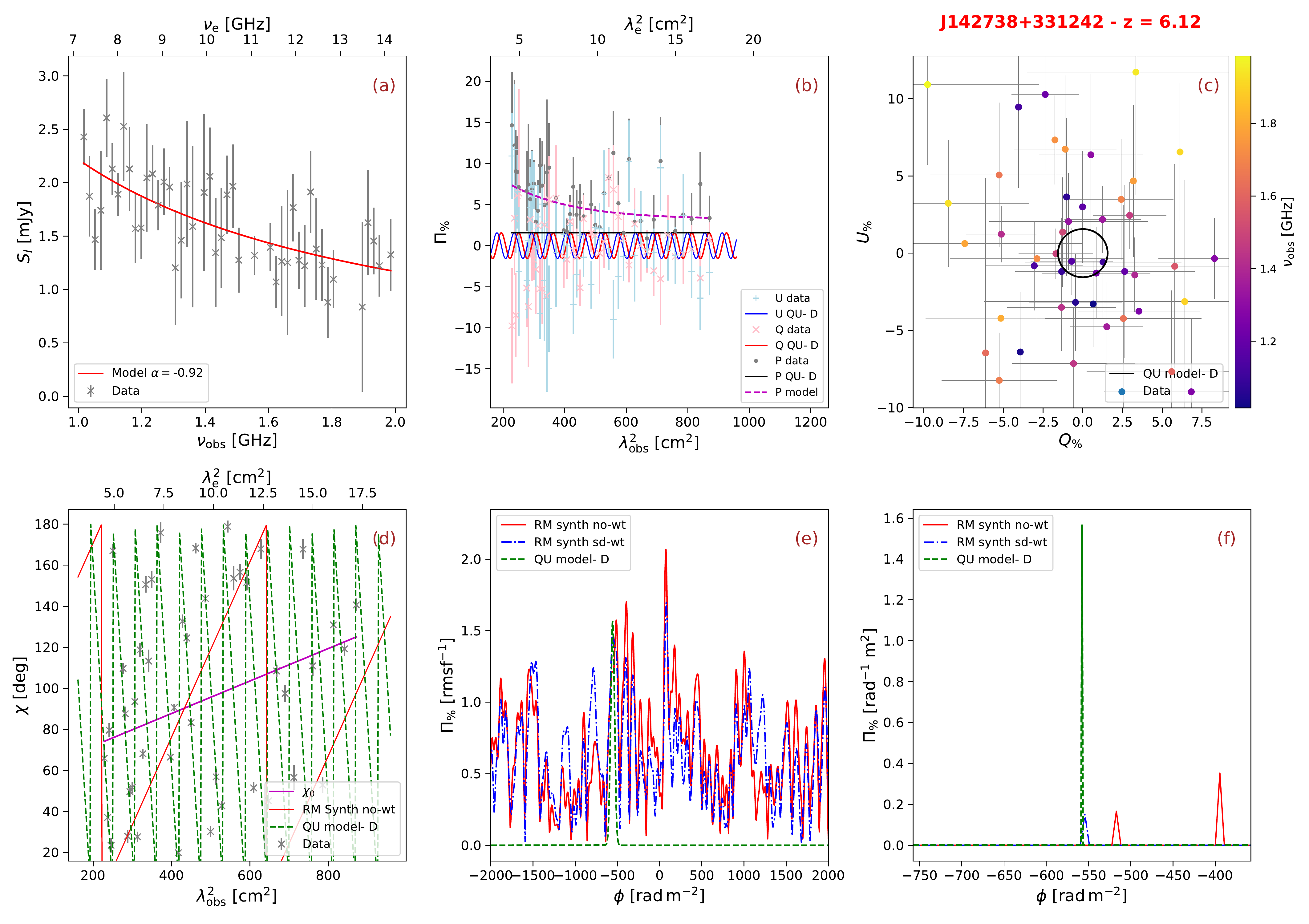}
\caption{As for Fig. 5. Source: J142738+331242}
\label{fig:spec25}
\end{figure*}

\begin{figure*}
\includegraphics[scale=0.35]{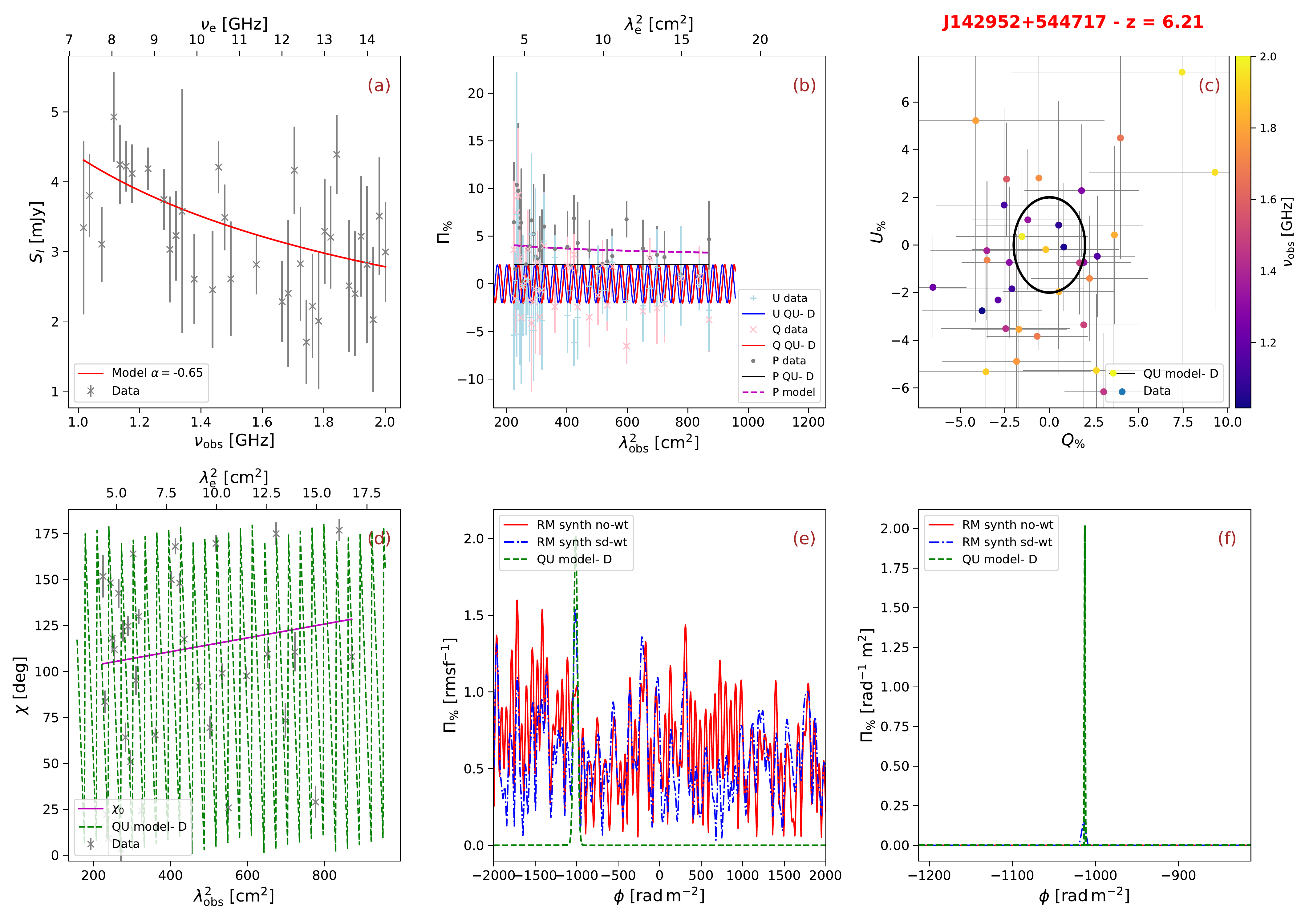}
\caption{As for Fig. 5. Source: J142952+544717}
\label{fig:spec26}
\end{figure*}

\begin{figure*}
\includegraphics[scale=0.35]{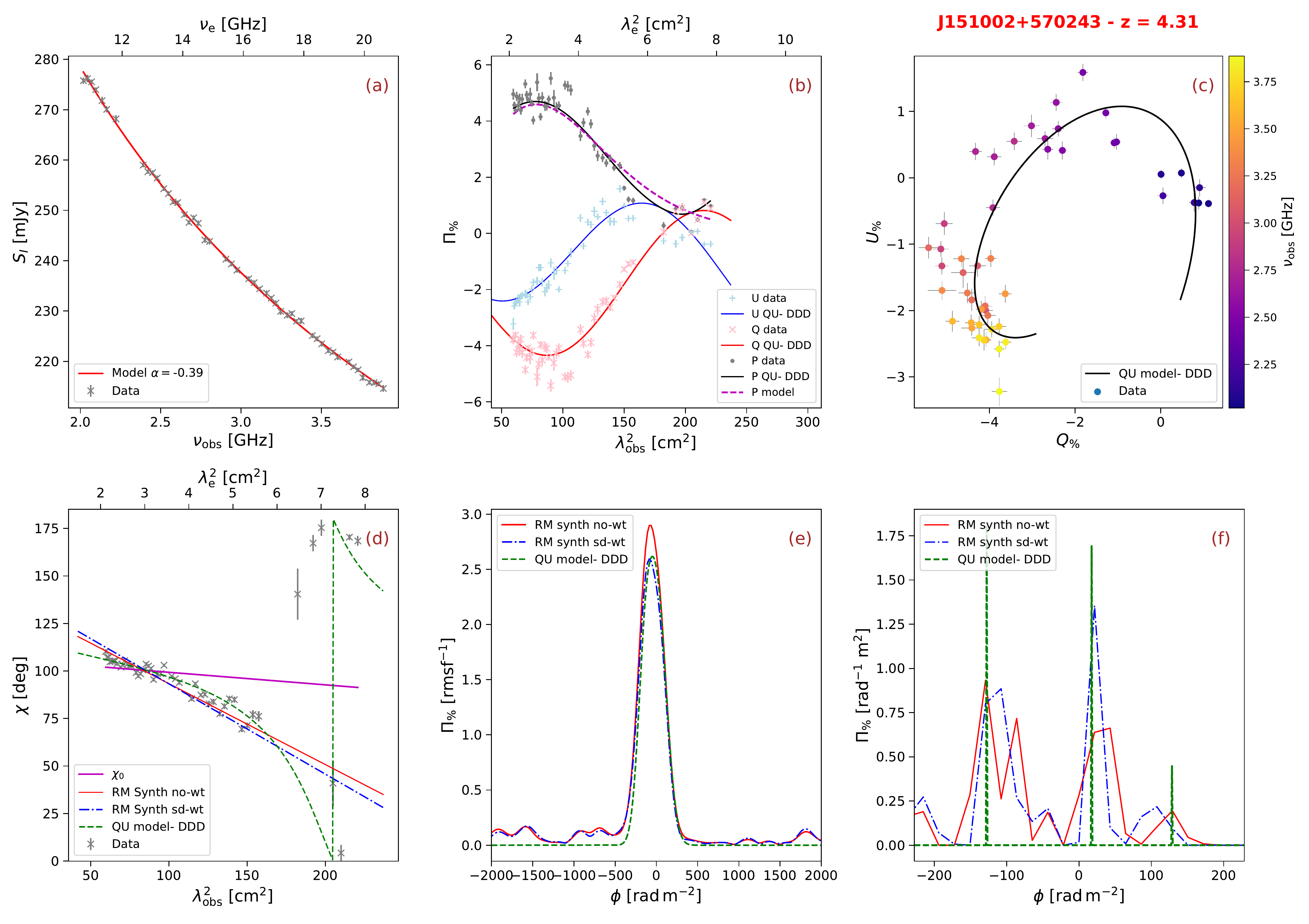}
\caption{As for Fig. 5. Source: J151002+570243}
\label{fig:spec27}
\end{figure*}

\begin{figure*}
\includegraphics[scale=0.35]{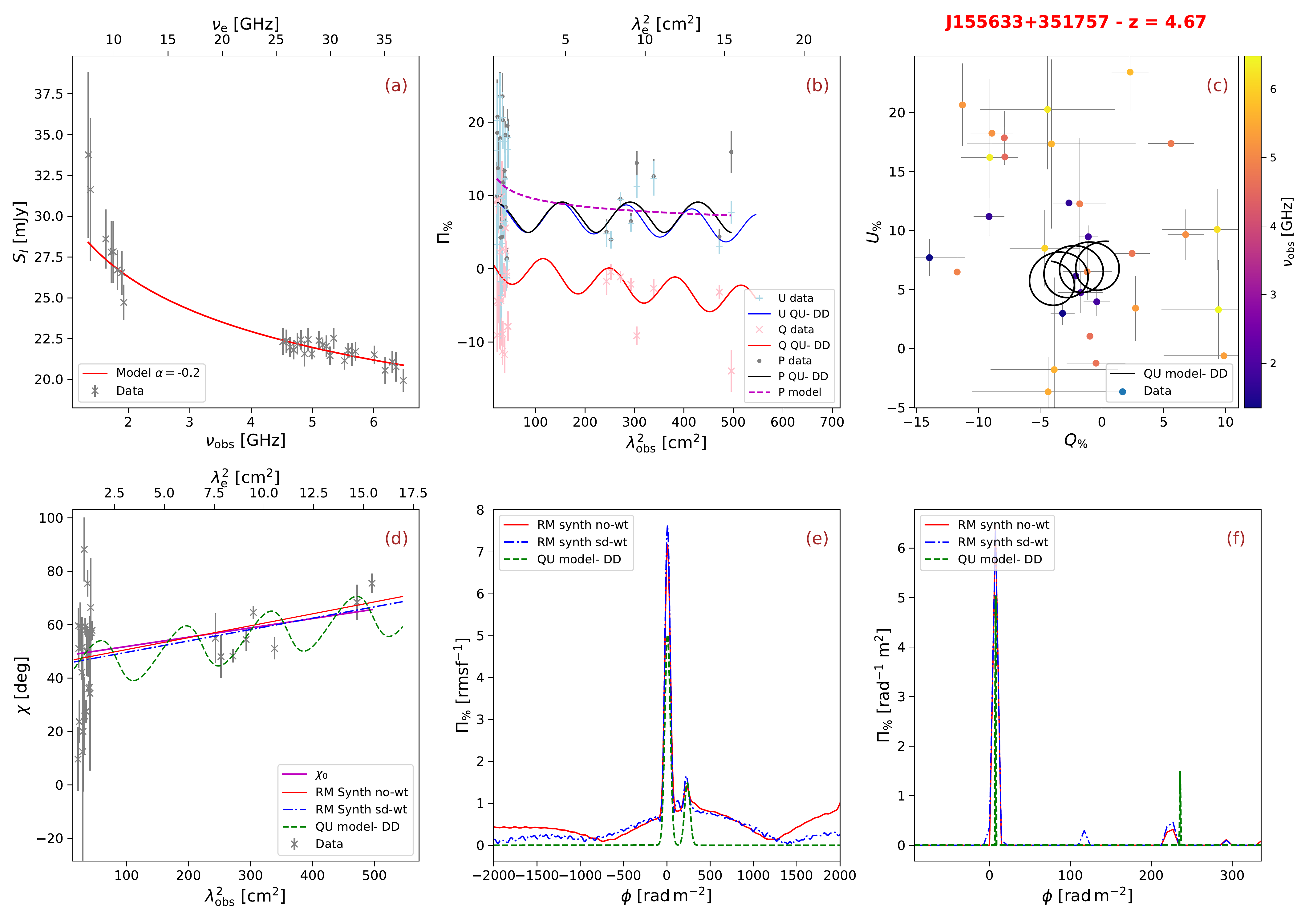}
\caption{As for Fig. 5. Source: J155633+351757}
\label{fig:spec28}
\end{figure*}

\begin{figure*}
\includegraphics[scale=0.35]{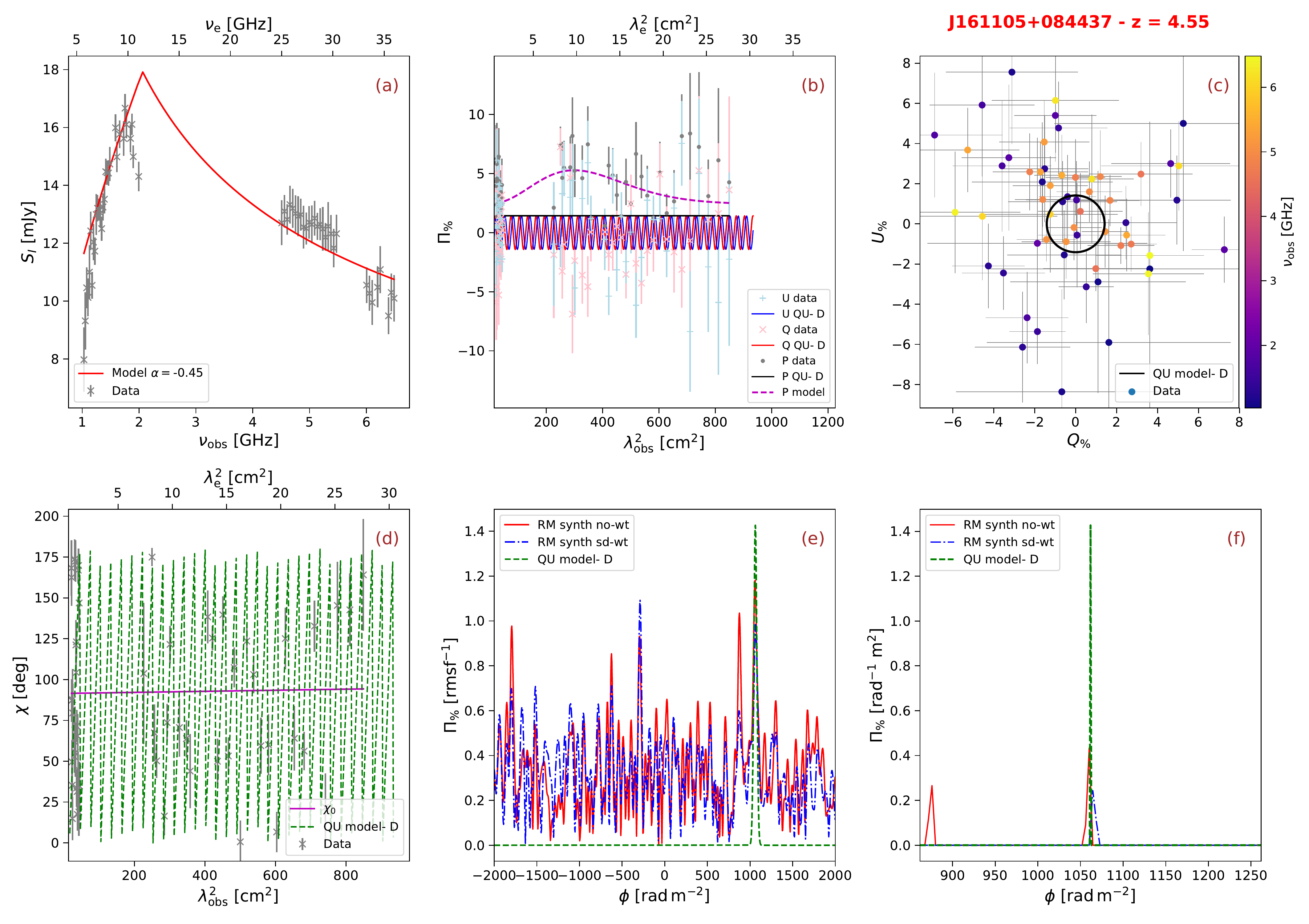}
\caption{As for Fig. 5. Source: J161105+084437}
\label{fig:spec29}
\end{figure*}

\begin{figure*}
\includegraphics[scale=0.35]{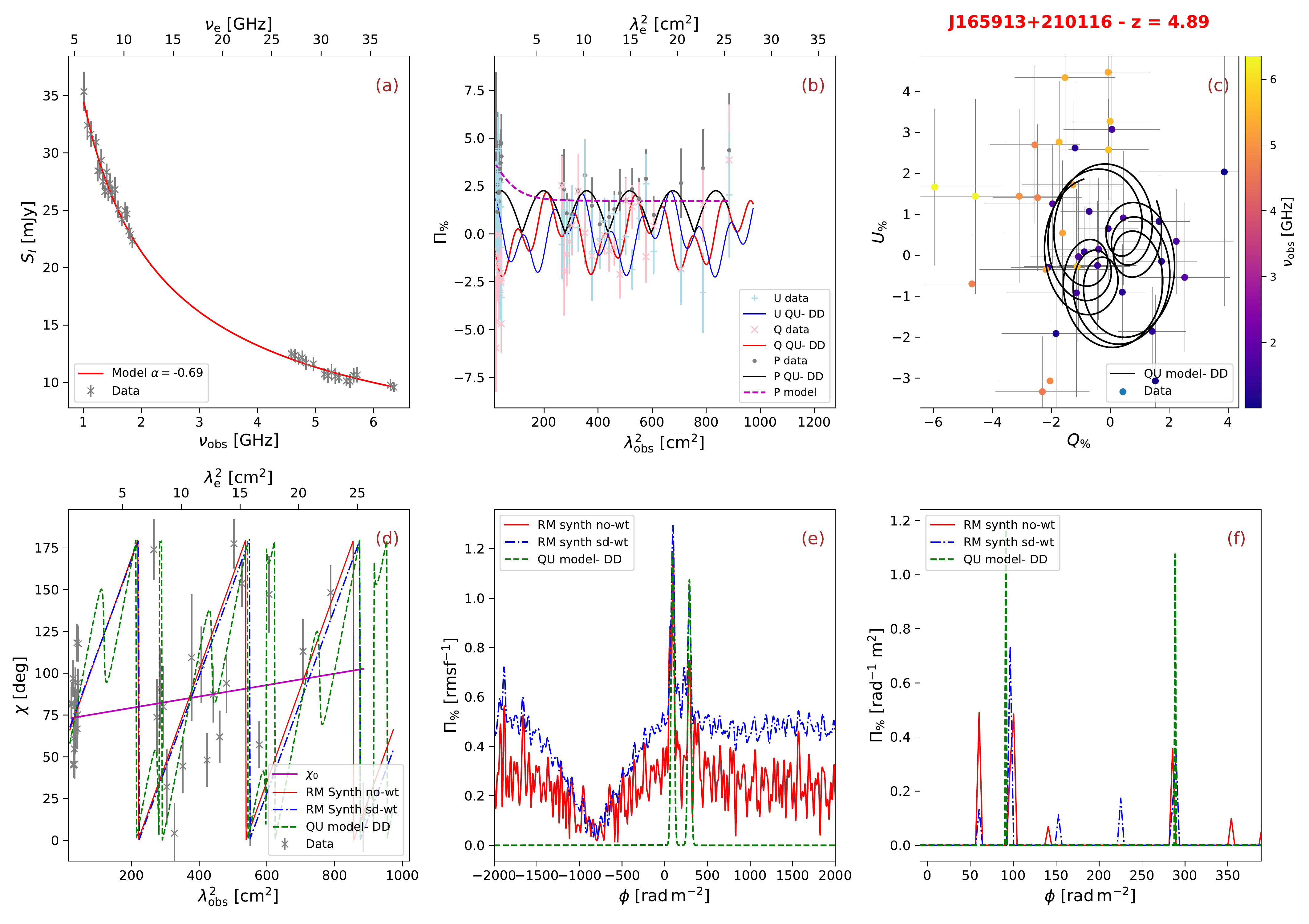}
\caption{As for Fig. 5. Source: J165913+210116}
\label{fig:spec30}
\end{figure*}

\begin{figure*}
\includegraphics[scale=0.35]{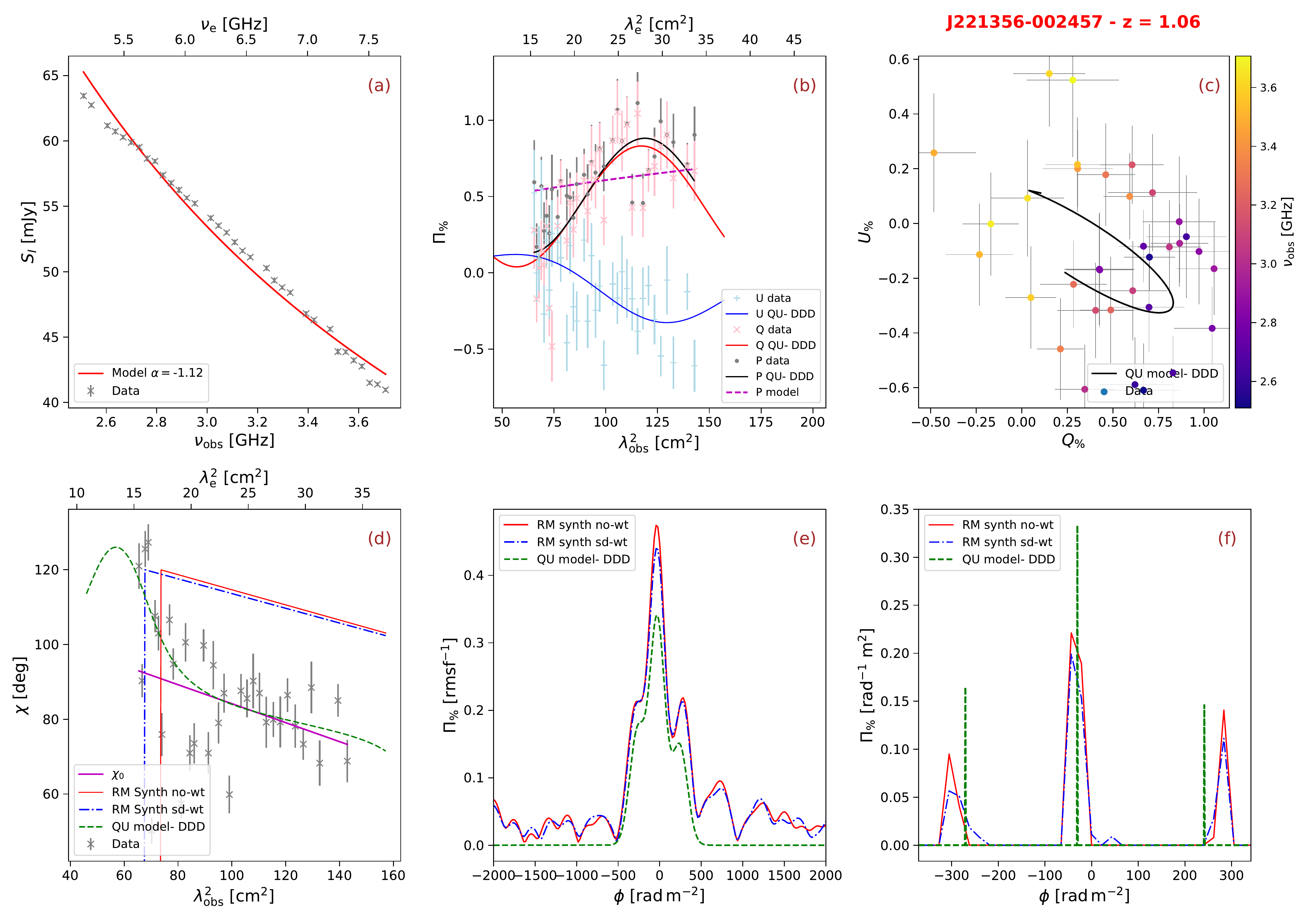}
\caption{As for Fig. 5. Source: J221356$-$002457}
\label{fig:spec31}
\end{figure*}

\begin{figure*}
\includegraphics[scale=0.35]{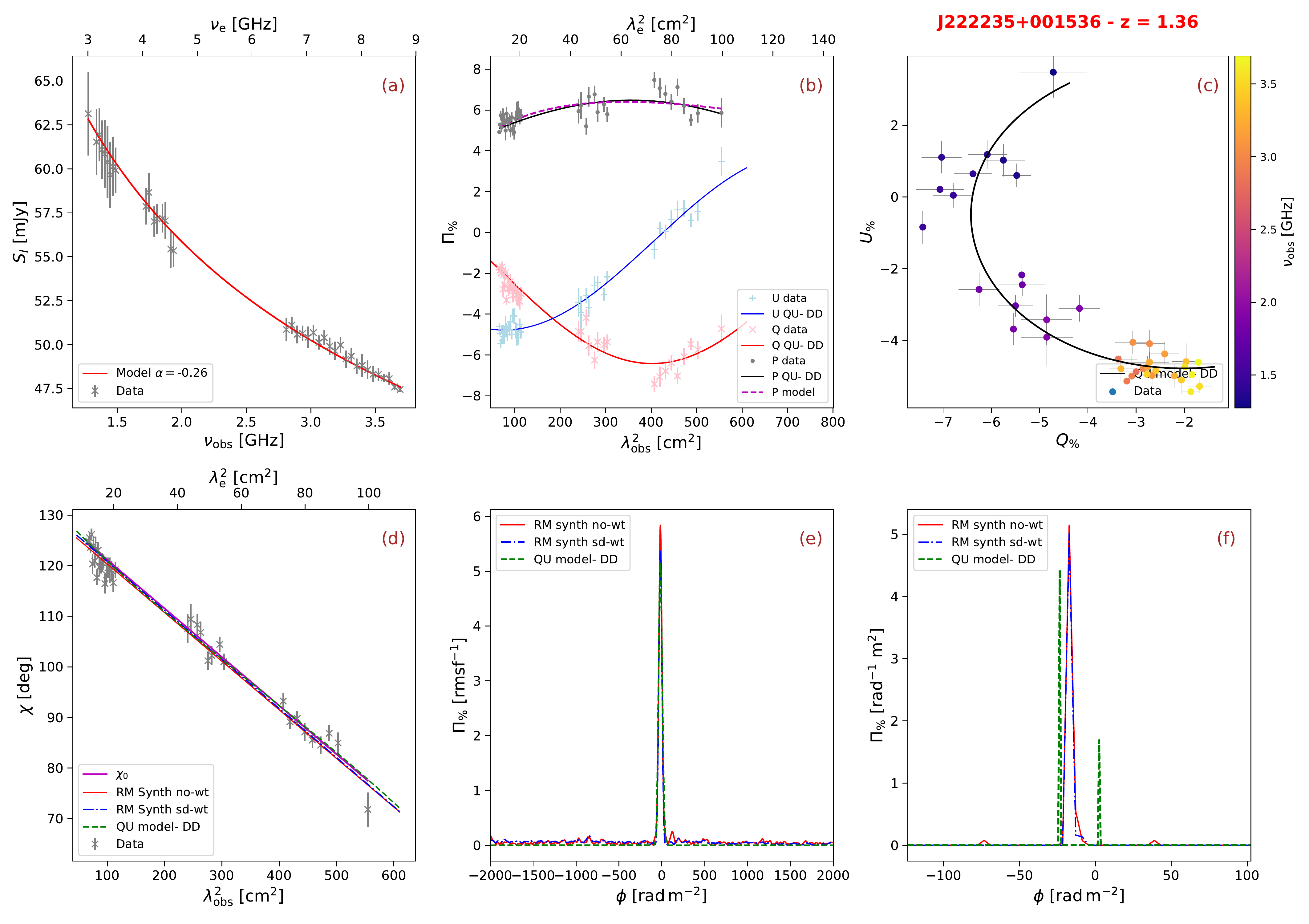}
\caption{As for Fig. 5. Source: J222235+001536}
\label{fig:spec33}
\end{figure*}

\begin{figure*}
\includegraphics[scale=0.35]{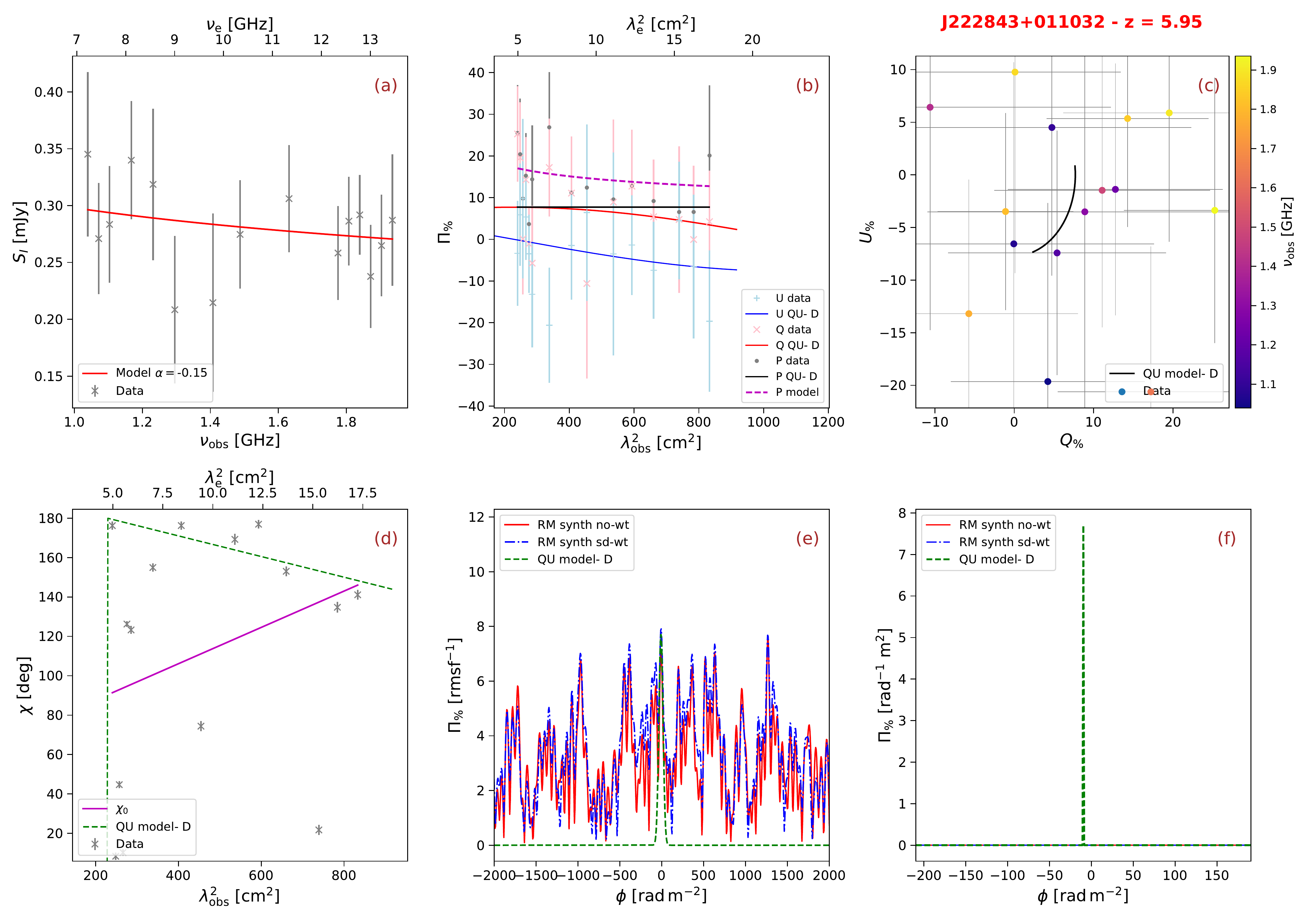}
\caption{As for Fig. 5. Source: J222843+011032}
\label{fig:spec34}
\end{figure*}

\begin{figure*}
\includegraphics[scale=0.35]{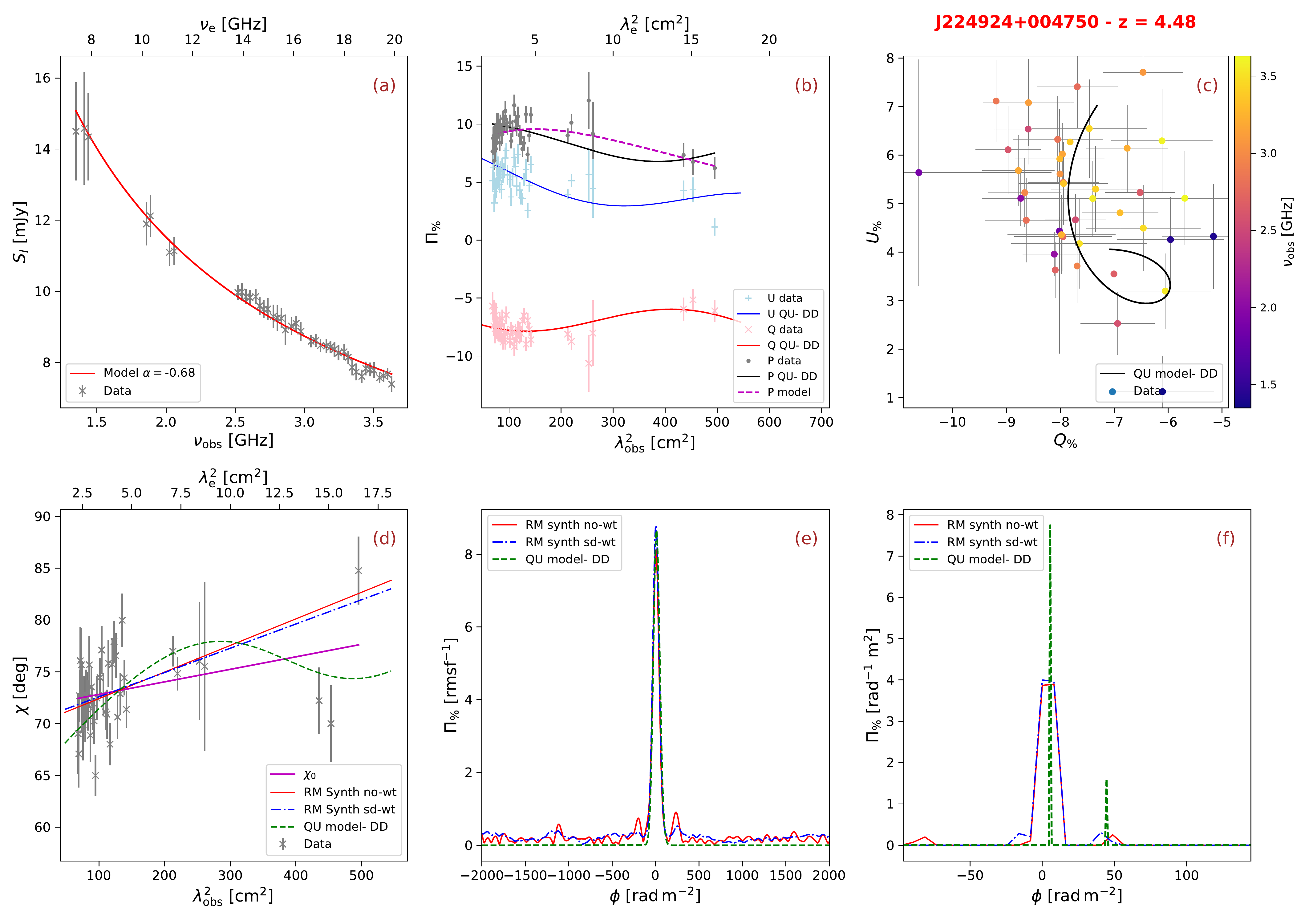}
\caption{As for Fig. 5. Source: J224924+004750}
\label{fig:spec35}
\end{figure*}

\begin{figure*}
\includegraphics[scale=0.35]{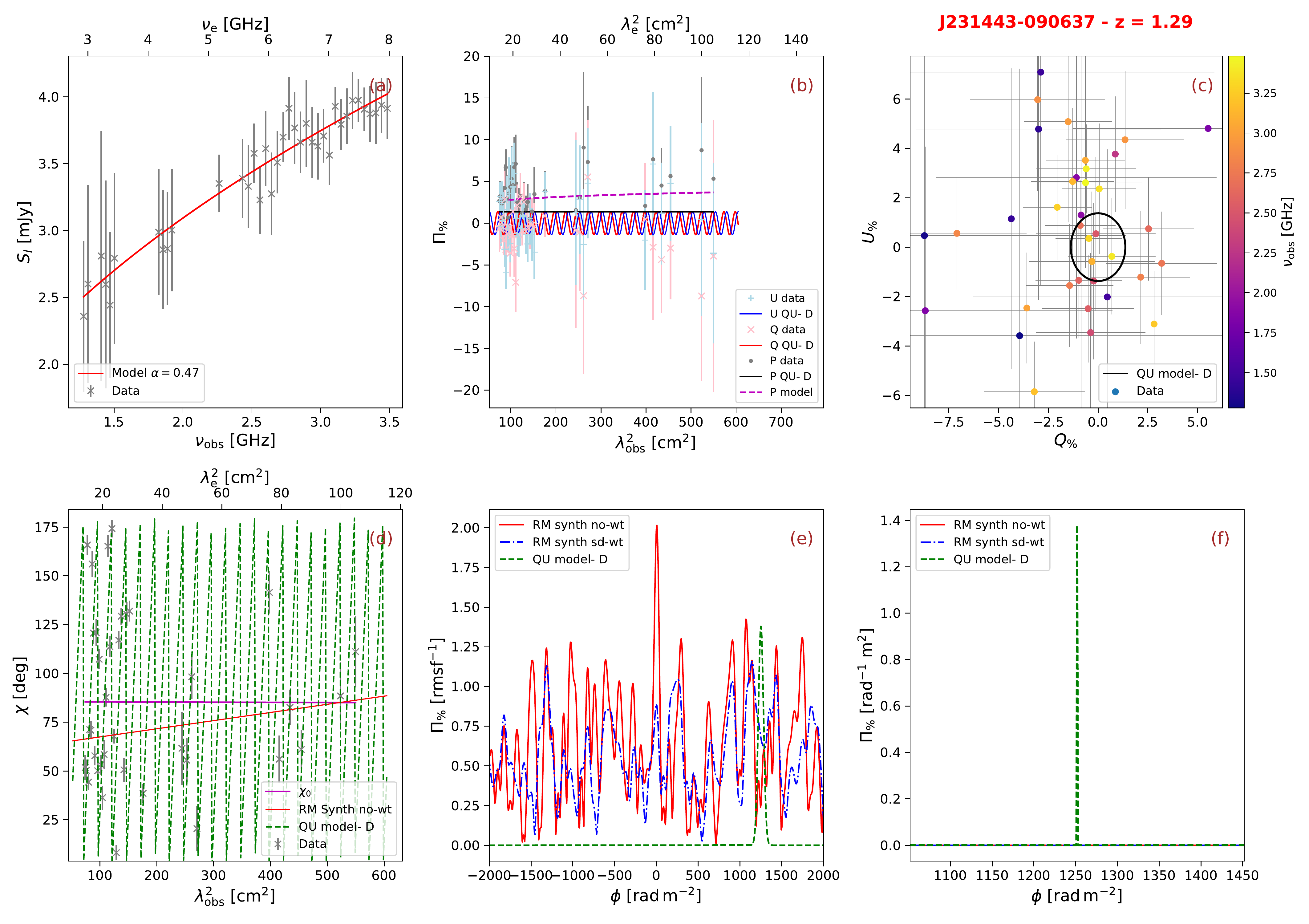}
\caption{As for Fig. 5. Source: J231443$-$090637}
\label{fig:spec36}
\end{figure*}

\begin{figure*}
\includegraphics[scale=0.35]{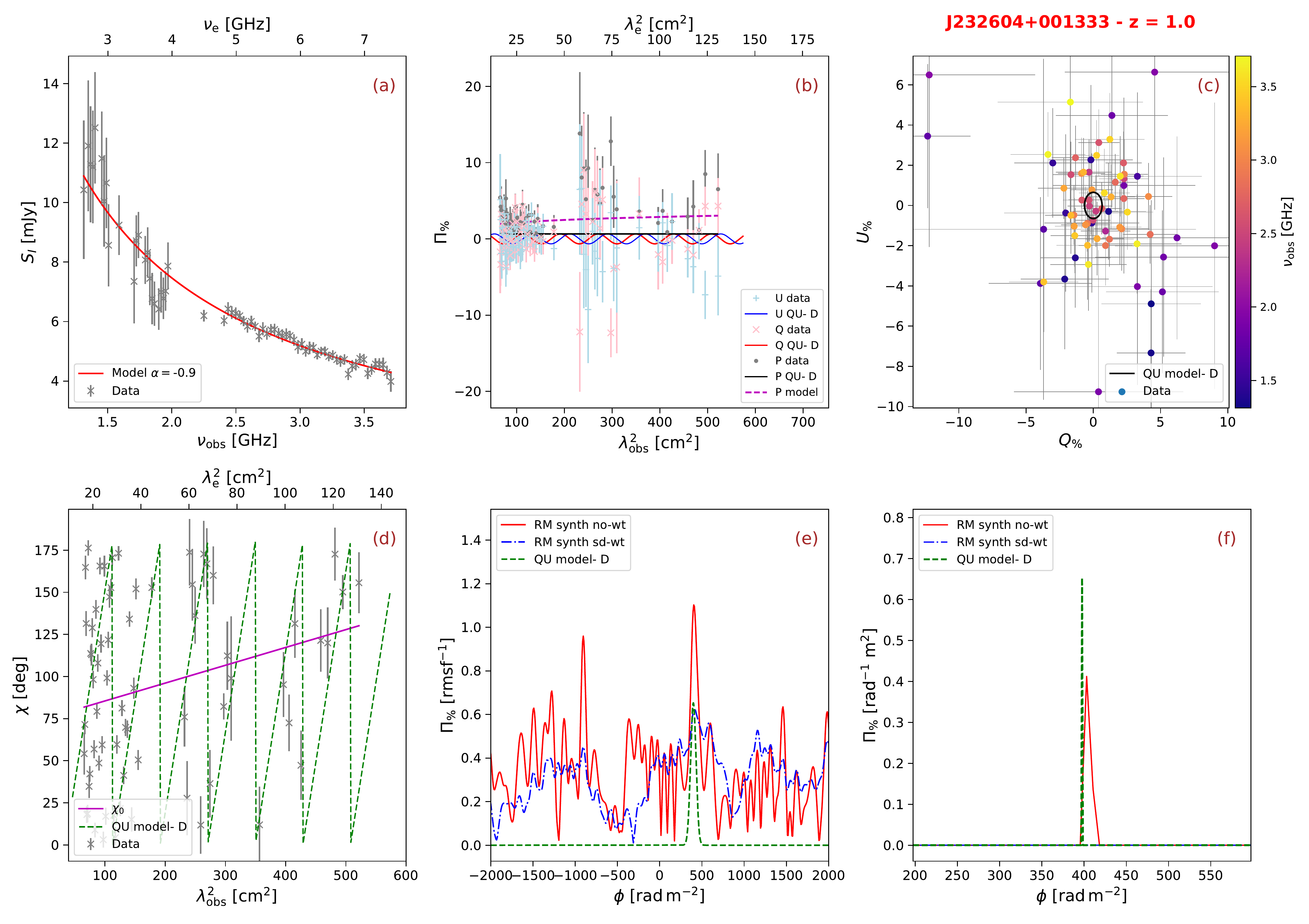}
\caption{As for Fig. 5. Source: J232604+001333}
\label{fig:spec37}
\end{figure*}

\begin{figure*}
\includegraphics[scale=0.35]{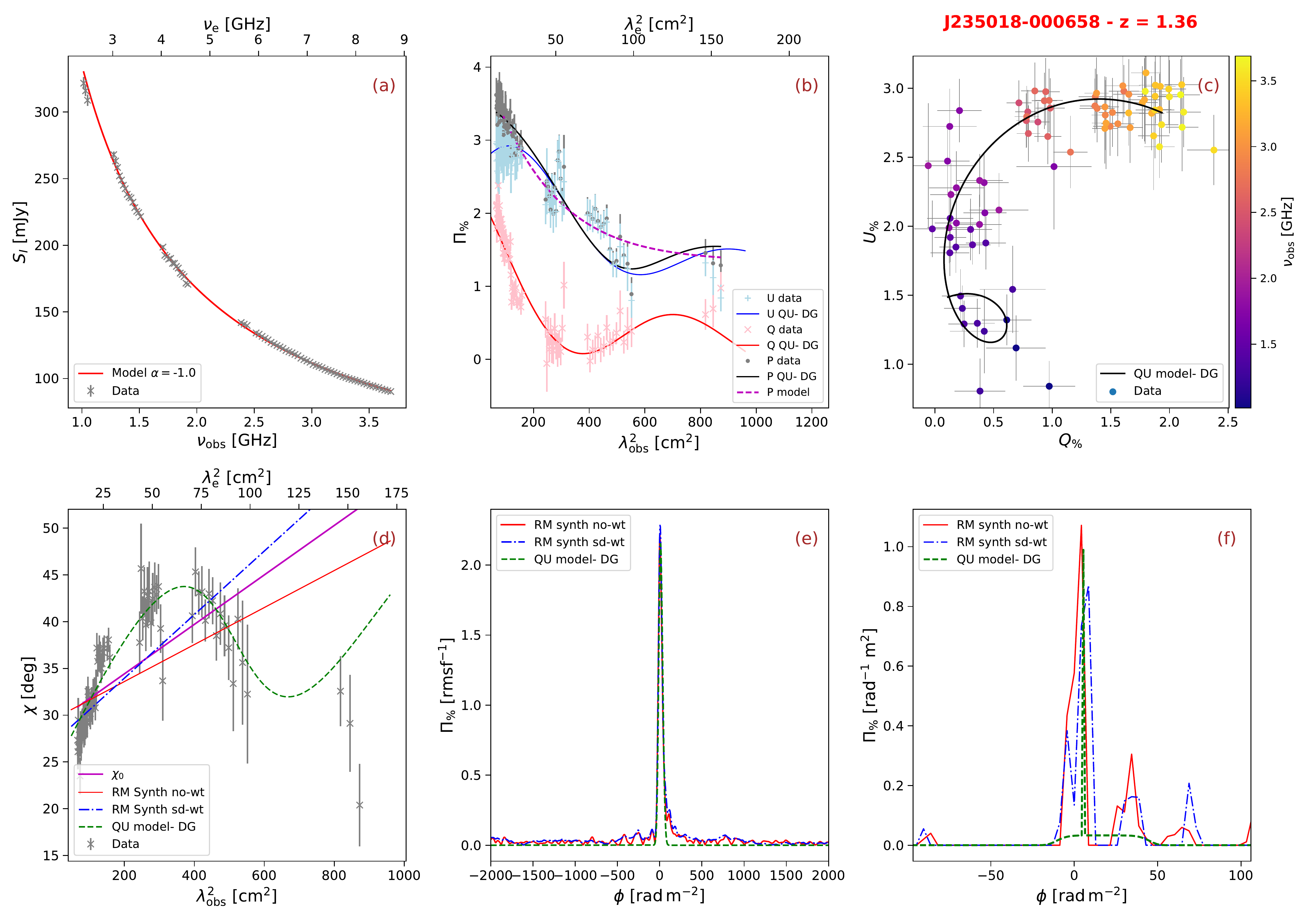}
\caption{As for Fig. 5. Source: J235018$-$000658}
\label{fig:spec38}
\end{figure*}

\label{lastpage}
\end{document}